\documentclass{article}

\usepackage[english]{babel}

\usepackage[a4paper, top=2cm,bottom=2cm,left=3cm,right=3cm,marginparwidth=1.75cm]{geometry}

\usepackage{amsmath}
\usepackage{graphicx}
\usepackage{mathptmx}
\usepackage[utf8]{inputenc}
\usepackage[T1]{fontenc}
\usepackage{tabularx}
\usepackage{xltabular}
\usepackage{float}
\usepackage{amsfonts}
\usepackage{amssymb}
\usepackage{dsfont}
\usepackage[style=apa,backend=biber]{biblatex}
\usepackage{hyperref}
\hypersetup{hidelinks, colorlinks=false}
\usepackage{cleveref}
\usepackage{booktabs}
\usepackage{makecell}
\usepackage{rotating}
\usepackage{multirow}
\usepackage{subcaption}
\usepackage{newunicodechar}
\usepackage{xcolor}
\usepackage{caption}
\usepackage{rotating}
\usepackage{csquotes}
\usepackage{authblk} 
\newunicodechar{̈}{\"{}}

\title{Anticipate, Simulate, Reason (ASR): A Comprehensive Generative AI Framework for Combating Messaging Scams}

\author[1]{Xue Wen Tan}
\author[1]{Kenneth See}
\author[2]{Stanley Kok}
\affil[1]{Asian Institute of Digital Finance, National University of Singapore}
\affil[2]{Department of Information Systems \& Analytics, National University of Singapore}

\begin{document}
\maketitle

\begin{abstract}
The rapid growth of messaging scams creates an escalating challenge for user security and financial safety. In this paper, we present the \textit{Anticipate, Simulate, Reason} (ASR) generative AI framework to enable users to proactively identify and comprehend scams within instant messaging platforms. Using large language models, ASR predicts scammer responses and delivers real-time, interpretable support to end-users. We also develop ScamGPT-J, a domain-specific language model fine-tuned on a new, high-quality dataset of scam conversations covering multiple scam types. Thorough experimental evaluation shows that the ASR framework substantially enhances scam detection, particularly in challenging contexts such as job scams, and uncovers important demographic patterns in user vulnerability and perceptions of AI-generated assistance. Our findings reveal a contradiction where those most at risk are often least receptive to AI support, emphasizing the importance of user-centered design in AI-driven fraud prevention. This work advances both the practical and theoretical foundations for interpretable and human-centered AI systems in combating evolving digital threats.
\end{abstract}

\section{Introduction}
\label{sec:ch3-introduction}
The proliferation of internet connectivity has dramatically increased the frequency of online fraud, with criminals exploiting the convenience and anonymity that digital platforms provide. Recent years have witnessed a particularly alarming rise in scams conducted through instant messaging platforms, resulting in an estimated global loss of US\$1.026 trillion in 2023 alone. Approximately 58\% of these losses are attributed to messaging platform fraud \parencite{GASA2023}. Moreover, these deceptive schemes are becoming increasingly sophisticated, with criminal syndicates adopting cutting-edge technology \parencite{Cross2022} and exploiting current trends \parencite{Wan2022} to enhance both the effectiveness and speed of their fraudulent operations. This escalating threat presents a critical challenge while simultaneously offering a significant opportunity for the Human-Computer Interaction (HCI) field to develop and implement technological countermeasures.

Current HCI literature on fraud prevention reveals several limitations that reinforce the urgency of this challenge. First, simply predicting whether a message constitutes a scam proves insufficient, as most attacks cannot be automatically detected with complete certainty \parencite{zhuo2023sok}. Second, while awareness campaigns and educational initiatives are essential for empowering users to identify and avoid fraudulent schemes \parencite{reinheimer2020investigation, wash2020experts}, their effectiveness remains questionable due to limited empirical evidence supporting best practices \parencite{shillair2022cybersecurity}.

Large-scale awareness campaigns, such as those implemented through public policies or community outreach programs, face additional constraints compared to individual education efforts. Although these campaigns aim to inform the general public about fraud prevention on a broader scale, their impact is often limited. Individuals may struggle to fully comprehend the advice provided or lack sufficient motivation to apply it effectively \parencite{bada2020}. This challenge is further compounded by people's tendency to misjudge their own vulnerability to deceptive schemes \parencite{downs2006decision}. Additionally, while \textcite{aneke2021help} propose an approach that assists users in recognizing phishing attempts by generating explanation messages when threats are detected, this method faces significant practical limitations. Most of these messages and detections remain static due to the unavailability of advanced large language models during that period. Furthermore, the subjective nature of these explanations presents another critical concern, as interpretations can vary significantly among different users and experts, particularly when explanations are designed for scenarios that may not match the user's specific context, potentially leading to misunderstandings or inaccuracies.

The ongoing exploitation of individuals through increasingly sophisticated fraudulent techniques highlights the urgent need for enhanced defensive measures. Recent studies reveal a notable shift in the psychological state of fraud victims \parencite{Dove2020a} which is a dimension that current HCI frameworks inadequately address. Existing HCI methods fall short in helping victims recognize their engagement with fraudulent schemes. Criminals skillfully employ psychological manipulation strategies that disrupt victims' capacity for rational thought, causing affected individuals to become emotionally compromised \parencite{Dove2020a} or trapped in cognitive biases such as the sunk cost fallacy \parencite{VanDijk2003}. In such mental states, logical reasoning becomes impaired \parencite{Nadine2014}, rendering straightforward advice less effective. Documentary evidence exists of cases where bank employees' attempts to discourage clients from transferring money to probable fraudsters proved unsuccessful \parencite{bankscam}.

To address these challenges, we introduce the \textbf{\textit{Anticipate, Simulate, Reason (ASR)} framework}, a novel generative AI framework designed to help potential victims identify and defend against messaging scams. The advancement of Large Language Models (LLMs) presents a unique opportunity to counter scam operations through their ability to generate human-like conversations, specifically to simulate realistic scammer responses. Furthermore, the emergence of reasoning models like DeepSeek R1 \parencite{guo2025deepseek} and ChatGPT reasoning models like o3 and o4-mini \parencite{openai_o3o4} enable these systems to provide comprehensive explanations for why a given interaction exhibits scammer behavior. Our framework is grounded in established psychological principles, specifically Heuristic and Systematic Information Processing \parencite{chaiken1980heuristic} and the Anchoring Effect \parencite{furnham2011literature}, which we will discuss in detail in Section \ref{sec:ch3-solution}. 

This paper investigates three fundamental research questions aligned with the components of our proposed framework: (1) \textbf{Anticipate}: Does providing users with a generative capability to predict scammer responses and a real-time scam classification score enhance their ability to identify scam conversations and what is the perceived helpfulness of the scam ? (2) \textbf{Simulate}: How can we create or fine-tune a large language model (LLM) to effectively simulate scammer behavior? (3) \textbf{Reason}: Do reasoning model explanations that articulate their thought processes help educate users about scammer tactics and improve their scam detection skills? We address these questions through qualitative and quantitative assessments through surveys and interactive platform experiments. The key contributions of our paper are as follows:

\begin{itemize}
    \item We present the \textit{Anticipate, Simulate, Reason} (ASR) framework, a novel user-centered generative AI approach that leverages large language models to anticipate scammer responses, simulate realistic scams, and provide real-time, explainable support for scam detection.
    \item We introduce ScamGPT-J, a fine-tuned language model for scam simulation, together with a high-quality, synthetic dataset of scam conversations that addresses critical gaps in conversational fraud detection.
    \item Our comprehensive experiments demonstrate that the ASR framework significantly improves scam detection, especially in challenging scenarios such as job scams. We also uncover important demographic insights, including a troubling paradox where the most vulnerable users are often the least receptive to AI assistance.
\end{itemize}

\section{Literature Review}
\label{sec:chp3-literature}
\subsection{Scam Detection and Prevention}

While email spam detection has evolved significantly, with modern solutions achieving over 90\% accuracy \parencite{Dedeturk2020, Saidaini2020}, detecting fraudulent schemes within instant messaging contexts presents unique challenges. A major obstacle is the lack of high-quality datasets, which hampers the development of effective detection methods \parencite{shafi2017review}. Email spam detection faced similar challenges, leading to the development of resources such as the NUS SMS Corpus \parencite{NUS-Scam-Data} and the UCI SMS Spam Collection \parencite{UCI-Scam-Data}. These datasets have facilitated preliminary applications of machine learning techniques for distinguishing spam text messages \parencite{kumar2024legitimate, wijaya2023spam}. However, it is paramount to recognize the fundamental differences between spam and fraudulent messages. Spam messages are generally unsolicited communications that, while annoying, are relatively harmless. In contrast, fraudulent messages are designed to manipulate victims into actions that result in financial loss, with consequences that extend far beyond monetary damage. Financial losses can lead to significant emotional and psychological stress, strain relationships, and even cause family breakdowns \parencite{kassem2024fraud}. In extreme cases, these stresses can trigger mental health issues such as anxiety, depression, or suicidal thoughts \parencite{whitty2016online}.

Fraudulent schemes continually evolve as perpetrators adapt to new trends and vulnerabilities, often employing sophisticated social engineering techniques \parencite{zhang2022vulnerability}. This adaptability leads to extensive socio-economic damage, with global financial losses from such schemes now exceeding half a trillion US dollars annually \parencite{GASA2023}. Accurately predicting whether a text conversation constitutes fraud is further complicated by the prevalent use of abbreviations, emoticons, and misspellings in instant messaging, which hinders effective detection \parencite{mishra2023dsmishsms}. While \textcite{mishra2023dsmishsms} propose a model for phishing SMS detection, the broader spectrum of fraudulent messages remains largely unexplored, highlighting a significant gap in current research and underscoring the need for comprehensive fraud message datasets. Furthermore, many existing detection approaches overlook the conversational context in which fraudulent messages occur. Scam messages might appear innocuous in isolation but reveal their deceptive nature when viewed as part of an ongoing conversation. The importance of analyzing entire conversations to detect and understand fraudulent schemes is highlighted by \textcite{edwards2017scamming}. \textcite{derakhshan2021detecting} demonstrate that telephone-based fraud can be identified with high accuracy using fraud signatures which is defined as sets of text phrases that collectively fulfill the perpetrator's goal. These signatures can be formulated manually through known fraud characteristics or identified through clustering of known fraudulent conversations. Beyond detection, the literature explores various prevention methods. \textcite{martin2018signal} conduct a field study showing that individuals are more vulnerable to targeted phishing attacks compared to general ones. They utilize equal-variance signal detection theory to model susceptibility to phishing scams, suggesting the need for differentiated prevention techniques based on individual risk profiles. Another countermeasure involves using social engineering against perpetrators \parencite{canham2022planting}. \textcite{edwards2017scamming} explore this approach, highlighting the trend of "scambaiting," where individuals engage with criminals to waste their resources.

Existing scam detection literature predominantly focuses on classifying texts in isolation, failing to account for conversational dynamics. Furthermore, current prevention methods rely on prescriptive approaches that simply inform users whether a message is fraudulent or not, rather than empowering users to develop their own detection capabilities. These limitations highlight the need for more sophisticated, user-centered approaches that consider the conversational nature of modern fraud schemes. Our proposed ASR framework addresses these challenges through a fundamentally different approach. Unlike existing methods that analyze isolated messages, our framework analyzes entire conversations. The LLM model learns to anticipate criminal behavior patterns, understanding not just what fraudulent messages look like, but how they evolve within conversations. This generative approach contrasts with existing methods that rely on keywords or isolated sentences to predict fraud. By using historical conversation context to generate subsequent dialogue, our approach more effectively captures the interactive nature of fraudulent schemes. This user-centered solution empowers individuals to perform fraud detection themselves, moving beyond prescriptive approaches toward a more interactive and educational framework.

\subsection{Employing LLMs to Guide Human Behavior}

\textcite{fogg1998persuasive} suggests that computers can function as persuasive tools to influence individual behavior. The deployment of such tools is not without merit, as they are capable of shaping desirable behavior. However, this influence raises ethical considerations that must not be overlooked. \textcite{acemoglu2025big} formalize \textcite{fogg1998persuasive} theory in a model that shows how artificial intelligence (AI) tools could offer consumer benefits under limited usage but can result in behavior manipulation when abused.

An emerging perspective in the field of HCI concerning AI's persuasive abilities is the study of reflection theory. \textcite{abdel2023ai} explore how AI-based systems can encourage reflective thinking among medical practitioners. Their findings suggest that when an AI system presents information that challenges a practitioner's existing beliefs and sustains engagement over time, it can create deep cognitive dissonance, leading to reflection rather than defensiveness. They introduce a Machine-Induced Reflection Model, which demonstrates how AI can guide users through a reflective process, ultimately helping them make more confident and thoughtful decisions, thereby enhancing their professional effectiveness. Their work contributes to the "massive rethinking" of Information Systems theory \parencite{teodorescu2021failures} by highlighting the potential of machine-induced reflections to foster a more contemplative approach to work. They also observe that while decision support and recommender systems are typically designed to aid decision-making and persuasion, they can sometimes introduce biases that undermine decision quality, such as automation bias \parencite{goddard2012automation}. In contrast, AI systems explicitly designed to promote reflection offer a promising alternative. Although their study focuses on the healthcare sector, their theoretical framework can be extended to scenarios such as scam detection, where individuals must determine if they are interacting with a scammer. Our system could similarly prompt users to "rethink" their decisions when they are at risk of falling for a scam, thereby reducing the likelihood of becoming victims and increasing their confidence in their judgment.

In a similar vein, LLMs can be conceptualized as an extension of these persuasive technologies, guiding individual actions through sophisticated language processing capabilities. Despite their inability to mimic the full spectrum of human reasoning, the potential of LLMs is noteworthy in their capacity to accurately discern sentiment in text \parencite{tabone2023using}. This capability enables the generation of contextually appropriate responses, making them powerful tools in conversational and behavioral guidance contexts. While originally designed for general use, LLMs can be tailored for specific applications \parencite{demszky2023using}. It is important to note that off-the-shelf LLMs may fall short in specialized scenarios, necessitating further training via methods such as fine-tuning or prompt-tuning.

The capacity of LLMs to influence societal behavior has been exploited for both beneficial and malevolent purposes. Documented instances reveal the use of bots leveraging LLMs to produce human-like content on platforms like Twitter (now known as X), and to orchestrate campaigns for questionable websites \parencite{yang2023anatomy}. Additionally, there have been instances where LLMs have played a key role in generating misinformation, with state-sponsored entities utilizing them for disinformation campaigns to advance national agendas \parencite{ezzeddine2023exposing}. Despite these nefarious applications, LLMs also hold the potential for positive impact. \textcite{chen2024combating} recognize the detrimental effects of LLM-generated misinformation but argue that these models can act as a double-edged sword. They propose that LLMs can be employed to counteract misinformation by facilitating rapid detection of false information. Ongoing research highlights the growing emphasis on using LLMs for fact verification, offering individuals timely and accurate feedback on the information they encounter \parencite{cheung2023factllama, zhang2023towards}.

In line with the theories presented by \textcite{fogg1998persuasive} and \textcite{acemoglu2025big}, we propose the ASR framework as a practical embodiment of LLMs as persuasive tools, with a specific focus on guiding human behavior in the context of scam prevention. Our work adds to the literature on specific-use LLMs by offering an LLM as a persuasive technology for increasing individual resilience against scams.
\section{Dataset Construction and Curation}
\label{sec:ch3-datasets}

\begin{figure}[htb!]
\centering
\includegraphics[width=0.8\linewidth]{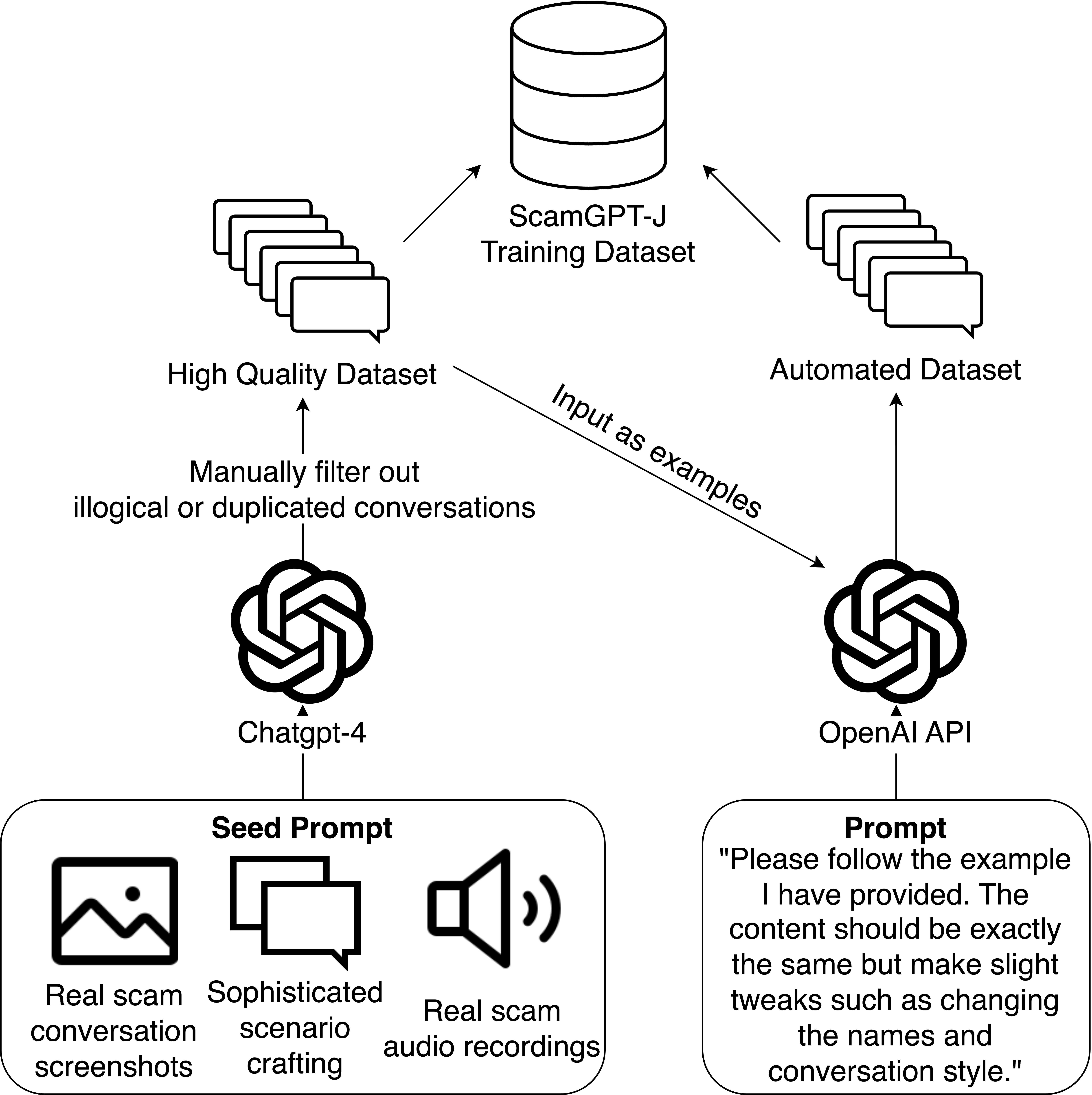}
\caption{ScamGPT-J training dataset generation pipeline.}
\label{fig:dataset}
\end{figure}

Before introducing our ASR framework, we first detail the process used to construct and curate our scam dataset. The primary objective was to create a high-quality corpus of scam-related conversations for fine-tuning in the \textit{Simulate} component. Our dataset development involved both synthetic generation and the careful incorporation of real-world data.

We initiated the process by employing ChatGPT-4 to generate an initial set of 90 conversation samples. These samples were crafted using a combination of prompts: (1) screenshots of authentic scam messages sourced from news articles and police advisories, and (2) sophisticated, hand-constructed scenarios inspired by real scam interactions that we manually collected. The synthetic dialogues were designed to capture four major scam categories: \textit{(1) Authority, (2) Job, (3) Love, and (4) Investment}.

To expand the dataset and enhance its diversity, we leveraged OpenAI’s GPT-3.5 Turbo via its API. Each of the original 90 conversations was used as a seed to generate 10 additional variants, systematically altering features such as names, conversational tone, and style while preserving the essential context and scam mechanism. This automated augmentation helped enrich the dataset, capturing subtle variations in scam tactics and linguistic nuance. Nevertheless, not all outputs met our quality standards due to occasional inconsistencies inherent to large language model generations. To address this, a manual vetting process was conducted: low-quality or unrealistic conversations were either corrected or discarded, ensuring only high-quality data were retained.

Although we opted to generate the majority of the dataset, this decision does not preclude the manual collection of real scam messages. We collected the real scam messages by receiving text messages from scammers and prompting them to converse further to capture more of their conversation patterns and nuances. These real-world conversations were collected through direct engagement with scammers, extending the dialogue to capture natural conversational patterns. However, the volume of real scam messages was still insufficient for large-scale model training. Furthermore, the use of genuine communications introduces significant privacy and ethical considerations, as such data is often sensitive or confidential. To address both scarcity and privacy concerns \parencite{reiter2011data}, these real conversations were primarily used as seed data to guide the generation of synthetic variants \parencite{endres2022synthetic}. This approach enabled us to build a comprehensive, diverse dataset while mitigating ethical and legal risks associated with using real-world personal data.

In total, our dataset comprises 902 scam-related conversations, of which 812 are allocated for training and 90 are reserved for validation, thereby supporting accurate assessment of model performance and generalization.

\section{Our "Anticipate, Simulate, Reason" Generative AI Framework}
\label{sec:ch3-solution}
We will be presenting our \textit{Anticipate, Simulate, Reason} (ASR) Generative AI Framework in this section. It comprises of three interconnected components that work together to enhance scam resilience. Our strategy focuses on implementing proactive and non-intrusive protective measures that strengthen individuals' ability to recognize and resist fraudulent schemes. By leveraging the capabilities of generative AI and foundational principles of persuasive technology, our approach serves dual purposes: providing robust safeguards against deceptive practices while simultaneously offering educational resources that empower users with knowledge and awareness. The following subsections examine each component of the ASR framework in detail, demonstrating how this integrated approach creates a comprehensive defense against increasingly sophisticated scam tactics.

\subsection{Anticipate}
\label{subsec:anticipate}
\subsubsection{Overview \& Motivation}

The \textit{Anticipate} component creates an interactive system involving the user, a scambot (a Large Language Model trained or prompted to replicate scammer behavior), and a potential scammer. This system leverages the generative capabilities and language understanding of LLMs to mimic scammer responses in real-time while providing real-time scam classification based on the ongoing conversation. During an actual scam attempt, the scambot generates believable replies that closely match the scammer's responses, resulting in a high scam classification score. In contrast, during legitimate interactions, the scambot continues to produce scam-like replies that appear unnatural and contrast sharply with the genuine responses, resulting in a low scam classification score and thereby revealing the authentic nature of the conversation.

\begin{figure}[htb!]
  \centering
  \includegraphics[width=0.6\linewidth]{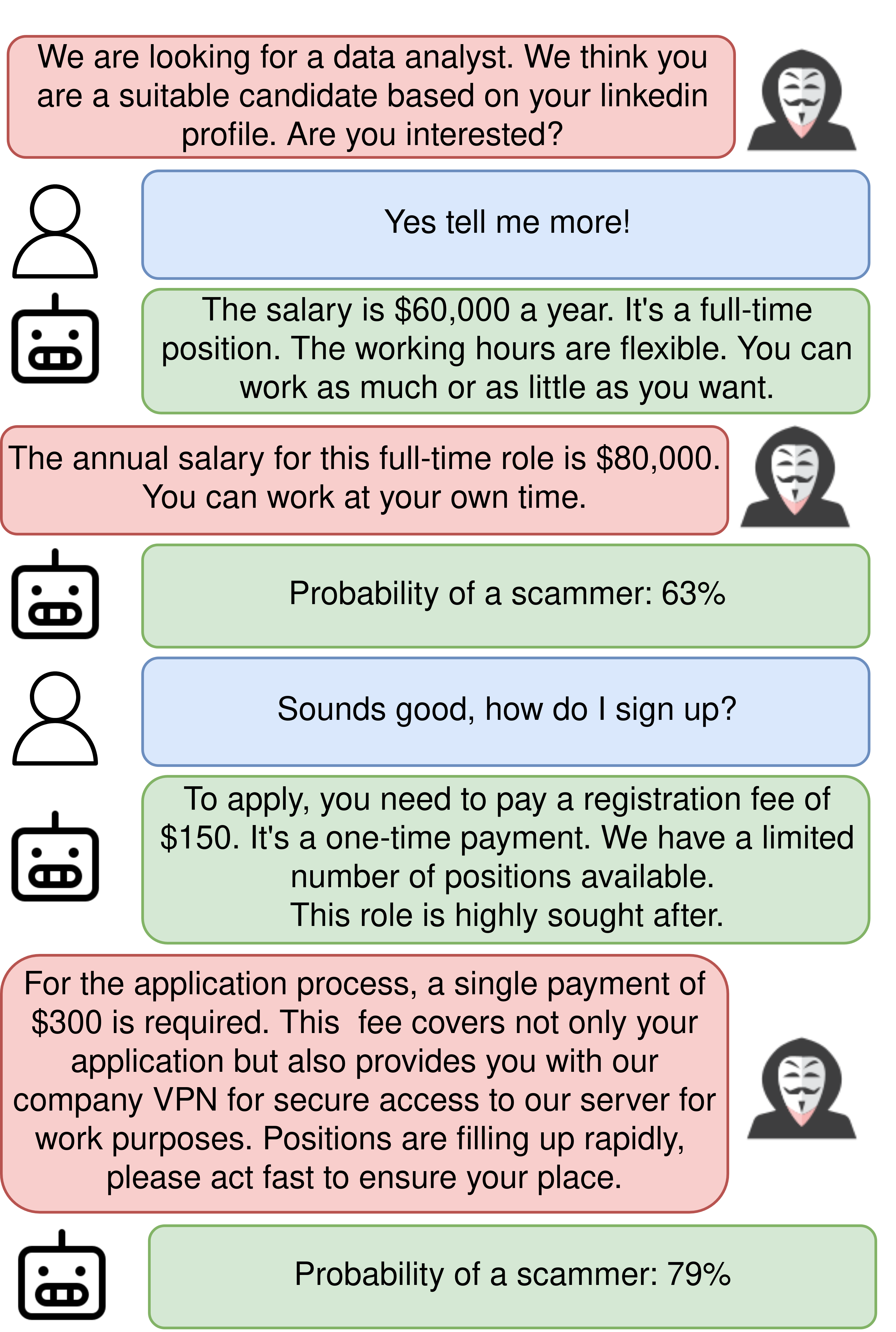}
  \caption{An illustrative dialogue demonstrating a job scam operation. The scammer's messages are shown in red, the potential victim's responses in blue, and the scambot's interjections in green. The scambot responses are hidden from the scammer and visible only to the user (potential victim).}
  \label{fig:overview}
\end{figure}

Since users are aware that the scambot is designed to simulate scammer behavior, they can carefully compare responses from both the potential scammer and the scambot. This informed perspective enables users to make more accurate judgments about whether a conversation is likely to be a scam. Figure \ref{fig:overview} graphically illustrates the communication dynamics among a scammer, a potential victim, and the scambot. The scambot's generated content is \textit{visible only to the user}.

We hypothesize that the scambot's predictive capabilities will significantly enhance the user's ability to identify deceptive conversations. By providing users with predictions of what the scammer may say next and continuously offering reminders through high scam classification scores, our approach addresses the limitations of traditional models that merely label interactions as "scam" or "not a scam" or abruptly terminate conversations once scam detection is triggered. Such abrupt interventions can leave users confused and unprepared for future encounters, whereas our approach empowers users to exercise their own judgment while being guided by real-time insights. Additionally, these conventional approaches often fail to capture the nuanced dynamics of evolving scam tactics \parencite{kawintiranon2022traditional}, providing victims with only generalized warnings that lack the specific contextual information needed (as discussed in Section \ref{sec:ch3-introduction}) to recognize sophisticated deceptive strategies.

In contrast, our scambot framework \textbf{anticipates} the scammer's likely responses, which provides tangible evidence of a scam in progress. This forward-looking approach does more than simply signal potential danger, as it aligns with the \textit{Heuristic-Systematic Model of Information Processing} theory from social psychology \parencite{chaiken1980heuristic}. This theory suggests that individuals are more susceptible to heuristic processing, which requires less cognitive effort and relies on non-content cues rather than message content, particularly when they exhibit low involvement with the subject matter. Non-content cues refer to factors such as emotional appeals and presentation style, which influence decision-making independently of the message's actual content. By providing concrete examples of anticipated scammer responses, the scambot enables users to engage in systematic processing, thereby increasing their awareness of deceptive tactics and catalyzing their shift toward content-focused analysis. Content cues pertain to the logical, factual, and argumentative quality of a message, including its factual data and relevance. These cues require detailed cognitive engagement, as they focus on critically analyzing the substance and validity of the information presented. When individuals engage in systematic processing, they focus on these content cues, analyzing and critically evaluating the message based on its merits and logical coherence. This contrasts with heuristic processing, which relies primarily on non-content cues. Under this framework, individuals are better equipped to analyze and detect evidence of deception, enabling them to recognize when they are potentially being scammed.

Additionally, we believe this anticipatory component plays an important role in counteracting the anchoring effect \parencite{furnham2011literature} that scammers may instill in their victims. The anchoring effect is a cognitive bias wherein individuals latch onto initial and incomplete information presented to them \parencite{tversky1974judgment}, forming an "anchor" that skews their perception of plausible realities. This effect is often exacerbated by \textit{Behavioral Confirmation Bias} \parencite{rosenthal1968pygmalion}, where individuals' expectations shape their perceptions and actions, leading them to interpret ambiguous information in ways that confirm their initial beliefs. As a result, individuals seek ways to validate what they believe to be the anchored reality, further cementing the effect \parencite{chapman1999anchoring, mussweiler2001semantics}. In the context of instant messaging scams, the anchoring effect manifests as victims forming beliefs from early interactions that their messaging counterpart is genuine. This explains why victims often refuse to believe they are involved in a scam, even when presented with contradictory evidence \parencite{straitsrefused2, straitsrefused}.

The scambot mitigates this anchoring effect by employing a variation of the \textit{consider-the-opposite} strategy proposed by \textcite{mussweiler2000overcoming}. This strategy involves consistently presenting individuals with anchor-inconsistent knowledge, which is credible information that contradicts the initial beliefs forming the anchor, thereby reducing their confidence in their initial assumptions. As individuals observe that the scambot which is designed to mimic scammers, consistently produces predictions that accurately reflect scammer behavior, they may gradually reconsider the genuineness of the person they are interacting with. Consequently, their defenses, which may have been lowered due to the anchoring effect, are raised, providing better protection against potential scams. 

\begin{figure}[ht]
    \centering
    \includegraphics[width=\linewidth]{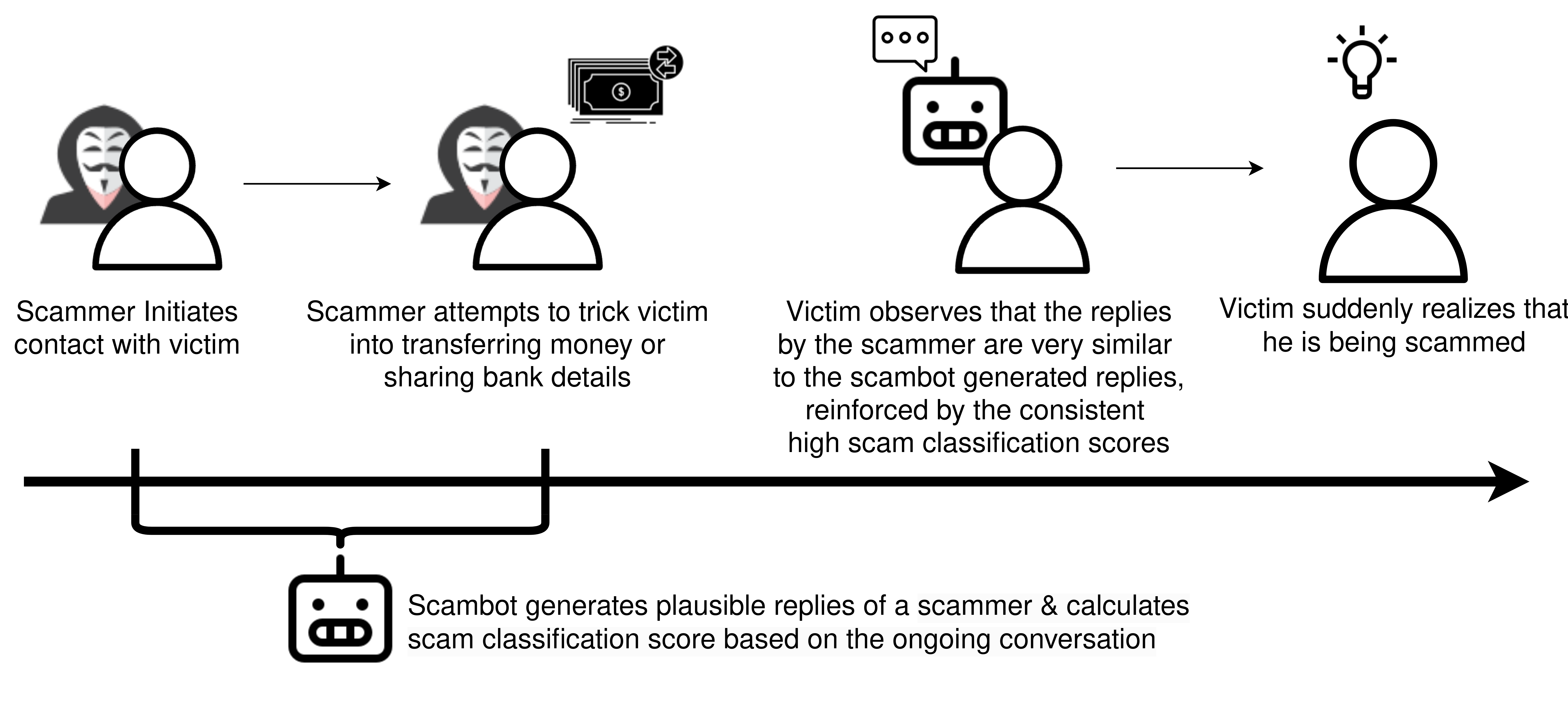}
    \caption{Expected outcome in which a potential victim becomes aware, with the help of the scambot, that they are being targeted by a scammer.}
    \label{fig:happyflow}
\end{figure}

In essence, our hypothesis posits that the anticipation component adds significant depth to scam analysis, transforming how potential victims perceive and react to deceptive tactics. It moves beyond mere detection to offer a persuasive, educational experience that empowers individuals to recognize and evade scams. This approach could have a profound psychological impact, shifting users from a state of passive awareness to active defense against online fraud. Figure \ref{fig:happyflow} provides a visual narrative of the interaction between a scammer, a potential victim, and the Generative AI-based intervention tool. It illustrates a four-step process beginning with a scammer initiating contact with a potential victim.
\vspace{-0.8em}
\begin{itemize}
  \item The scammer initiates contact and builds deceptive rapport with the victim.
  \item The scammer attempts to have the victim transfer money or divulge sensitive financial information.
  \item The victim, with the assistance of the scambot, observes that the scammer's replies closely resemble the responses predicted by the AI system, reinforced by consistently high scam classification scores.
  \item The victim recognizes the shared patterns between the scammer's and scambot's responses and realizes they are being targeted by a scam.
\end{itemize}

\subsubsection{Experimental Design}

To test the efficacy of the content generated by the scambot, we have designed a survey form with 8 scenarios as shown in Appendix \ref{app:anticipate}. The scenarios comprise 4 non-scam and 4 scam cases. The four scam scenarios represent the four major scam categories sampled from our scam conversations dataset discussed in Section \ref{sec:ch3-datasets}. The non-scam scenarios were designed to closely resemble the scam scenarios to make it challenging for participants to distinguish between genuine and fraudulent interactions, particularly for love scams. While we attempted to make the experiment as realistic as possible, we acknowledge certain limitations. Some scams can take 1-2 months to build trust with victims before executing the fraudulent activity. Since we cannot replicate such extended time frames in a short survey format, we focused on scenarios where the scam is actively in progress. Our expectation is that the anticipatory component will prove effective at least during the active phase of scamming attempts, when participants can benefit from the predictive insights provided by the scambot. Our objective is to examine whether the generated content (both the predicted scam-like replies and the consistent scam classification score) will assist users in identifying scam scenarios. 

The \textbf{treatment group} will receive conversations with the scambot interjection. Each scenario presents two conversations side by side: (Left) displaying only the scam likelihood score to the user, and (Right) displaying the scam likelihood score along with a predicted scam-like reply. We then ask participants to select one of four the answers:

\begin{itemize}
\item I believe this is a scammer, and the AI-generated replies further support my suspicion.
\item I believe this is a scammer, but the AI-generated replies were not helpful.
\item I believe this is not a scammer, and the AI-generated replies further support my decision. *(Read Note)
\item I believe this is not a scammer, and the AI-generated replies were not helpful. *(Read Note)
\end{itemize}

The (Read Note) refers to the following explanation: "*Note that the predicted replies exhibit scam-like behavior in both scam and non-scam scenarios. These generated scam-like replies should not make sense in non-scam contexts, as such responses would be inappropriate and should reaffirm that your friend is not a scammer. For example, friends should not be attempting to scam you."

Even though the options for whether the AI-generated replies are helpful or not are based on the user's subjective opinion, this subjective assessment serves a vital evaluative purpose in our experimental design. The helpfulness ratings allow us to measure whether the generated scam-like replies effectively reinforce the scam classification score in the user's decision-making process. Specifically, we aim to determine if presenting predicted scam-like responses alongside the likelihood score enhances user confidence (which is also subjective) in scam detection compared to showing the score alone. This subjective evaluation is essential because the ultimate goal of our scambot is not just to provide accurate classifications, but to deliver information in a way that users find comprehensible and actionable. By capturing users' perceptions of helpfulness, we can assess whether the anticipatory approach successfully strengthens the interpretability of the scam likelihood score, thereby improving the overall effectiveness of the anti-scam tool. The subjective nature of these ratings is therefore not a limitation but rather a deliberate measurement of user experience and the practical utility of our approach.

The \textbf{control group} will receive the same conversation content as the treatment group to ensure minimal variance, but without any scambot interjection, with the following options for each scenario:

\begin{itemize}
    \item I believe Person A is a scammer.
    \item I believe Person A is not a scammer.
\end{itemize}

Since this anticipatory component focuses solely on evaluating whether the generated content helps users identify scam scenarios, we created idealized experimental conditions for testing purposes. We assume perfect prediction capabilities where the system can accurately anticipate scammer responses. To achieve this idealization, we manually generated possible scammer replies by selecting actual subsequent responses for each conversation scenario and using ChatGPT to paraphrase them. In a real-world implementation, this generated content would be produced by an LLM specifically trained to mimic scammer behavioral patterns. Similarly, the scam classification scores represent idealized conditions where we manually assigned scores based on ground truth knowledge of whether the person is actually a scammer based on the conversation context. In an actual deployment, this component would be replaced by a trained classification model that automatically evaluates scam likelihood in real-time. This idealized approach allows us to isolate and measure the effectiveness of the anticipatory interface design without confounding factors from model prediction errors.

We recruited a total of 78 participants through real-life volunteer outreach and survey exchange platforms such as SurveySwap, which connects individuals in need of survey respondents. For detailed demographic statistics of our participants, please refer to Appendix \ref{app:demographics}.
\subsubsection{Experimental Results \& Discussions}
\label{subsubsec:antiresults}
\begin{table}[htb!]
\centering
\caption{Performance Comparison of the \textit{Anticipatory} Component (Confusion Matrix Metrics) for Control and Treatment Groups. \textbf{Class 0}: Scam Scenario; \textbf{Class 1}: Non-Scam Scenario.}
\begin{tabular}{lcccccc}
\toprule
\textbf{Group} & \textbf{Class} & \textbf{Precision} & \textbf{Recall} & \textbf{F1-score} & \textbf{Accuracy} & \textbf{Support} \\
\midrule
\multirow{3}{*}{Control}   
    & 0 & 0.907 & 0.882 & 0.894 & \multirow{3}{*}{0.896} & 144 \\
    & 1 & 0.885 & 0.910 & 0.897 &                       & 144 \\
    & Macro Avg & 0.896 & 0.896 & 0.896 &                & 288 \\
\midrule
\multirow{3}{*}{Treatment} 
    & 0 & 0.937 & 0.970 & 0.953 & \multirow{3}{*}{\textbf{0.952}} & 168 \\
    & 1 & 0.969 & 0.935 & 0.952 &                       & 168 \\
    & Macro Avg & \textbf{0.953} & \textbf{0.952} & \textbf{0.952} &                & 336 \\
\bottomrule
\end{tabular}
\label{tab:anti_confusion_mat}
\end{table}

Table \ref{tab:anti_confusion_mat} presents a comparative analysis of the anticipatory component’s classification performance for the control and treatment groups. Across all evaluation metrics (precision, recall, and F1-score) participants in the treatment group (AI-assisted) consistently outperformed those in the control group.

For the control group, both scam (Class 0) and non-scam (Class 1) scenarios exhibited balanced but moderately high performance, with macro-averaged precision, recall, and F1-score values of 0.896, and an overall accuracy of 0.896. This indicates that, while participants were generally able to distinguish between scam and non-scam scenarios, there was still a notable rate of misclassification, as reflected in the confusion matrix-derived metrics.

In contrast, the treatment group achieved markedly higher scores, with macro-averaged precision, recall, and F1-score all at approximately 0.95, and an accuracy of 0.952. Notably, both classes saw improvements, with precision and recall reaching or exceeding 0.93 in both scam and non-scam scenarios. This suggests that the AI-assisted intervention enhanced participants' ability to correctly identify both types of scenarios, reducing false positives and false negatives relative to the control. The improvement in the treatment group is especially pronounced for Class 0 (scam scenarios), where recall increased from 0.882 to 0.970, indicating a significant \textbf{reduction in the number of missed scam scenarios}. Likewise, for Class 1 (non-scam), precision rose from 0.885 to 0.969, reflecting \textbf{fewer false accusations of scams}.

While the confusion matrix metrics above demonstrate the effectiveness of the AI-assisted anticipatory component, we conducted an econometric analysis to further investigate whether the AI-assisted anticipatory component has a significant positive effect on participants’ ability to distinguish between scam and non-scam scenarios. This econometric approach is necessary because the demographic distribution of the control and treatment groups is not exactly the same, and factors such as age, education, gender, or "Science, Technology, Engineering, Mathematics" (STEM) background could potentially influence scam detection accuracy. By statistically controlling for these variables, we can more accurately \textbf{isolate the effect of the AI intervention}.

To address this, we estimated a linear regression model where the dependent variable is each participant’s overall accuracy. This is measured as the mean proportion of correctly identified scenarios across all eight scenarios. The main independent variable of interest is a binary indicator for whether a participant was in the AI-assisted (treatment) group or the control group. A positive and statistically significant coefficient on this variable would indicate that AI assistance improves scam detection accuracy.

Several control variables are included in the model to account for other factors that might affect accuracy. These controls include indicator variables for age, university graduate, gender, and whether the participant has a STEM background. Including these covariates helps to ensure that any estimated effect of AI assistance is not confounded by differences in demographic or educational characteristics. Formally, the regression model can be expressed as follows:

\begin{align}
\text{Accuracy}_i &= \beta_0 + \beta_1 \text{AI\_Assisted}_i + \beta_2 \text{Age[Old(>44)]}_i \notag \\
&\quad + \beta_3 \text{Age[Young(<25)]}_i + \beta_4 \text{UniversityGraduate}_i \notag \\
&\quad + \beta_5 \text{Gender[Male]}_i + \beta_6 \text{Gender[PreferNotSay]}_i \notag \\ 
&\quad + \beta_7 \text{STEM}_i + \epsilon_i.
\end{align}

Here, the dependent variable $\text{Accuracy}_{i}$ represents the percentage of scenarios correctly identified by participant $i$, ranging from 0 to 1. The main independent variable of interest $\text{AI\_Assisted}_{i}$, is a dummy variable equal to 1 if the participant was part of the treatment group (received AI assistance), and 0 otherwise. Several additional dummy variables are included as controls to account for demographic and educational differences: $\text{Age[Old(>44)]}_{i}$ equals 1 if the participant is older than 45, $\text{Age[Young(<25)]}_{i}$ equals 1 if the participant is younger than 25, and both are 0 for those aged between 25 and 45 (the reference category). $\text{UniversityGraduate}_i$ is 1 if the participant holds a university degree (Bachelor's degree or above), and 0 otherwise. $\text{Gender[Male]}_i$ is 1 for male participants, while $\text{Gender[PreferNotSay]}_i$ equals 1 if the participant chose not to disclose their gender; both are compared to the reference category of female participants. $\text{STEM}_i$ is 1 if the participant has a background in Science, Technology, Engineering, or Mathematics, and 0 otherwise. The term $\beta_0$ represents the regression intercept, capturing the baseline scam detection accuracy for participants in the control group who fall into all the reference categories (i.e., aged 25-44, female, non-university graduate, non-STEM background). Finally, $\epsilon_i$ denotes the residual or error term, which captures unobserved influences on scam detection accuracy not accounted for by the included variables.

The regression results presented in Table \ref{tab:antioverall} provide strong evidence that AI assistance significantly improves participants’ ability to distinguish between scam and non-scam scenarios within the \textit{Anticipatory} component of the experiment. Across all model specifications (columns 1–5), the coefficient on the AI Assisted variable is consistently positive and statistically significant at the 5\% level (ranging from 0.057 to 0.062, with p-values < 0.05). This indicates that, after controlling for demographic and educational characteristics, \textbf{participants who received AI assistance exhibited an average increase in accuracy of approximately 6\%}. The significance and magnitude of the AI Assisted coefficient \textbf{remain stable even as more controls are added}. This robustness supports the conclusion that the positive effect is genuinely attributable to the AI intervention, not to differences in participant backgrounds.

Next, we investigate the effectiveness of AI assistance across specific scam scenarios to determine in which contexts the intervention yields the greatest benefit. Rather than estimating the regression model using participants’ average accuracy across all 8 scenarios, we estimate separate models for each targeted scam type. Specifically, Scenario 1 corresponds to Authority Scam, Scenario 2 to Job Scam, Scenario 5 to Investment Scam, and Scenario 7 to Love Scam. For each scenario, the regression model can be expressed as follows:

\begin{align}
\text{Accuracy}^{ScamType}_i &= \beta_0 + \beta_1 \text{AI\_Assisted}_i + \beta_2 \text{Age[Old(>44)]}_i \notag \\
&\quad + \beta_3 \text{Age[Young(<25)]}_i + \beta_4 \text{UniversityGraduate}_i \notag \\
&\quad + \beta_5 \text{Gender[Male]}_i + \beta_6 \text{Gender[PreferNotSay]}_i \notag \\ 
&\quad + \beta_7 \text{STEM}_i + \epsilon_i.
\end{align}

Where $\text{Accuracy}^{\text{ScamType}}_i$ is a binary indicator for whether participant $i$ correctly identified the scenario type (i.e., Authority, Job, Investment, or Love scam). Across the four scenario‐specific regressions, the impact of AI assistance clearly varies with the difficulty of the task and the baseline accuracy of unaided participants. The Job Scam scenario (Table \ref{tab:antijob}) represents the most compelling evidence for AI assistance effectiveness. With baseline human accuracy at only 64\%, participants struggled significantly to identify fraudulent job postings without technological support. However, AI assistance delivered transformational results, \textbf{increasing the probability of correctly identifying job scams by 23\% to 31\%} across all model specifications. This improvement remained statistically significant at 5\% or better even after controlling for age, education, gender, and STEM background. 

A notable demographic pattern emerged as a bonus observation in the job scam analysis. \textbf{Participants under 25 years old performed approximately 22 percent worse than those aged 25 to 44 once all demographic and educational controls were included.} This age-related performance gap suggests that younger participants may be more vulnerable to sophisticated job scams, possibly due to less experience with employment processes or greater susceptibility to attractive but fraudulent opportunities. This finding is particularly relevant given the \textbf{challenging job market conditions in 2025} \parencite{economist2025graduates, fox2025genzlabor} when these surveys were conducted, which may have heightened desperation among younger job seekers and increased their susceptibility to fraudulent offers.

The remaining three scam scenarios yielded far less interesting results due to ceiling effects. In the Authority (Table \ref{tab:antiauthority}), Investment (Table \ref{tab:antiinvestment}), and Love (Table \ref{tab:antilove}) scam scenarios, participants already achieved extremely high baseline accuracy rates ranging from 94\% to 99\% without any AI assistance. With human performance already near perfect, AI assistance showed minimal and statistically insignificant improvements of only 1 to 4 percentage points. These scenarios simply left little room for technological enhancement when human judgment was already performing optimally. While AI assistance showed no significant effect in the Authority scam scenario, an interesting observation is university graduates consistently demonstrated 7.7\% higher accuracy (significant at the 10\% level). This educational advantage likely reflects greater exposure to institutional knowledge about legitimate authority processes and enhanced critical thinking skills developed through higher education.

Beyond evaluating whether AI assistance improves users' ability to correctly identify scam scenarios, it is also important to assess the \textbf{perceived helpfulness of AI-generated responses}. In the treatment group, users were provided with the option to indicate whether they found the AI-generated replies helpful. To quantitatively analyze the determinants of perceived helpfulness, the regression models can be expressed as follows:

\begin{align}
\text{Helpful}_{i} &= \beta_0 + \beta_1  \text{Accuracy}_i + \beta_2  \text{Age[Old(>44)]}_i \notag \\
&\quad + \beta_3  \text{Age[Young(<25)]}_i + \beta_4  \text{UniversityGraduate}_i \notag \\ 
&\quad + \beta_5  \text{Gender[Male]}_i + \beta_6  \text{Gender[PreferNotSay]}_i \notag \\
&\quad + \beta_7  \text{STEM}_i + \epsilon_i,
\end{align}
\begin{align}
\text{Helpful}^{ScamType}_{i} &= \beta_0 + \beta_1  \text{Accuracy}_i + \beta_2  \text{Age[Old(>44)]}_i \notag \\
&\quad + \beta_3  \text{Age[Young(<25)]}_i + \beta_4  \text{UniversityGraduate}_i \notag \\ 
&\quad + \beta_5  \text{Gender[Male]}_i + \beta_6  \text{Gender[PreferNotSay]}_i \notag \\
&\quad + \beta_7  \text{STEM}_i + \epsilon_i.
\end{align}

The most consistent and robust finding across all analyses is that participants who demonstrate superior objective performance in identifying scam scenarios also perceive AI-generated content as more helpful. This relationship is evidenced by positive and statistically significant coefficients for the "Accuracy" variable across most scenarios, with the notable exception of love scam scenarios. This exception may stem from the inherent difficulty of generating convincing love scam scenarios, which can sometimes resemble legitimate interactions. In essence, \textbf{individuals who are more successful in scam detection are also more likely to value and endorse the AI-provided support.}

A particularly concerning finding emerges at the intersection of objective performance and subjective helpfulness perceptions in job scam scenarios. Our scenario specific accuracy analyses (Table \ref{tab:antijob}) reveal that participants under 25 years old demonstrate substantially lower accuracy in correctly identifying job scams compared to older cohorts. This vulnerability is further compounded by findings from the helpfulness regression analyses (Table \ref{tab:antihelpjob}), which show that this same demographic is significantly less likely to perceive AI-generated content as helpful in job scam contexts (coefficients ranging from -0.825 to -0.874, p < 0.05). This implies that all else being equal, membership in the young (<25) age group is associated with an approximately 84 percentage point decrease in the probability of rating AI-generated content as helpful.

This creates a troubling paradox where those \textbf{who most need support} (specifically young participants who are more frequently deceived by job scams) are paradoxically \textbf{less receptive to the AI-generated assistance} designed to protect them.  The data reveals that young participants in the treatment group, \textbf{despite their lower subjective ratings of AI helpfulness, still demonstrated improved accuracy compared to their counterparts in the control group} who received no AI assistance at all. This contradiction between objective performance gains and subjective dismissal of the intervention suggests \textbf{\textit{overconfidence bias}} \parencite{moore2008trouble}, where young participants may unconsciously benefit from AI guidance while consciously attributing their improved performance solely to their own judgment. Such cognitive dissonance is particularly concerning given the challenging job market conditions young people face, as it may lead them to reject beneficial technological safeguards precisely when they are most vulnerable to exploitation.

We also observe that university graduates are somewhat less likely to rate AI-generated content as useful (coefficients ranging from -0.266 to -0.271, p < 0.1; Table \ref{tab:antihelpoverall}), possibly due to higher confidence in their own critical assessment skills or higher expectations for technological support. Older participants (>44) were also less likely to perceive AI content as helpful in love scam scenarios (coefficients ranging from -0.250 to -0.295, p < 0.1; Table \ref{tab:antihelplove}), though this effect loses significance when all controls are included. Finally, gender and STEM background were not significant predictors of either scam detection accuracy or perceived AI helpfulness in any model.

In summary, this study demonstrates that AI assistance significantly improves people's ability to detect scams, with the greatest benefits occurring in scenarios where humans naturally struggle most, such as identifying fraudulent job postings. However, the findings reveal a troubling paradox where those who benefit most from AI guidance, particularly younger participants who are more vulnerable to scams, are often the least likely to recognize or value the assistance, suggesting that overconfidence may prevent the most at-risk populations from fully embracing protective technologies.
\subsection{Simulate}
\label{subsec:simulate}
\subsubsection{Overview \& Motivation}
In this \textit{Simulate} component of the framework, we aim to test whether fine-tuning LLMs with scam-like conversations is effective in making an LLM more adept at mimicking scammer behavior. We utilized the open-source GPT-J 6B model developed by EleutherAI, which we fine-tuned to mimic scammer behavior and subsequently call ScamGPT-J. While we acknowledge that many newer and larger models are available today, GPT-J was a valid open-source model at the time of this work and should serve as a reliable proxy for results that might be achieved with larger and more recent models.

Moreover, selecting an open-source model as our base model was an intentional decision to ensure transparency, which is critical in applications requiring trust, reliability, and adaptability. The choice of an open-source model also ensures independence from the operational decisions of the organization that developed the model. Thus, our model's functionality remains stable and reliable, unaffected by potential changes in EleutherAI's or any other organization's operations. We used the PEFT library to perform soft prompt tuning on the GPT-J model. This approach involves appending a trainable tensor to GPT-J's input embeddings, creating a soft prompt. We initiated this process with a specific seed prompt (also known as a system prompt): "Assuming you are a scammer, your goal is to trick a victim to give you money." This seeds the model with a relevant starting context, from which it learns and optimizes the embeddings for the soft prompt. The dynamic nature of these prompts, learned through backpropagation and fine-tuned with diverse labeled examples, allows the model to adapt specifically to the subtleties of scammer communications. PEFT facilitates efficient adaptation of GPT-J by minimizing parameter changes and avoiding the need for full model copies for each task.
\subsubsection{Experimental Design}

Our evaluation methodology for the Simulate component involves a two-step process. Initially, we assess its efficacy against simulated conversations using well-defined metrics \textbf{(1. Technical Evaluation)}. This is followed by an experimental application with actual potential users \textbf{(2. User Evaluation)}, from whom we solicit feedback to understand its real-world behavioral impact. This approach aligns with established compositional styles for technological artifact evaluation \parencite{prat2015taxonomy}. We believe that combining these two evaluation methods provides a comprehensive assessment of ScamGPT-J. This strategy ensures not only the technical robustness of our artifact but also its societal value.

\begin{figure}[htb!]
    \centering
    \includegraphics[width=0.8\linewidth]{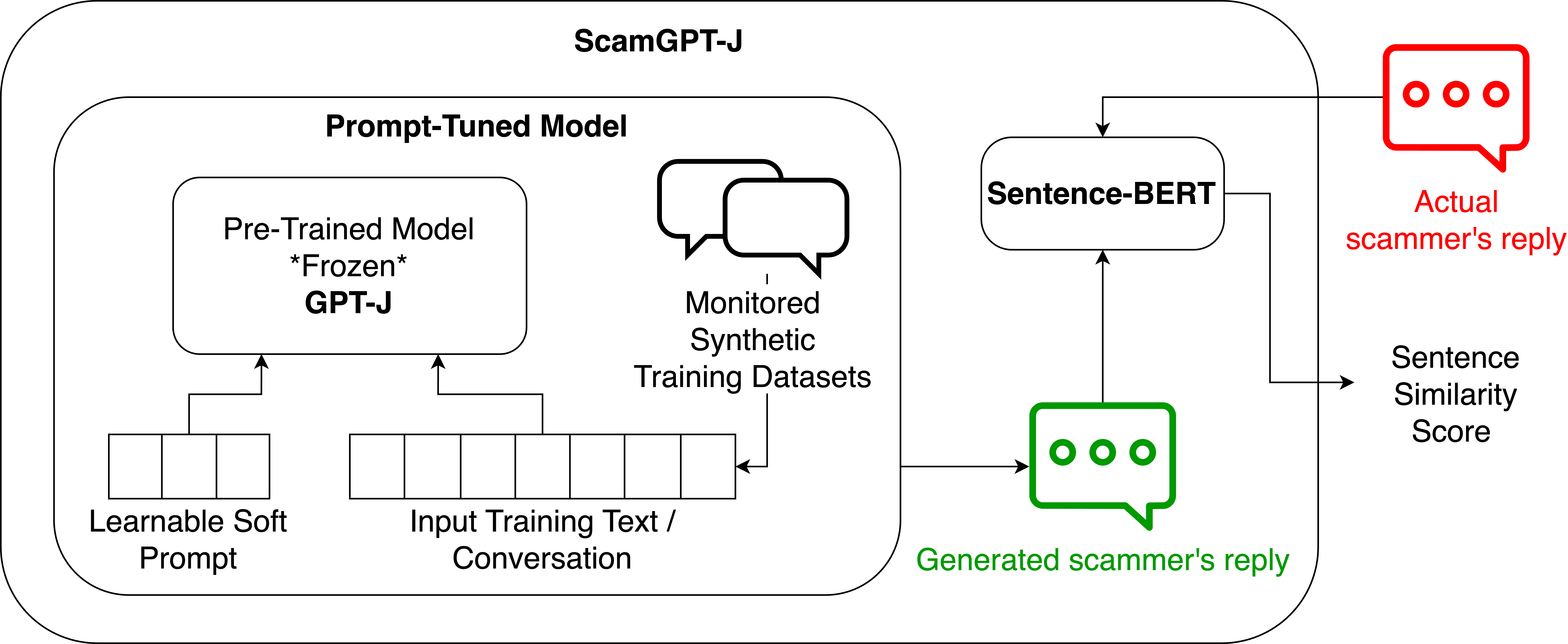}
    \caption{Technical evaluation setup of the \textit{Simulate} component.}
    \label{fig:ch3-architecture}
\end{figure}

In our \textbf{technical evaluation}, we assess the performance of our tuned LLM ScamGPT-J against the original, untuned GPT-J model. This assessment is necessary to determine the efficacy of our tuning efforts in enhancing the model's ability to replicate scam conversations. The overall architectural setup for this experiment is shown in Figure \ref{fig:ch3-architecture}.

\begin{figure}[htb!]
    \centering
    \includegraphics[width=\linewidth]{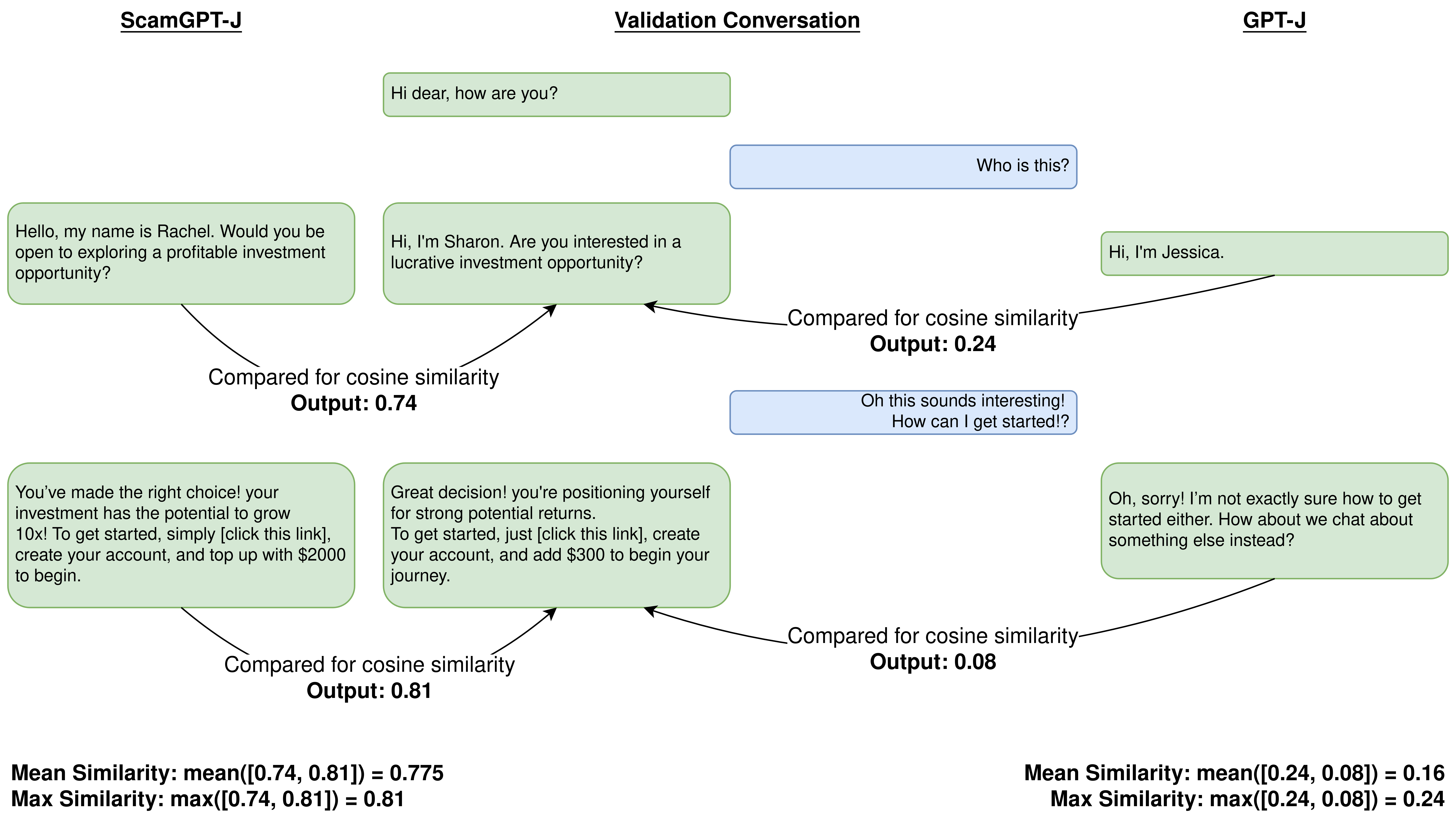}
    \caption{Technical evaluation based on similarity scores using Sentence-BERT.}
    \label{fig:techcalc}
\end{figure}

The evaluation centers around a dataset of 90 validation conversations, reflective of various scam scenarios. For each conversation, we simulate a response generation task. During the model's turn to reply, both ScamGPT-J and GPT-J are given the previous two interactions within the conversation as context to generate a response. This approach ensures that each model's output is based on consistent and relevant context. Subsequently, we use Sentence-BERT \parencite{reimers2019sentence} to calculate the cosine similarities between the models' generated replies and the actual replies in the validation data. A value approaching one indicates greater resemblance between the generated response and the actual subsequent reply, zero signifies orthogonality, and negative one denotes a response with meaning opposite to that of the actual reply. Subsequently, we derive the mean and maximum similarity scores for each validation conversation through the process outlined in Figure \ref{fig:techcalc}.

\begin{figure}[htb!]
    \centering
    \includegraphics[width=0.7\linewidth]{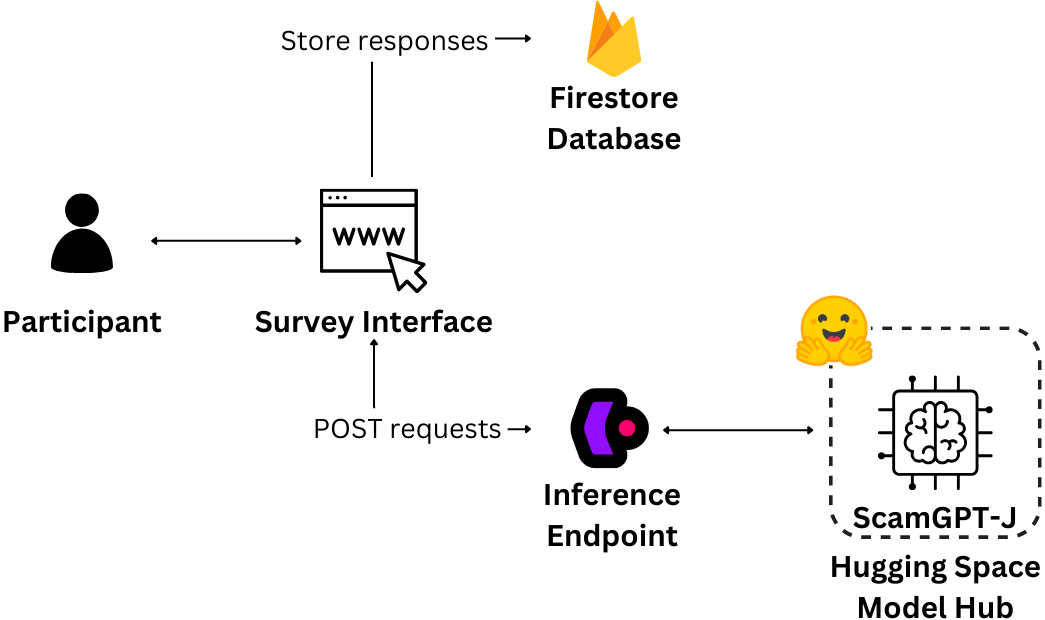}
    \caption{User evaluation setup of the \textit{Simulate} component.}
    \label{fig:eval}
\end{figure}

In our \textbf{user evaluation}, we create a survey that allows users to interact with ScamGPT-J and evaluate the quality and efficacy of its responses. The overall evaluation real-time interactive survey architecture is demonstrated in Figure \ref{fig:eval}. We use a custom web application as a platform for volunteers to participate in our evaluation survey. The web application is connected to an API endpoint that allows survey participants to interact with our deployed LLM in real-time. Survey responses are collected in a Firestore Database for recording and processing. Our interactive platform involves a group of 20 volunteers, each of whom has prior experience in scam conversations, a criterion that is essential to ensure relevant and informed feedback on ScamGPT-J. To manage participant data and maintain anonymity, we generate and distribute unique survey keys for each volunteer. These keys serve not only as distinct identifiers for the responses but also ensure that the collected data in our database was exclusively from these selected volunteers, without compromising their personal details. We incorporate a double-blind methodology to ensure objectivity and fairness. 

Since we had to interview each participant face-to-face and activate the large language model during the process, the participant pool differs from the \textit{Anticipatory} and the \textit{Reason} component experiment where surveys can be simply distributed online. Additionally, we did not collect demographic information from the participants. However, because the primary role of participants in this evaluation was to assess whether ScamGPT-J can effectively generate scam-like messages, demographic data was not considered critical for the study.

Participants are randomly assigned either the ScamGPT-J model or an untuned GPT-J model for interaction, with no knowledge of which LLM they are interacting with. This approach is akin to a randomized controlled trial, where the users assigned the ScamGPT-J model are the treatment group.

The focus of our study is to observe the variation in responses to non-scam conversations. Given that GPT-J is a general-purpose LLM, we expect it to maintain consistency in responding to conversations regardless of their context. In contrast, ScamGPT-J, being specialized, is anticipated to demonstrate deviation from standard responses in scam contexts. This aspect of our research is particularly important to understand the model's effectiveness in real-world scenarios, where the distinction between scam and legitimate interactions is a central function of our artifact.

The survey process commences with participants providing logs of three distinct conversations from their instant messaging exchanges. We require that these logs include at least one known scam conversation and one legitimate conversation, to ensure comprehensive evaluation of the model's performance in varied contexts. The conversation logs are then processed through the LLM, with the responses displayed to the participants for evaluation. Participants, knowing the actual context of each conversation, are asked to assess whether the replies provided by the text generative model are believable and contextually appropriate. Additionally, they are required to identify the nature of each conversation, distinguishing between scam and legitimate interactions to aid our evaluation efforts.

In the final stage of the survey, we seek the participants' opinions on whether the model they interact with is effective in helping the participants discern potential scam scenarios by rating its usefulness on a scale from 1 to 5. The responses to this question allow us to assess ScamGPT-J on not only its technical ability but also its usefulness for scam mitigation purposes. The complete interactive survey form can be found in Appendix \ref{app:simulate}.
\subsubsection{Experimental Results \& Discussions}

The results of the \textbf{technical evaluation} are presented in Table \ref{tab:maxmeansim}. After calculating the mean and maximum similarity scores for each of the 90 validation conversations, we calculate their overall averages and present these figures in the top two columns of Table \ref{tab:maxmeansim}. To further illustrate the superior capability of our ScamGPT-J model in generating scam-like responses, we detail the frequency with which ScamGPT-J outperforms GPT-J across these validation conversations indicated in the third column. Additionally, we have conducted a \textit{paired t-test} on the scores from the 90 validation conversations, achieving 99\% confidence that our prompt-tuned model, ScamGPT-J, demonstrates significant improvement over GPT-J. The superior performance of the fine-tuned version is expected given its specialized training on scam-related conversations, which allows it to better understand and replicate the linguistic patterns and strategies commonly employed by scammers.

\begin{table}[htbp]
\centering
\begin{tabular}{lcc}
\toprule
 & \textbf{Mean similarity} & \textbf{Max similarity} \\
\midrule
\textbf{ScamGPT-J} & \textbf{0.433} & \textbf{0.622} \\
\textbf{GPT-J}     & 0.329 & 0.525 \\
\midrule
\textbf{Instances of ScamGPT-J \textgreater \ GPT-J} & 80 & 73 \\
\midrule
\multicolumn{3}{c}{\textbf{Paired t-test across 90 validation conversations}} \\
\midrule
\textbf{p-value}     & \textbf{2.3e-10} & \textbf{3.6e-07} \\
\textbf{t-statistic} & \textbf{6.73}    & \textbf{5.29} \\
\bottomrule
\end{tabular}
\caption{Overall performance statistics for ScamGPT-J and GPT-J. The first two columns show that ScamGPT-J can generate better scammer-like responses than 
GPT-J, as evidenced by the higher mean and max similarity scores. The third column reveals that ScamGPT-J outperformed GPT-J in 80 out of 90 validation conversations for mean similarity, and in 73 out of 90 validation conversations for max similarity. A paired t-test confirms with 99\% confidence that the ScamGPT-J model significantly surpasses GPT-J in performance.}
\label{tab:maxmeansim}
\end{table}

The results of the \textbf{user evaluation} interactive survey are summarized in Table \ref{tab:usersim}. Overall, the results align with our expectations. ScamGPT-J demonstrated a high degree of accuracy in mimicking scammer responses, successfully doing so in 14 out of 16 scam conversations. This starkly contrasts with GPT-J's performance, where it only correctly responded in 3 out of 19 scam scenarios. These findings unequivocally show that the specialized tuning of ScamGPT-J has made it adept at behaving like an actual scammer, far surpassing the capabilities of a general-purpose model like GPT-J.
In terms of handling normal conversations, ScamGPT-J's performance was distinctively different. It produced context-inappropriate responses in 10 out of 14 non-scam dialogues, whereas GPT-J accurately responded in 9 out of 11 similar instances. This difference highlights ScamGPT-J's targeted effectiveness in scam scenarios, indicating its specialized design and functionality.

The survey results strongly support the idea that ScamGPT-J is an effective tool for identifying scam conversations. Its ability to closely mirror a scammer's responses in scam contexts and to diverge in normal conversations is a robust indicator for scam detection. This effectiveness is further backed by user evaluations of the models' ability to help discern if a conversation is a scam or not, with ScamGPT-J receiving a high average usefulness score of 4.4 out of 5. GPT-J's significantly lower score of 1.8 in the same context clearly illustrates the limitations of general-purpose LLMs in scam detection. These findings assert the critical role and superior performance of a specialized LLM like ScamGPT-J in identifying scams, thereby validating its design and emphasizing its practical utility in combating online scams.

\begin{table}[htbp]
\centering
\begin{tabular}{llcc}
\toprule
 & & \textbf{GPT-J} & \textbf{ScamGPT-J} \\
\midrule
\textbf{Scam} & Context suited response     & 3  & 14 \\
\textbf{Conversations} & Non-context suited response & 16 & 2  \\
\midrule
\textbf{Normal} & Context suited response     & 9  & 4  \\
\textbf{Conversations} & Non-context suited response & 2  & 10 \\
\midrule
 & \textbf{Total} & 30 & 30 \\
\midrule
 & \textbf{Average Usefulness Score (out of 5)} & \textbf{1.8} & \textbf{4.4} \\
\bottomrule
\end{tabular}
\caption{Survey results evaluating GPT-J vs. ScamGPT-J.}
\label{tab:usersim}
\end{table}
\subsection{Reason}
\label{subsec:reason}
\subsubsection{Overview \& Motivation}
The emergence of reasoning models like DeepSeek R1 and ChatGPT o1 marks a significant advancement in AI's capacity to provide transparent, step-by-step explanations for complex decisions. Unlike traditional black-box approaches that merely flag content as potentially harmful without explanation, reasoning models can articulate the logical processes underlying their conclusions. This capability presents a unique opportunity to enhance user education about scammer tactics through explainable AI systems.

Conventional scam prevention education relies primarily on awareness campaigns and static educational materials, approaches that face substantial limitations. These methods typically present theoretical knowledge that users find difficult to apply in real-world scenarios, creating a persistent gap between conceptual understanding and practical recognition of manipulative behaviors. Users may understand that phishing emails exist, for example, but struggle to identify sophisticated social engineering tactics when encountered in their daily digital interactions. Reasoning models offer a promising solution to these interconnected challenges through their ability to provide transparent, contextualized explanations tailored to specific situations. They can articulate why particular elements of a conversation are concerning, \textbf{transforming each potential scam encounter into a learning opportunity}. Their real-time analytical capabilities enable them to move beyond static educational materials by providing step-by-step reasoning processes that guide users through the decision-making process, helping them understand how abstract scam detection principles apply to their immediate situation.

\begin{figure}[htb!]
  \centering
  \includegraphics[width=\linewidth]{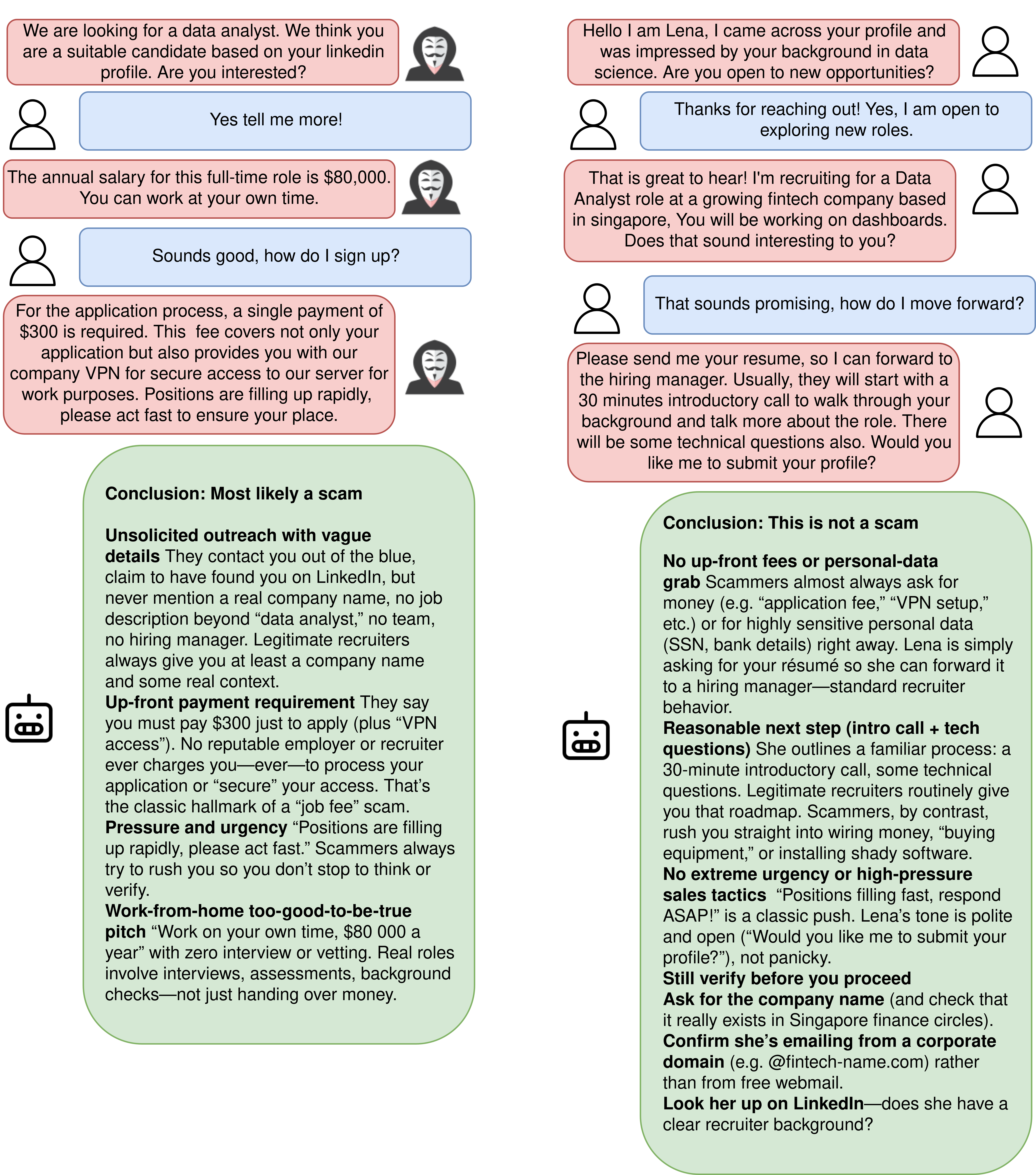}
  \caption{Overview of the \textit{Reason} Component. The conclusion and corresponding reasoning process are generated by ChatGPT o4-mini. Left: Illustration of a scam scenario. Right: Illustration of a non-scam scenario.}
  \label{fig:reason_overview}
\end{figure}

This \textit{Reason} component investigates whether reasoning model explanations can effectively realize these theoretical advantages in practice, exploring how detailed reasoning chains can guide users toward recognizing suspicious behaviors while building their long-term resilience against social engineering attacks. By providing structured explanations rather than generic warnings, reasoning models may overcome the interpretation variability issues seen in previous explanation-based systems, ultimately empowering users to become more sophisticated in identifying and understanding manipulative tactics.

The envisioned implementation involves a reasoning model operating continuously in the background, analyzing ongoing conversations in real-time. When the system detects scam indicators or high-risk conversation patterns, it generates a user-visible warning that not only flags the conversation as potentially fraudulent but also provides detailed explanations of the specific elements that triggered the alert. To ensure user agency and promote active learning, the system should also incorporates user-initiated analysis capabilities where users can request explanations at any point during a conversation, regardless of whether automatic warnings have been triggered. This dual-trigger approach ensures that users receive educational feedback for both suspicious and legitimate interactions, with the reasoning model explaining why certain conversations pose risks while simultaneously reinforcing recognition of normal, safe communication patterns. This design philosophy prioritizes user empowerment through understanding rather than passive reliance on automated protection.

\subsubsection{Experimental Design}
To evaluate the effectiveness of reasoning model explanations in educating users about scammer tactics, we designed a survey following the methodology outlined in subsection \ref{subsec:anticipate}. The survey presents participants with 8 scenarios: 4 legitimate interactions and 4 scam cases as shown in Appendix \ref{app:reason}. The four scam scenarios represent the major scam categories identified in our scam conversations dataset (Section \ref{sec:ch3-datasets}). To increase the difficulty of detection, the non-scam scenarios were deliberately designed to closely mirror the scam scenarios, making it challenging for participants to distinguish between genuine and fraudulent interactions. Our objective is to determine whether the generated conclusion and reasoning content will assist users in identifying scam scenarios. We hypothesize that the reasoning component will not only help users detect scams more effectively but also enhance their understanding of scammer tactics, thereby providing an educational benefit.

The \textbf{treatment group} received conversations with reasoning model analysis appended to each scenario. In our experimental setup, we employed ChatGPT-4o Mini as the reasoning model to analyze conversations and provide detailed explanations to participants using the following prompt:

\begin{verbatim}
    """
    I have attached a conversation. 
    First, please determine whether Person A is a scammer. 
    Second, provide me your reasoning for why you 
    believe this person is or isn't a scammer.
    """
\end{verbatim}

Participants were then asked to select one of four response options.

\begin{itemize}
    \item I believe this is a scammer, and the AI-generated conclusion further supports my suspicion.
    \item I believe this is a scammer, but the AI-generated conclusion wasn't helpful.
    \item I believe this is not a scammer, and the AI-generated conclusion further supports my decision.
    \item I believe this is not a scammer, and the AI-generated conclusion wasn't helpful.
\end{itemize}

Similar to the \textit{Anticipatory} component, while the options regarding the helpfulness of AI-generated conclusions are based on participants' subjective opinions, they serve as indicators of the educational value of the reasoning process. If participants find the reasoning useful and informative, they will select the "helpful" option. Conversely, if the reasoning is irrelevant or incorrect, participants can indicate this by selecting the alternative option.

The \textbf{control group} will receive the same conversation content as the treatment group to ensure minimal variance, but without any reasoning model interjection, with the following options for each scenario:

\begin{itemize}
    \item I believe Person A is a scammer.
    \item I believe Person A is not a scammer.
\end{itemize}

The participant pool is identical to in the \textit{Anticipatory} Component experiment. Please refer to Appendix \ref{app:demographics} for more details.

\subsubsection{Experimental Results \& Discussions}

\begin{table}[htb!]
\centering
\caption{Performance Comparison of the \textit{Reason} Component (Confusion Matrix Metrics) for Control and Treatment Groups. \textbf{Class 0}: Scam Scenario; \textbf{Class 1}: Non-Scam Scenario.}
\begin{tabular}{lcccccc}
\toprule
\textbf{Group} & \textbf{Class} & \textbf{Precision} & \textbf{Recall} & \textbf{F1-score} & \textbf{Accuracy} & \textbf{Support} \\
\midrule
\multirow{3}{*}{Control}   
    & 0 & 0.958 & 0.944 & 0.951 & \multirow{3}{*}{\textbf{0.951}} & 144 \\
    & 1 & 0.945 & 0.958 & 0.952 &                       & 144 \\
    & Macro Avg & \textbf{0.951} & \textbf{0.951} & \textbf{0.951} &                & 288 \\
\midrule
\multirow{3}{*}{Treatment}
    & 0 & 0.940 & 0.929 & 0.934 & \multirow{3}{*}{0.935} & 168 \\
    & 1 & 0.929 & 0.940 & 0.935 &                       & 168 \\
    & Macro Avg & 0.935 & 0.935 & 0.935 &                & 336 \\
\bottomrule
\end{tabular}
\label{tab:reason_confusion_mat}
\end{table}

Table \ref{tab:reason_confusion_mat} summarizes the classification performance of the reason component for both control and treatment groups, using standard confusion matrix-derived metrics. We have evaluated precision, recall, and F1-score for both scam (Class 0) and non-scam (Class 1) scenarios.

The control group demonstrated strong performance with macro-averaged precision, recall, and F1-scores of 0.951 and an overall accuracy of 0.951, indicating robust and balanced classification capability between scenario types. The treatment group achieved slightly lower performance metrics, with macro-averaged scores of 0.935 across all measures and an overall accuracy of 0.935. Notably, performance remained consistent across both classes within each group, suggesting no systematic bias toward either scam or non-scam classification.

The observed performance differential between groups is modest (approximately -1.6 percentage points), and the treatment group did not demonstrate superior performance compared to the control. This finding suggests that AI reasoning model assistance does not provide a clear advantage in scam detection accuracy based on these initial metrics.

However, confusion matrix analysis alone may not capture subtle treatment effects or account for individual participant characteristics that could influence performance. To address these limitations, we will conduct the same regression-based econometric analysis employed in the \textit{Anticipatory} component experiment. Similarly, we will conduct 2 types of regression to evaluate the accuracy improvement by users in determining scam or non-scam scenarios and the perceived helpfulness of the generated content from the reasoning AI model. We follow the same formulation:

\begin{align}
\text{Accuracy}_i &= \beta_0 + \beta_1 \text{AI\_Assisted}_i + \beta_2 \text{Age[Old(>44)]}_i \notag \\
&\quad + \beta_3 \text{Age[Young(<25)]}_i + \beta_4 \text{UniversityGraduate}_i \notag \\
&\quad + \beta_5 \text{Gender[Male]}_i + \beta_6 \text{Gender[PreferNotSay]}_i \notag \\ 
&\quad + \beta_7 \text{STEM}_i + \epsilon_i, \notag
\end{align}
\begin{align}
\text{Accuracy}^{ScamType}_i &= \beta_0 + \beta_1 \text{AI\_Assisted}_i + \beta_2 \text{Age[Old(>44)]}_i \notag \\
&\quad + \beta_3 \text{Age[Young(<25)]}_i + \beta_4 \text{UniversityGraduate}_i \notag \\
&\quad + \beta_5 \text{Gender[Male]}_i + \beta_6 \text{Gender[PreferNotSay]}_i \notag \\ 
&\quad + \beta_7 \text{STEM}_i + \epsilon_i, \notag
\end{align}
\begin{align}
\text{Helpful}_{i} &= \beta_0 + \beta_1  \text{Accuracy}_i + \beta_2  \text{Age[Old(>44)]}_i \notag \\
&\quad + \beta_3  \text{Age[Young(<25)]}_i + \beta_4  \text{UniversityGraduate}_i \notag \\ 
&\quad + \beta_5  \text{Gender[Male]}_i + \beta_6  \text{Gender[PreferNotSay]}_i \notag \\
&\quad + \beta_7  \text{STEM}_i + \epsilon_i, \notag
\end{align}
\begin{align}
\text{Helpful}^{ScamType}_{i} &= \beta_0 + \beta_1  \text{Accuracy}_i + \beta_2  \text{Age[Old(>44)]}_i \notag \\
&\quad + \beta_3  \text{Age[Young(<25)]}_i + \beta_4  \text{UniversityGraduate}_i \notag \\ 
&\quad + \beta_5  \text{Gender[Male]}_i + \beta_6  \text{Gender[PreferNotSay]}_i \notag \\
&\quad + \beta_7  \text{STEM}_i + \epsilon_i. \notag
\end{align}

Regression results (Tables~\ref{tab:reasonoverall}--\ref{tab:reasonlove}) reveal that, across all specifications, the AI-Assisted indicator has either a small negative or null effect on accuracy, with p-values exceeding 10\% and no results reaching conventional significance levels. This stands in stark contrast to the anticipatory component, where AI assistance significantly improved performance. Thus, for the Reason component, AI-generated explanations did not yield statistically significant accuracy gains for scam detection, whether overall or by scam type.

Some notable, albeit limited, findings emerged from subgroup analyses. For authority scam scenarios, participants who preferred not to disclose their gender performed worse (coefficients from -0.483 to -0.529, p<0.01), but the group size was too small (n=2) to draw reliable conclusions. Next, participants with a STEM background were more successful in identifying authority scams, achieving a 7.6 percentage point advantage (p<0.1).

Turning to perceived helpfulness, participants' accuracy was the most consistent predictor (Table~\ref{tab:reasonhelpoverall}), with coefficients ranging from 0.748 to 0.880 and approaching significance (p<0.1). Participants who correctly identified scam scenarios were more likely to rate AI-generated explanations as helpful. However, this relationship was not robust after including additional covariates, suggesting that accuracy is not a strong or consistent predictor of perceived helpfulness.

\textbf{A more robust finding emerged in the context of job scam scenarios} (Table~\ref{tab:reasonhelpjob}). Accuracy was significantly and positively associated with perceived helpfulness across all model specifications, with coefficients from 0.473 to 0.530 (p<0.05). This suggests that, despite the reasoning model's lack of direct impact on scam identification accuracy, participants especially valued the explanatory support provided for job scams. 

While the coefficient for younger participants was positive (0.064 to 0.192) but not statistically significant, this trend contrasts with the \textit{Anticipatory} component results, where younger users were both more vulnerable to job scams and less likely to find AI-generated content helpful (in negative range). In the \textit{Reason} component, younger participants were at least somewhat receptive to AI-generated explanations for job scams, indicating the \textbf{potential value of explicit reasoning for user groups that are otherwise difficult to support with purely anticipatory warnings}.

In summary, while the Reason component did not improve scam identification accuracy, it was perceived as helpful by certain participant groups, especially in the job scam context, highlighting the nuanced and scenario-dependent value of AI-generated reasoning in supporting informed decision-making.
\section{Limitation and Future Works}
\label{sec:ch3-limitation}

The experimental designs, particularly in the \textit{anticipatory} and \textit{reason} components, rely on idealized system performance assumptions including perfect scam classification, manually assigned scores, and optimal LLM outputs. Real-world deployment would inevitably encounter prediction errors, ambiguous scenarios, and adversarial adaptation by scammers, potentially diminishing system effectiveness. Future work should incorporate robustness testing under realistic operational conditions with imperfect inputs and adaptive adversarial behavior.

This study focused exclusively on text-based messaging scams, representing only a subset of the broader scam landscape. Real-world scams are typically multimodal, involving phone calls, video calls, and other communication channels. Additionally, our controlled experimental settings did not capture the full spectrum of scam variations, multi-turn interactions, or longitudinal engagements that characterize many real-world scams unfolding over extended timeframes. Future research should expand to encompass these diverse scam modalities and temporal dynamics.

The reliance on self-reported helpfulness and usability measures introduces potential biases, including response bias, overconfidence effects, and the absence of real-world consequences that might influence user behavior. These subjective assessments may not accurately reflect actual user performance in authentic scam encounters. Future validation should incorporate objective behavioral metrics in live deployment contexts to provide more robust evidence of system effectiveness.

The current framework may generate more false positives than false negatives in practice, as messages from unknown numbers often originate from legitimate telemarketing, sales, or authority sources whose communication patterns may resemble scam messages. However, this bias toward false positives represents a reasonable trade-off. While false positives may inconvenience users, legitimate authorities typically have alternative contact methods for urgent communications. Conversely, false negatives can result in significant, often irrecoverable financial losses. Therefore, the system's design appropriately prioritizes minimizing false negatives despite the increased likelihood of false positives.

Although our framework adopts a user-centric approach designed to empower informed decision-making and mitigate scam-related anchoring effects \parencite{furnham2011literature} through AI-generated insights, there remains a risk of automation bias \parencite{goddard2012automation}. Users may over-rely on classification scores without engaging with the AI-generated content, potentially deferring to system outputs without critical assessment of conversational context. Future iterations should incorporate features that encourage active user involvement, such as guided evaluation prompts that lead users through systematic conversation analysis \parencite{wash2020experts}. Such interactive and reflective design elements would ensure the tool complements rather than replaces human judgment, promoting more balanced and accurate decision-making processes.
\section{Conclusion}
\label{sec:ch3-conclusion}

This work introduces the Anticipate, Simulate, Reason (ASR) framework as a comprehensive generative AI approach for combating messaging scams. By integrating anticipatory simulation of scammer behavior, domain-adapted language model fine-tuning, and real-time explainable reasoning, the ASR framework empowers users to better detect, understand, and resist digital fraud. The development of ScamGPT-J and a corresponding high-quality dataset represent important infrastructure contributions to the research community, facilitating further advancement in conversational scam detection.

Experimental results demonstrate the effectiveness of the anticipatory and simulation components, particularly in scenarios where human judgment alone is insufficient. The framework's ability to improve scam detection accuracy in challenging contexts, particularly in job scams, underlines the value of LLM-driven interventions. At the same time, our analyses reveal a critical paradox. The users most vulnerable to scams may be the least likely to appreciate or utilize AI-generated assistance, highlighting the importance of designing not just accurate but also user-trusted and engaging AI support systems.

While the reasoning component did not significantly improve objective detection accuracy, it offered educational value, especially for difficult scam types and the younger user group who is resistant to the anticipatory warnings. These findings suggest that explainable AI holds promise for long-term user education, even where immediate accuracy gains are modest.

Looking ahead, the ASR framework lays the groundwork for more adaptive, transparent, and user-centered approaches to scam prevention. Future work should focus on expanding dataset diversity, improving real-world robustness, addressing adversarial adaptation, and optimizing user engagement strategies. Ultimately, the synergy between generative AI, human-centered design, and empirical evaluation offers a promising path forward in safeguarding individuals against the evolving threat of online scams.

{\captionsetup{justification=centering}
\begin{sidewaystable}[!htbp] \centering
\caption{Participant Accuracy in Identifying Scam vs. Non-Scam Scenarios within the \textit{Anticipatory} Component\\(Mean Across All 8 Scenarios).}
\label{tab:antioverall}
\begin{tabular}{@{\extracolsep{5pt}}lccccc}
\\[-1.8ex]\hline
\hline \\[-1.8ex]
& \multicolumn{5}{c}{\textit{Dependent variable: Participants Accuracy (Overall)}} \
\cr \cline{2-6}
\\[-1.8ex] & (1) & (2) & (3) & (4) & (5) \\
\hline \\[-1.8ex]
 AI Assisted & 0.057$^{**}$ & 0.062$^{**}$ & 0.059$^{*}$ & 0.062$^{**}$ & 0.060$^{**}$ \\
& (0.025) & (0.030) & (0.030) & (0.030) & (0.030) \\
 Age[Old (>44)] & & -0.018$^{}$ & -0.033$^{}$ & -0.041$^{}$ & -0.051$^{}$ \\
& & (0.031) & (0.034) & (0.034) & (0.034) \\
 Age[Young (<25)] & & 0.003$^{}$ & -0.016$^{}$ & -0.012$^{}$ & -0.020$^{}$ \\
& & (0.036) & (0.039) & (0.039) & (0.039) \\
 University Graduate & & & -0.043$^{}$ & -0.051$^{}$ & -0.050$^{}$ \\
& & & (0.038) & (0.038) & (0.037) \\
 Gender[Male] & & & & -0.017$^{}$ & -0.006$^{}$ \\
& & & & (0.025) & (0.026) \\
 Gender[Prefer not to say] & & & & -0.147$^{*}$ & -0.121$^{}$ \\
& & & & (0.081) & (0.082) \\
 STEM & & & & & -0.043$^{}$ \\
& & & & & (0.027) \\
 Intercept & 0.896$^{***}$ & 0.897$^{***}$ & 0.942$^{***}$ & 0.960$^{***}$ & 0.981$^{***}$ \\
& (0.018) & (0.026) & (0.047) & (0.048) & (0.050) \\
\hline \\[-1.8ex]
 Observations & 78 & 78 & 78 & 78 & 78 \\
 $R^2$ & 0.063 & 0.068 & 0.084 & 0.126 & 0.157 \\
 Adjusted $R^2$ & 0.051 & 0.030 & 0.034 & 0.052 & 0.073 \\
 Residual Std. Error & 0.110 (df=76) & 0.111 (df=74) & 0.111 (df=73) & 0.110 (df=71) & 0.109 (df=70) \\
 F Statistic & 5.118$^{**}$ (df=1; 76) & 1.793$^{}$ (df=3; 74) & 1.672$^{}$ (df=4; 73) & 1.707$^{}$ (df=6; 71) & 1.862$^{*}$ (df=7; 70) \\
\hline
\hline \\[-1.8ex]
\textit{Note:} & \multicolumn{5}{r}{$^{*}$p$<$0.1; $^{**}$p$<$0.05; $^{***}$p$<$0.01} \\
\end{tabular}
\end{sidewaystable}}

\begin{sidewaystable}[!htbp] \centering
\caption{Participant Accuracy in Identifying Authority Scam Scenario within the \textit{Anticipatory} Component.}
\label{tab:antiauthority}
\begin{tabular}{@{\extracolsep{5pt}}lccccc}
\\[-1.8ex]\hline
\hline \\[-1.8ex]
& \multicolumn{5}{c}{\textit{Dependent variable: Participants Accuracy (Authority)}} \
\cr \cline{2-6}
\\[-1.8ex] & (1) & (2) & (3) & (4) & (5) \\
\hline \\[-1.8ex]
 AI Assisted & 0.028$^{}$ & 0.004$^{}$ & 0.010$^{}$ & 0.011$^{}$ & 0.009$^{}$ \\
& (0.026) & (0.031) & (0.030) & (0.030) & (0.031) \\
 Age[Old (>44)] & & -0.001$^{}$ & 0.025$^{}$ & 0.027$^{}$ & 0.021$^{}$ \\
& & (0.032) & (0.034) & (0.034) & (0.035) \\
 Age[Young (<25)] & & -0.053$^{}$ & -0.022$^{}$ & -0.020$^{}$ & -0.025$^{}$ \\
& & (0.037) & (0.039) & (0.040) & (0.040) \\
 University Graduate & & & 0.071$^{*}$ & 0.077$^{**}$ & 0.077$^{**}$ \\
& & & (0.038) & (0.038) & (0.038) \\
 Gender[Male] & & & & -0.027$^{}$ & -0.021$^{}$ \\
& & & & (0.026) & (0.027) \\
 Gender[Prefer not to say] & & & & 0.038$^{}$ & 0.052$^{}$ \\
& & & & (0.082) & (0.084) \\
 STEM & & & & & -0.025$^{}$ \\
& & & & & (0.027) \\
 Intercept & 0.972$^{***}$ & 0.997$^{***}$ & 0.922$^{***}$ & 0.929$^{***}$ & 0.941$^{***}$ \\
& (0.019) & (0.026) & (0.047) & (0.049) & (0.051) \\
\hline \\[-1.8ex]
 Observations & 78 & 78 & 78 & 78 & 78 \\
 $R^2$ & 0.015 & 0.044 & 0.088 & 0.107 & 0.118 \\
 Adjusted $R^2$ & 0.002 & 0.005 & 0.038 & 0.032 & 0.029 \\
 Residual Std. Error & 0.113 (df=76) & 0.113 (df=74) & 0.111 (df=73) & 0.111 (df=71) & 0.112 (df=70) \\
 F Statistic & 1.169$^{}$ (df=1; 76) & 1.122$^{}$ (df=3; 74) & 1.763$^{}$ (df=4; 73) & 1.424$^{}$ (df=6; 71) & 1.333$^{}$ (df=7; 70) \\
\hline
\hline \\[-1.8ex]
\textit{Note:} & \multicolumn{5}{r}{$^{*}$p$<$0.1; $^{**}$p$<$0.05; $^{***}$p$<$0.01} \\
\end{tabular}
\end{sidewaystable}

\begin{sidewaystable}[!htbp] \centering
\caption{Participant Accuracy in Identifying Job Scam Scenario within the \textit{Anticipatory} Component.}
\label{tab:antijob}
\begin{tabular}{@{\extracolsep{5pt}}lccccc}
\\[-1.8ex]\hline
\hline \\[-1.8ex]
& \multicolumn{5}{c}{\textit{Dependent variable: Participants Accuracy (Job)}} \
\cr \cline{2-6}
\\[-1.8ex] & (1) & (2) & (3) & (4) & (5) \\
\hline \\[-1.8ex]
 AI Assisted & 0.313$^{***}$ & 0.235$^{**}$ & 0.229$^{**}$ & 0.238$^{**}$ & 0.232$^{**}$ \\
& (0.083) & (0.099) & (0.100) & (0.101) & (0.101) \\
 Age[Old (>44)] & & -0.020$^{}$ & -0.044$^{}$ & -0.060$^{}$ & -0.080$^{}$ \\
& & (0.103) & (0.112) & (0.113) & (0.115) \\
 Age[Young (<25)] & & -0.187$^{}$ & -0.217$^{}$ & -0.207$^{}$ & -0.224$^{*}$ \\
& & (0.118) & (0.131) & (0.131) & (0.132) \\
 University Graduate & & & -0.069$^{}$ & -0.085$^{}$ & -0.084$^{}$ \\
& & & (0.125) & (0.127) & (0.127) \\
 Gender[Male] & & & & -0.053$^{}$ & -0.030$^{}$ \\
& & & & (0.085) & (0.088) \\
 Gender[Prefer not to say] & & & & -0.319$^{}$ & -0.266$^{}$ \\
& & & & (0.272) & (0.278) \\
 STEM & & & & & -0.087$^{}$ \\
& & & & & (0.090) \\
 Intercept & 0.639$^{***}$ & 0.729$^{***}$ & 0.802$^{***}$ & 0.846$^{***}$ & 0.891$^{***}$ \\
& (0.061) & (0.085) & (0.157) & (0.162) & (0.169) \\
\hline \\[-1.8ex]
 Observations & 78 & 78 & 78 & 78 & 78 \\
 $R^2$ & 0.157 & 0.185 & 0.188 & 0.206 & 0.216 \\
 Adjusted $R^2$ & 0.146 & 0.152 & 0.144 & 0.139 & 0.138 \\
 Residual Std. Error & 0.367 (df=76) & 0.365 (df=74) & 0.367 (df=73) & 0.368 (df=71) & 0.368 (df=70) \\
 F Statistic & 14.180$^{***}$ (df=1; 76) & 5.597$^{***}$ (df=3; 74) & 4.234$^{***}$ (df=4; 73) & 3.067$^{**}$ (df=6; 71) & 2.759$^{**}$ (df=7; 70) \\
\hline
\hline \\[-1.8ex]
\textit{Note:} & \multicolumn{5}{r}{$^{*}$p$<$0.1; $^{**}$p$<$0.05; $^{***}$p$<$0.01} \\
\end{tabular}
\end{sidewaystable}

\begin{sidewaystable}[!htbp] \centering
\caption{Participant Accuracy in Identifying Investment Scam Scenario within the \textit{Anticipatory} Component.}
\label{tab:antiinvestment}
\begin{tabular}{@{\extracolsep{5pt}}lccccc}
\\[-1.8ex]\hline
\hline \\[-1.8ex]
& \multicolumn{5}{c}{\textit{Dependent variable: Participants Accuracy (Investment)}} \
\cr \cline{2-6}
\\[-1.8ex] & (1) & (2) & (3) & (4) & (5) \\
\hline \\[-1.8ex]
 AI Assisted & 0.004$^{}$ & 0.020$^{}$ & 0.020$^{}$ & 0.031$^{}$ & 0.030$^{}$ \\
& (0.036) & (0.043) & (0.044) & (0.038) & (0.039) \\
 Age[Old (>44)] & & 0.046$^{}$ & 0.047$^{}$ & 0.020$^{}$ & 0.018$^{}$ \\
& & (0.045) & (0.049) & (0.043) & (0.044) \\
 Age[Young (<25)] & & 0.060$^{}$ & 0.061$^{}$ & 0.071$^{}$ & 0.070$^{}$ \\
& & (0.052) & (0.057) & (0.050) & (0.051) \\
 University Graduate & & & 0.002$^{}$ & -0.030$^{}$ & -0.030$^{}$ \\
& & & (0.055) & (0.048) & (0.049) \\
 Gender[Male] & & & & -0.028$^{}$ & -0.026$^{}$ \\
& & & & (0.032) & (0.034) \\
 Gender[Prefer not to say] & & & & -0.524$^{***}$ & -0.519$^{***}$ \\
& & & & (0.104) & (0.106) \\
 STEM & & & & & -0.008$^{}$ \\
& & & & & (0.035) \\
 Intercept & 0.972$^{***}$ & 0.939$^{***}$ & 0.936$^{***}$ & 0.988$^{***}$ & 0.993$^{***}$ \\
& (0.027) & (0.037) & (0.069) & (0.062) & (0.065) \\
\hline \\[-1.8ex]
 Observations & 78 & 78 & 78 & 78 & 78 \\
 $R^2$ & 0.000 & 0.026 & 0.026 & 0.284 & 0.285 \\
 Adjusted $R^2$ & -0.013 & -0.013 & -0.027 & 0.224 & 0.213 \\
 Residual Std. Error & 0.160 (df=76) & 0.160 (df=74) & 0.161 (df=73) & 0.140 (df=71) & 0.141 (df=70) \\
 F Statistic & 0.012$^{}$ (df=1; 76) & 0.670$^{}$ (df=3; 74) & 0.497$^{}$ (df=4; 73) & 4.699$^{***}$ (df=6; 71) & 3.982$^{***}$ (df=7; 70) \\
\hline
\hline \\[-1.8ex]
\textit{Note:} & \multicolumn{5}{r}{$^{*}$p$<$0.1; $^{**}$p$<$0.05; $^{***}$p$<$0.01} \\
\end{tabular}
\end{sidewaystable}

\begin{sidewaystable}[!htbp] \centering
\caption{Participant Accuracy in Identifying Love Scam Scenario within the \textit{Anticipatory} Component.}
\label{tab:antilove}
\begin{tabular}{@{\extracolsep{5pt}}lccccc}
\\[-1.8ex]\hline
\hline \\[-1.8ex]
& \multicolumn{5}{c}{\textit{Dependent variable: Participants Accuracy (Love)}} \
\cr \cline{2-6}
\\[-1.8ex] & (1) & (2) & (3) & (4) & (5) \\
\hline \\[-1.8ex]
 AI Assisted & 0.008$^{}$ & 0.039$^{}$ & 0.040$^{}$ & 0.039$^{}$ & 0.041$^{}$ \\
& (0.051) & (0.060) & (0.060) & (0.061) & (0.062) \\
 Age[Old (>44)] & & 0.091$^{}$ & 0.093$^{}$ & 0.096$^{}$ & 0.103$^{}$ \\
& & (0.062) & (0.068) & (0.069) & (0.071) \\
 Age[Young (<25)] & & 0.120$^{*}$ & 0.123$^{}$ & 0.122$^{}$ & 0.128$^{}$ \\
& & (0.071) & (0.079) & (0.080) & (0.081) \\
 University Graduate & & & 0.005$^{}$ & 0.008$^{}$ & 0.008$^{}$ \\
& & & (0.076) & (0.077) & (0.078) \\
 Gender[Male] & & & & -0.002$^{}$ & -0.010$^{}$ \\
& & & & (0.052) & (0.054) \\
 Gender[Prefer not to say] & & & & 0.046$^{}$ & 0.027$^{}$ \\
& & & & (0.166) & (0.170) \\
 STEM & & & & & 0.031$^{}$ \\
& & & & & (0.055) \\
 Intercept & 0.944$^{***}$ & 0.877$^{***}$ & 0.872$^{***}$ & 0.869$^{***}$ & 0.853$^{***}$ \\
& (0.037) & (0.051) & (0.095) & (0.099) & (0.103) \\
\hline \\[-1.8ex]
 Observations & 78 & 78 & 78 & 78 & 78 \\
 $R^2$ & 0.000 & 0.054 & 0.054 & 0.056 & 0.060 \\
 Adjusted $R^2$ & -0.013 & 0.016 & 0.003 & -0.024 & -0.034 \\
 Residual Std. Error & 0.223 (df=76) & 0.220 (df=74) & 0.222 (df=73) & 0.225 (df=71) & 0.226 (df=70) \\
 F Statistic & 0.024$^{}$ (df=1; 76) & 1.418$^{}$ (df=3; 74) & 1.050$^{}$ (df=4; 73) & 0.696$^{}$ (df=6; 71) & 0.637$^{}$ (df=7; 70) \\
\hline
\hline \\[-1.8ex]
\textit{Note:} & \multicolumn{5}{r}{$^{*}$p$<$0.1; $^{**}$p$<$0.05; $^{***}$p$<$0.01} \\
\end{tabular}
\end{sidewaystable}

\begin{sidewaystable}[!htbp] \centering
\caption{Helpfulness of AI-Generated content within the \textit{Anticipatory} Component\\(Mean Across All 8 Scenarios).}
\label{tab:antihelpoverall}
\begin{tabular}{@{\extracolsep{5pt}}lccccc}
\\[-1.8ex]\hline
\hline \\[-1.8ex]
& \multicolumn{5}{c}{\textit{Dependent variable: Helpful (Overall)}} \
\cr \cline{2-6}
\\[-1.8ex] & (1) & (2) & (3) & (4) & (5) \\
\hline \\[-1.8ex]
 Accuracy & 0.962$^{*}$ & 0.990$^{*}$ & 0.792$^{}$ & 0.795$^{}$ & 0.774$^{}$ \\
& (0.499) & (0.511) & (0.507) & (0.665) & (0.670) \\
 Age[Old (>44)] & & -0.004$^{}$ & -0.131$^{}$ & -0.130$^{}$ & -0.112$^{}$ \\
& & (0.092) & (0.112) & (0.119) & (0.122) \\
 Age[Young (<25)] & & -0.187$^{}$ & -0.177$^{}$ & -0.186$^{}$ & -0.146$^{}$ \\
& & (0.291) & (0.282) & (0.295) & (0.301) \\
 University Graduate & & & -0.271$^{*}$ & -0.266$^{*}$ & -0.270$^{*}$ \\
& & & (0.146) & (0.156) & (0.157) \\
 Gender[Male] & & & & -0.014$^{}$ & -0.026$^{}$ \\
& & & & (0.093) & (0.095) \\
 Gender[Prefer not to say] & & & & -0.013$^{}$ & -0.052$^{}$ \\
& & & & (0.366) & (0.372) \\
 STEM & & & & & 0.071$^{}$ \\
& & & & & (0.096) \\
 Intercept & -0.282$^{}$ & -0.303$^{}$ & 0.157$^{}$ & 0.157$^{}$ & 0.142$^{}$ \\
& (0.477) & (0.490) & (0.535) & (0.697) & (0.702) \\
\hline \\[-1.8ex]
 Observations & 42 & 42 & 42 & 42 & 42 \\
 $R^2$ & 0.085 & 0.095 & 0.172 & 0.173 & 0.186 \\
 Adjusted $R^2$ & 0.062 & 0.023 & 0.083 & 0.031 & 0.019 \\
 Residual Std. Error & 0.278 (df=40) & 0.284 (df=38) & 0.275 (df=37) & 0.283 (df=35) & 0.285 (df=34) \\
 F Statistic & 3.712$^{*}$ (df=1; 40) & 1.327$^{}$ (df=3; 38) & 1.928$^{}$ (df=4; 37) & 1.221$^{}$ (df=6; 35) & 1.111$^{}$ (df=7; 34) \\
\hline
\hline \\[-1.8ex]
\textit{Note:} & \multicolumn{5}{r}{$^{*}$p$<$0.1; $^{**}$p$<$0.05; $^{***}$p$<$0.01} \\
\end{tabular}
\end{sidewaystable}

\begin{sidewaystable}[!htbp] \centering
\caption{Helpfulness of AI-Generated content in Authority Scam within the \textit{Anticipatory} Component.}
\label{tab:antihelpauthority}
\begin{tabular}{@{\extracolsep{5pt}}lccccc}
\\[-1.8ex]\hline
\hline \\[-1.8ex]
& \multicolumn{5}{c}{\textit{Dependent variable: Helpful (Authority)}} \
\cr \cline{2-6}
\\[-1.8ex] & (1) & (2) & (3) & (4) & (5) \\
\hline \\[-1.8ex]
 Accuracy & 0.452$^{***}$ & 0.462$^{***}$ & 0.587$^{***}$ & 0.592$^{***}$ & 0.618$^{***}$ \\
& (0.023) & (0.030) & (0.082) & (0.085) & (0.089) \\
 Age[Old (>44)] & & -0.056$^{}$ & -0.173$^{}$ & -0.166$^{}$ & -0.193$^{}$ \\
& & (0.098) & (0.120) & (0.123) & (0.126) \\
 Age[Young (<25)] & & 0.077$^{}$ & 0.077$^{}$ & 0.043$^{}$ & -0.011$^{}$ \\
& & (0.309) & (0.302) & (0.314) & (0.319) \\
 University Graduate & & & -0.250$^{}$ & -0.227$^{}$ & -0.224$^{}$ \\
& & & (0.154) & (0.160) & (0.160) \\
 Gender[Male] & & & & -0.067$^{}$ & -0.050$^{}$ \\
& & & & (0.098) & (0.100) \\
 Gender[Prefer not to say] & & & & 0.043$^{}$ & 0.087$^{}$ \\
& & & & (0.314) & (0.317) \\
 STEM & & & & & -0.098$^{}$ \\
& & & & & (0.102) \\
 Intercept & 0.452$^{***}$ & 0.462$^{***}$ & 0.587$^{***}$ & 0.592$^{***}$ & 0.618$^{***}$ \\
& (0.023) & (0.030) & (0.082) & (0.085) & (0.089) \\
\hline \\[-1.8ex]
 Observations & 42 & 42 & 42 & 42 & 42 \\
 $R^2$ & 0.000 & 0.011 & 0.075 & 0.089 & 0.112 \\
 Adjusted $R^2$ & 0.000 & -0.040 & 0.002 & -0.038 & -0.040 \\
 Residual Std. Error & 0.297 (df=41) & 0.303 (df=39) & 0.297 (df=38) & 0.303 (df=36) & 0.303 (df=35) \\
 F Statistic & nan$^{***}$ (df=0; 41) & 0.216$^{}$ (df=2; 39) & 1.033$^{}$ (df=3; 38) & 0.701$^{}$ (df=5; 36) & 0.737$^{}$ (df=6; 35) \\
\hline
\hline \\[-1.8ex]
\textit{Note:} & \multicolumn{5}{r}{$^{*}$p$<$0.1; $^{**}$p$<$0.05; $^{***}$p$<$0.01} \\
\end{tabular}
\end{sidewaystable}

\begin{sidewaystable}[!htbp] \centering
\caption{Helpfulness of AI-Generated content in Job Scam within the \textit{Anticipatory} Component.}
\label{tab:antihelpjob}
\begin{tabular}{@{\extracolsep{5pt}}lccccc}
\\[-1.8ex]\hline
\hline \\[-1.8ex]
& \multicolumn{5}{c}{\textit{Dependent variable: Helpful (Job)}} \
\cr \cline{2-6}
\\[-1.8ex] & (1) & (2) & (3) & (4) & (5) \\
\hline \\[-1.8ex]
 Accuracy & 0.825$^{***}$ & 0.848$^{***}$ & 0.848$^{***}$ & 0.838$^{*}$ & 0.866$^{**}$ \\
& (0.275) & (0.265) & (0.273) & (0.417) & (0.422) \\
 Age[Old (>44)] & & 0.016$^{}$ & 0.016$^{}$ & 0.013$^{}$ & -0.007$^{}$ \\
& & (0.119) & (0.152) & (0.164) & (0.168) \\
 Age[Young (<25)] & & -0.840$^{**}$ & -0.840$^{**}$ & -0.825$^{**}$ & -0.874$^{**}$ \\
& & (0.373) & (0.378) & (0.394) & (0.404) \\
 University Graduate & & & -0.001$^{}$ & -0.011$^{}$ & -0.004$^{}$ \\
& & & (0.195) & (0.211) & (0.212) \\
 Gender[Male] & & & & 0.027$^{}$ & 0.039$^{}$ \\
& & & & (0.127) & (0.129) \\
 Gender[Prefer not to say] & & & & 0.013$^{}$ & 0.080$^{}$ \\
& & & & (0.563) & (0.576) \\
 STEM & & & & & -0.087$^{}$ \\
& & & & & (0.130) \\
 Intercept & 0.000$^{}$ & -0.008$^{}$ & -0.007$^{}$ & -0.002$^{}$ & 0.012$^{}$ \\
& (0.269) & (0.265) & (0.361) & (0.509) & (0.514) \\
\hline \\[-1.8ex]
 Observations & 42 & 42 & 42 & 42 & 42 \\
 $R^2$ & 0.183 & 0.282 & 0.282 & 0.283 & 0.293 \\
 Adjusted $R^2$ & 0.163 & 0.226 & 0.205 & 0.160 & 0.147 \\
 Residual Std. Error & 0.380 (df=40) & 0.365 (df=38) & 0.370 (df=37) & 0.381 (df=35) & 0.384 (df=34) \\
 F Statistic & 8.980$^{***}$ (df=1; 40) & 4.985$^{***}$ (df=3; 38) & 3.640$^{**}$ (df=4; 37) & 2.306$^{*}$ (df=6; 35) & 2.010$^{*}$ (df=7; 34) \\
\hline
\hline \\[-1.8ex]
\textit{Note:} & \multicolumn{5}{r}{$^{*}$p$<$0.1; $^{**}$p$<$0.05; $^{***}$p$<$0.01} \\
\end{tabular}
\end{sidewaystable}

\begin{sidewaystable}[!htbp] \centering
\caption{Helpfulness of AI-Generated content in Investment Scam within the \textit{Anticipatory} Component.}
\label{tab:antihelpinvestment}
\begin{tabular}{@{\extracolsep{5pt}}lccccc}
\\[-1.8ex]\hline
\hline \\[-1.8ex]
& \multicolumn{5}{c}{\textit{Dependent variable: Helpful (Investment)}} \
\cr \cline{2-6}
\\[-1.8ex] & (1) & (2) & (3) & (4) & (5) \\
\hline \\[-1.8ex]
 Accuracy & 0.756$^{*}$ & 0.800$^{*}$ & 0.800$^{*}$ & 0.584$^{***}$ & 0.586$^{***}$ \\
& (0.440) & (0.448) & (0.453) & (0.183) & (0.184) \\
 Age[Old (>44)] & & -0.133$^{}$ & -0.175$^{}$ & -0.170$^{}$ & -0.200$^{}$ \\
& & (0.143) & (0.180) & (0.182) & (0.188) \\
 Age[Young (<25)] & & 0.200$^{}$ & 0.200$^{}$ & 0.155$^{}$ & 0.095$^{}$ \\
& & (0.448) & (0.453) & (0.465) & (0.475) \\
 University Graduate & & & -0.089$^{}$ & -0.062$^{}$ & -0.059$^{}$ \\
& & & (0.230) & (0.237) & (0.239) \\
 Gender[Male] & & & & -0.080$^{}$ & -0.061$^{}$ \\
& & & & (0.146) & (0.149) \\
 Gender[Prefer not to say] & & & & -0.261$^{}$ & -0.208$^{}$ \\
& & & & (0.313) & (0.324) \\
 STEM & & & & & -0.111$^{}$ \\
& & & & & (0.152) \\
 Intercept & 0.000$^{}$ & 0.000$^{}$ & 0.089$^{}$ & 0.323$^{}$ & 0.378$^{}$ \\
& (0.435) & (0.439) & (0.500) & (0.217) & (0.231) \\
\hline \\[-1.8ex]
 Observations & 42 & 42 & 42 & 42 & 42 \\
 $R^2$ & 0.069 & 0.097 & 0.100 & 0.108 & 0.121 \\
 Adjusted $R^2$ & 0.045 & 0.025 & 0.003 & -0.016 & -0.029 \\
 Residual Std. Error & 0.435 (df=40) & 0.439 (df=38) & 0.444 (df=37) & 0.449 (df=36) & 0.452 (df=35) \\
 F Statistic & 2.952$^{*}$ (df=1; 40) & 1.357$^{}$ (df=3; 38) & 1.033$^{}$ (df=4; 37) & 0.870$^{}$ (df=5; 36) & 0.804$^{}$ (df=6; 35) \\
\hline
\hline \\[-1.8ex]
\textit{Note:} & \multicolumn{5}{r}{$^{*}$p$<$0.1; $^{**}$p$<$0.05; $^{***}$p$<$0.01} \\
\end{tabular}
\end{sidewaystable}

\begin{sidewaystable}[!htbp] \centering
\caption{Helpfulness of AI-Generated content in Love Scam within the \textit{Anticipatory} Component.}
\label{tab:antihelplove}
\begin{tabular}{@{\extracolsep{5pt}}lccccc}
\\[-1.8ex]\hline
\hline \\[-1.8ex]
& \multicolumn{5}{c}{\textit{Dependent variable: Helpful (Love)}} \
\cr \cline{2-6}
\\[-1.8ex] & (1) & (2) & (3) & (4) & (5) \\
\hline \\[-1.8ex]
 Accuracy & 0.325$^{}$ & 0.417$^{}$ & 0.417$^{}$ & 0.406$^{}$ & 0.327$^{}$ \\
& (0.287) & (0.284) & (0.287) & (0.293) & (0.310) \\
 Age[Old (>44)] & & -0.250$^{*}$ & -0.292$^{*}$ & -0.295$^{*}$ & -0.257$^{}$ \\
& & (0.127) & (0.159) & (0.163) & (0.170) \\
 Age[Young (<25)] & & 0.083$^{}$ & 0.083$^{}$ & 0.151$^{}$ & 0.221$^{}$ \\
& & (0.394) & (0.398) & (0.412) & (0.422) \\
 University Graduate & & & -0.089$^{}$ & -0.128$^{}$ & -0.132$^{}$ \\
& & & (0.202) & (0.210) & (0.211) \\
 Gender[Male] & & & & 0.114$^{}$ & 0.095$^{}$ \\
& & & & (0.129) & (0.132) \\
 Gender[Prefer not to say] & & & & 0.151$^{}$ & 0.106$^{}$ \\
& & & & (0.412) & (0.417) \\
 STEM & & & & & 0.115$^{}$ \\
& & & & & (0.142) \\
 Intercept & 0.500$^{*}$ & 0.500$^{*}$ & 0.589$^{*}$ & 0.571$^{}$ & 0.584$^{}$ \\
& (0.280) & (0.273) & (0.342) & (0.348) & (0.350) \\
\hline \\[-1.8ex]
 Observations & 42 & 42 & 42 & 42 & 42 \\
 $R^2$ & 0.031 & 0.125 & 0.130 & 0.150 & 0.166 \\
 Adjusted $R^2$ & 0.007 & 0.056 & 0.035 & 0.004 & -0.006 \\
 Residual Std. Error & 0.396 (df=40) & 0.386 (df=38) & 0.390 (df=37) & 0.397 (df=35) & 0.399 (df=34) \\
 F Statistic & 1.282$^{}$ (df=1; 40) & 1.810$^{}$ (df=3; 38) & 1.377$^{}$ (df=4; 37) & 1.027$^{}$ (df=6; 35) & 0.967$^{}$ (df=7; 34) \\
\hline
\hline \\[-1.8ex]
\textit{Note:} & \multicolumn{5}{r}{$^{*}$p$<$0.1; $^{**}$p$<$0.05; $^{***}$p$<$0.01} \\
\end{tabular}
\end{sidewaystable}

{\captionsetup{justification=centering}
\begin{sidewaystable}[!htbp] \centering
\caption{Participant Accuracy in Identifying Scam vs. Non-Scam Scenarios within the \textit{Reason} Component\\(Mean Across All 8 Scenarios).}
\label{tab:reasonoverall}
\begin{tabular}{@{\extracolsep{5pt}}lccccc}
\\[-1.8ex]\hline
\hline \\[-1.8ex]
& \multicolumn{5}{c}{\textit{Dependent variable: Participants Accuracy (Overall)}} \
\cr \cline{2-6}
\\[-1.8ex] & (1) & (2) & (3) & (4) & (5) \\
\hline \\[-1.8ex]
 AI Assisted & -0.017$^{}$ & -0.012$^{}$ & -0.014$^{}$ & -0.014$^{}$ & -0.014$^{}$ \\
& (0.028) & (0.034) & (0.034) & (0.034) & (0.035) \\
 Age[Old (>44)] & & 0.048$^{}$ & 0.038$^{}$ & 0.034$^{}$ & 0.034$^{}$ \\
& & (0.035) & (0.038) & (0.039) & (0.040) \\
 Age[Young (<25)] & & 0.038$^{}$ & 0.026$^{}$ & 0.026$^{}$ & 0.025$^{}$ \\
& & (0.040) & (0.044) & (0.045) & (0.045) \\
 University Graduate & & & -0.027$^{}$ & -0.034$^{}$ & -0.034$^{}$ \\
& & & (0.042) & (0.043) & (0.043) \\
 Gender[Male] & & & & 0.015$^{}$ & 0.015$^{}$ \\
& & & & (0.029) & (0.030) \\
 Gender[Prefer not to say] & & & & -0.072$^{}$ & -0.071$^{}$ \\
& & & & (0.093) & (0.095) \\
 STEM & & & & & -0.002$^{}$ \\
& & & & & (0.031) \\
 Intercept & 0.951$^{***}$ & 0.928$^{***}$ & 0.957$^{***}$ & 0.958$^{***}$ & 0.959$^{***}$ \\
& (0.021) & (0.029) & (0.053) & (0.055) & (0.058) \\
\hline \\[-1.8ex]
 Observations & 78 & 78 & 78 & 78 & 78 \\
 $R^2$ & 0.005 & 0.036 & 0.041 & 0.055 & 0.055 \\
 Adjusted $R^2$ & -0.008 & -0.003 & -0.011 & -0.025 & -0.040 \\
 Residual Std. Error & 0.124 (df=76) & 0.124 (df=74) & 0.124 (df=73) & 0.125 (df=71) & 0.126 (df=70) \\
 F Statistic & 0.357$^{}$ (df=1; 76) & 0.916$^{}$ (df=3; 74) & 0.786$^{}$ (df=4; 73) & 0.685$^{}$ (df=6; 71) & 0.579$^{}$ (df=7; 70) \\
\hline
\hline \\[-1.8ex]
\textit{Note:} & \multicolumn{5}{r}{$^{*}$p$<$0.1; $^{**}$p$<$0.05; $^{***}$p$<$0.01} \\
\end{tabular}
\end{sidewaystable}}

\begin{sidewaystable}[!htbp] \centering
\caption{Participant Accuracy in Identifying Authority Scam Scenario within the \textit{Reason} Component.}
\label{tab:reasonauthority}
\begin{tabular}{@{\extracolsep{5pt}}lccccc}
\\[-1.8ex]\hline
\hline \\[-1.8ex]
& \multicolumn{5}{c}{\textit{Dependent variable: Participants Accuracy (Authority)}} \
\cr \cline{2-6}
\\[-1.8ex] & (1) & (2) & (3) & (4) & (5) \\
\hline \\[-1.8ex]
 AI Assisted & -0.071$^{}$ & -0.081$^{}$ & -0.081$^{}$ & -0.073$^{}$ & -0.068$^{}$ \\
& (0.043) & (0.052) & (0.052) & (0.049) & (0.048) \\
 Age[Old (>44)] & & 0.086$^{}$ & 0.082$^{}$ & 0.057$^{}$ & 0.074$^{}$ \\
& & (0.054) & (0.059) & (0.055) & (0.055) \\
 Age[Young (<25)] & & 0.027$^{}$ & 0.022$^{}$ & 0.029$^{}$ & 0.043$^{}$ \\
& & (0.062) & (0.068) & (0.063) & (0.063) \\
 University Graduate & & & -0.010$^{}$ & -0.043$^{}$ & -0.045$^{}$ \\
& & & (0.065) & (0.061) & (0.060) \\
 Gender[Male] & & & & 0.004$^{}$ & -0.016$^{}$ \\
& & & & (0.041) & (0.042) \\
 Gender[Prefer not to say] & & & & -0.483$^{***}$ & -0.529$^{***}$ \\
& & & & (0.132) & (0.132) \\
 STEM & & & & & 0.076$^{*}$ \\
& & & & & (0.043) \\
 Intercept & 1.000$^{***}$ & 0.978$^{***}$ & 0.989$^{***}$ & 1.027$^{***}$ & 0.988$^{***}$ \\
& (0.032) & (0.044) & (0.082) & (0.078) & (0.080) \\
\hline \\[-1.8ex]
 Observations & 78 & 78 & 78 & 78 & 78 \\
 $R^2$ & 0.034 & 0.067 & 0.067 & 0.220 & 0.253 \\
 Adjusted $R^2$ & 0.022 & 0.029 & 0.016 & 0.154 & 0.178 \\
 Residual Std. Error & 0.191 (df=76) & 0.191 (df=74) & 0.192 (df=73) & 0.178 (df=71) & 0.175 (df=70) \\
 F Statistic & 2.698$^{}$ (df=1; 76) & 1.765$^{}$ (df=3; 74) & 1.312$^{}$ (df=4; 73) & 3.332$^{***}$ (df=6; 71) & 3.389$^{***}$ (df=7; 70) \\
\hline
\hline \\[-1.8ex]
\textit{Note:} & \multicolumn{5}{r}{$^{*}$p$<$0.1; $^{**}$p$<$0.05; $^{***}$p$<$0.01} \\
\end{tabular}
\end{sidewaystable}

\begin{sidewaystable}[!htbp] \centering
\caption{Participant Accuracy in Identifying Job Scam Scenario within the \textit{Reason} Component.}
\label{tab:reasonjob}
\begin{tabular}{@{\extracolsep{5pt}}lccccc}
\\[-1.8ex]\hline
\hline \\[-1.8ex]
& \multicolumn{5}{c}{\textit{Dependent variable: Participants Accuracy (Job)}} \
\cr \cline{2-6}
\\[-1.8ex] & (1) & (2) & (3) & (4) & (5) \\
\hline \\[-1.8ex]
 AI Assisted & 0.040$^{}$ & -0.010$^{}$ & -0.006$^{}$ & -0.007$^{}$ & -0.005$^{}$ \\
& (0.066) & (0.079) & (0.079) & (0.077) & (0.078) \\
 Age[Old (>44)] & & 0.022$^{}$ & 0.041$^{}$ & 0.022$^{}$ & 0.030$^{}$ \\
& & (0.081) & (0.089) & (0.087) & (0.089) \\
 Age[Young (<25)] & & -0.100$^{}$ & -0.077$^{}$ & -0.081$^{}$ & -0.075$^{}$ \\
& & (0.094) & (0.104) & (0.101) & (0.102) \\
 University Graduate & & & 0.053$^{}$ & 0.019$^{}$ & 0.018$^{}$ \\
& & & (0.099) & (0.098) & (0.098) \\
 Gender[Male] & & & & 0.108$^{}$ & 0.099$^{}$ \\
& & & & (0.065) & (0.068) \\
 Gender[Prefer not to say] & & & & -0.333$^{}$ & -0.354$^{}$ \\
& & & & (0.210) & (0.215) \\
 STEM & & & & & 0.034$^{}$ \\
& & & & & (0.070) \\
 Intercept & 0.889$^{***}$ & 0.933$^{***}$ & 0.878$^{***}$ & 0.868$^{***}$ & 0.850$^{***}$ \\
& (0.048) & (0.067) & (0.124) & (0.125) & (0.130) \\
\hline \\[-1.8ex]
 Observations & 78 & 78 & 78 & 78 & 78 \\
 $R^2$ & 0.005 & 0.023 & 0.027 & 0.105 & 0.108 \\
 Adjusted $R^2$ & -0.008 & -0.017 & -0.027 & 0.029 & 0.019 \\
 Residual Std. Error & 0.289 (df=76) & 0.290 (df=74) & 0.291 (df=73) & 0.283 (df=71) & 0.285 (df=70) \\
 F Statistic & 0.366$^{}$ (df=1; 76) & 0.576$^{}$ (df=3; 74) & 0.498$^{}$ (df=4; 73) & 1.389$^{}$ (df=6; 71) & 1.212$^{}$ (df=7; 70) \\
\hline
\hline \\[-1.8ex]
\textit{Note:} & \multicolumn{5}{r}{$^{*}$p$<$0.1; $^{**}$p$<$0.05; $^{***}$p$<$0.01} \\
\end{tabular}
\end{sidewaystable}

\begin{sidewaystable}[!htbp] \centering
\caption{Participant Accuracy in Identifying Investment Scam Scenario within the \textit{Reason} Component.}
\label{tab:reasoninvestment}
\begin{tabular}{@{\extracolsep{5pt}}lccccc}
\\[-1.8ex]\hline
\hline \\[-1.8ex]
& \multicolumn{5}{c}{\textit{Dependent variable: Participants Accuracy (Investment)}} \
\cr \cline{2-6}
\\[-1.8ex] & (1) & (2) & (3) & (4) & (5) \\
\hline \\[-1.8ex]
 AI Assisted & -0.020$^{}$ & -0.007$^{}$ & -0.007$^{}$ & -0.010$^{}$ & -0.007$^{}$ \\
& (0.044) & (0.053) & (0.053) & (0.054) & (0.054) \\
 Age[Old (>44)] & & 0.074$^{}$ & 0.074$^{}$ & 0.076$^{}$ & 0.086$^{}$ \\
& & (0.054) & (0.060) & (0.061) & (0.062) \\
 Age[Young (<25)] & & 0.069$^{}$ & 0.069$^{}$ & 0.066$^{}$ & 0.074$^{}$ \\
& & (0.063) & (0.069) & (0.070) & (0.071) \\
 University Graduate & & & -0.001$^{}$ & 0.000$^{}$ & -0.001$^{}$ \\
& & & (0.066) & (0.068) & (0.068) \\
 Gender[Male] & & & & 0.027$^{}$ & 0.015$^{}$ \\
& & & & (0.045) & (0.047) \\
 Gender[Prefer not to say] & & & & 0.054$^{}$ & 0.027$^{}$ \\
& & & & (0.146) & (0.149) \\
 STEM & & & & & 0.044$^{}$ \\
& & & & & (0.048) \\
 Intercept & 0.972$^{***}$ & 0.931$^{***}$ & 0.932$^{***}$ & 0.919$^{***}$ & 0.896$^{***}$ \\
& (0.032) & (0.045) & (0.083) & (0.087) & (0.090) \\
\hline \\[-1.8ex]
 Observations & 78 & 78 & 78 & 78 & 78 \\
 $R^2$ & 0.003 & 0.036 & 0.036 & 0.042 & 0.053 \\
 Adjusted $R^2$ & -0.010 & -0.003 & -0.016 & -0.039 & -0.042 \\
 Residual Std. Error & 0.195 (df=76) & 0.194 (df=74) & 0.195 (df=73) & 0.197 (df=71) & 0.198 (df=70) \\
 F Statistic & 0.202$^{}$ (df=1; 76) & 0.930$^{}$ (df=3; 74) & 0.688$^{}$ (df=4; 73) & 0.520$^{}$ (df=6; 71) & 0.562$^{}$ (df=7; 70) \\
\hline
\hline \\[-1.8ex]
\textit{Note:} & \multicolumn{5}{r}{$^{*}$p$<$0.1; $^{**}$p$<$0.05; $^{***}$p$<$0.01} \\
\end{tabular}
\end{sidewaystable}

\begin{sidewaystable}[!htbp] \centering
\caption{Participant Accuracy in Identifying Love Scam Scenario within the \textit{Reason} Component.}
\label{tab:reasonlove}
\begin{tabular}{@{\extracolsep{5pt}}lccccc}
\\[-1.8ex]\hline
\hline \\[-1.8ex]
& \multicolumn{5}{c}{\textit{Dependent variable: Participants Accuracy (Love)}} \
\cr \cline{2-6}
\\[-1.8ex] & (1) & (2) & (3) & (4) & (5) \\
\hline \\[-1.8ex]
 AI Assisted & -0.012$^{}$ & -0.041$^{}$ & -0.052$^{}$ & -0.052$^{}$ & -0.049$^{}$ \\
& (0.066) & (0.079) & (0.079) & (0.080) & (0.081) \\
 Age[Old (>44)] & & 0.051$^{}$ & 0.002$^{}$ & 0.006$^{}$ & 0.018$^{}$ \\
& & (0.082) & (0.089) & (0.091) & (0.092) \\
 Age[Young (<25)] & & -0.037$^{}$ & -0.098$^{}$ & -0.097$^{}$ & -0.087$^{}$ \\
& & (0.094) & (0.103) & (0.105) & (0.106) \\
 University Graduate & & & -0.138$^{}$ & -0.132$^{}$ & -0.133$^{}$ \\
& & & (0.099) & (0.101) & (0.102) \\
 Gender[Male] & & & & -0.019$^{}$ & -0.033$^{}$ \\
& & & & (0.068) & (0.070) \\
 Gender[Prefer not to say] & & & & 0.065$^{}$ & 0.032$^{}$ \\
& & & & (0.217) & (0.222) \\
 STEM & & & & & 0.055$^{}$ \\
& & & & & (0.072) \\
 Intercept & 0.917$^{***}$ & 0.929$^{***}$ & 1.074$^{***}$ & 1.075$^{***}$ & 1.047$^{***}$ \\
& (0.048) & (0.068) & (0.124) & (0.129) & (0.135) \\
\hline \\[-1.8ex]
 Observations & 78 & 78 & 78 & 78 & 78 \\
 $R^2$ & 0.000 & 0.009 & 0.035 & 0.038 & 0.046 \\
 Adjusted $R^2$ & -0.013 & -0.031 & -0.018 & -0.043 & -0.050 \\
 Residual Std. Error & 0.289 (df=76) & 0.292 (df=74) & 0.290 (df=73) & 0.294 (df=71) & 0.295 (df=70) \\
 F Statistic & 0.033$^{}$ (df=1; 76) & 0.235$^{}$ (df=3; 74) & 0.668$^{}$ (df=4; 73) & 0.467$^{}$ (df=6; 71) & 0.481$^{}$ (df=7; 70) \\
\hline
\hline \\[-1.8ex]
\textit{Note:} & \multicolumn{5}{r}{$^{*}$p$<$0.1; $^{**}$p$<$0.05; $^{***}$p$<$0.01} \\
\end{tabular}
\end{sidewaystable}

\begin{sidewaystable}[!htbp] \centering
\caption{Helpfulness of AI-Generated content within the \textit{Reason} Component\\(Mean Across All 8 Scenarios).}
\label{tab:reasonhelpoverall}
\begin{tabular}{@{\extracolsep{5pt}}lccccc}
\\[-1.8ex]\hline
\hline \\[-1.8ex]
& \multicolumn{5}{c}{\textit{Dependent variable: Helpful (Overall)}} \
\cr \cline{2-6}
\\[-1.8ex] & (1) & (2) & (3) & (4) & (5) \\
\hline \\[-1.8ex]
 Accuracy & 0.880$^{*}$ & 0.841$^{}$ & 0.748$^{}$ & 0.772$^{}$ & 0.791$^{}$ \\
& (0.498) & (0.517) & (0.531) & (0.539) & (0.555) \\
 Age[Old (>44)] & & 0.005$^{}$ & -0.069$^{}$ & -0.065$^{}$ & -0.072$^{}$ \\
& & (0.123) & (0.151) & (0.154) & (0.160) \\
 Age[Young (<25)] & & 0.287$^{}$ & 0.295$^{}$ & 0.217$^{}$ & 0.202$^{}$ \\
& & (0.382) & (0.384) & (0.395) & (0.408) \\
 University Graduate & & & -0.167$^{}$ & -0.122$^{}$ & -0.120$^{}$ \\
& & & (0.198) & (0.205) & (0.208) \\
 Gender[Male] & & & & -0.127$^{}$ & -0.123$^{}$ \\
& & & & (0.123) & (0.127) \\
 Gender[Prefer not to say] & & & & -0.186$^{}$ & -0.174$^{}$ \\
& & & & (0.392) & (0.403) \\
 STEM & & & & & -0.025$^{}$ \\
& & & & & (0.131) \\
 Intercept & -0.156$^{}$ & -0.128$^{}$ & 0.125$^{}$ & 0.133$^{}$ & 0.127$^{}$ \\
& (0.469) & (0.480) & (0.568) & (0.577) & (0.586) \\
\hline \\[-1.8ex]
 Observations & 42 & 42 & 42 & 42 & 42 \\
 $R^2$ & 0.072 & 0.086 & 0.103 & 0.132 & 0.133 \\
 Adjusted $R^2$ & 0.049 & 0.014 & 0.006 & -0.017 & -0.046 \\
 Residual Std. Error & 0.366 (df=40) & 0.373 (df=38) & 0.374 (df=37) & 0.378 (df=35) & 0.384 (df=34) \\
 F Statistic & 3.116$^{*}$ (df=1; 40) & 1.191$^{}$ (df=3; 38) & 1.064$^{}$ (df=4; 37) & 0.886$^{}$ (df=6; 35) & 0.743$^{}$ (df=7; 34) \\
\hline
\hline \\[-1.8ex]
\textit{Note:} & \multicolumn{5}{r}{$^{*}$p$<$0.1; $^{**}$p$<$0.05; $^{***}$p$<$0.01} \\
\end{tabular}
\end{sidewaystable}

\begin{sidewaystable}[!htbp] \centering
\caption{Helpfulness of AI-Generated content in Authority Scam within the \textit{Reason} Component.}
\label{tab:reasonhelpauthority}
\label{tab:}
\begin{tabular}{@{\extracolsep{5pt}}lccccc}
\\[-1.8ex]\hline
\hline \\[-1.8ex]
& \multicolumn{5}{c}{\textit{Dependent variable: Helpful (Authority)}} \
\cr \cline{2-6}
\\[-1.8ex] & (1) & (2) & (3) & (4) & (5) \\
\hline \\[-1.8ex]
 Accuracy & 0.103$^{}$ & 0.072$^{}$ & 0.072$^{}$ & 0.255$^{}$ & 0.198$^{}$ \\
& (0.261) & (0.273) & (0.267) & (0.312) & (0.332) \\
 Age[Old (>44)] & & 0.061$^{}$ & -0.114$^{}$ & -0.100$^{}$ & -0.073$^{}$ \\
& & (0.147) & (0.178) & (0.174) & (0.183) \\
 Age[Young (<25)] & & 0.261$^{}$ & 0.261$^{}$ & 0.127$^{}$ & 0.177$^{}$ \\
& & (0.454) & (0.444) & (0.440) & (0.454) \\
 University Graduate & & & -0.375$^{}$ & -0.295$^{}$ & -0.298$^{}$ \\
& & & (0.225) & (0.224) & (0.227) \\
 Gender[Male] & & & & -0.236$^{*}$ & -0.249$^{*}$ \\
& & & & (0.138) & (0.141) \\
 Gender[Prefer not to say] & & & & 0.382$^{}$ & 0.293$^{}$ \\
& & & & (0.523) & (0.554) \\
 STEM & & & & & 0.083$^{}$ \\
& & & & & (0.152) \\
 Intercept & 0.667$^{**}$ & 0.667$^{**}$ & 1.042$^{***}$ & 0.913$^{**}$ & 0.922$^{**}$ \\
& (0.251) & (0.257) & (0.337) & (0.372) & (0.376) \\
\hline \\[-1.8ex]
 Observations & 42 & 42 & 42 & 42 & 42 \\
 $R^2$ & 0.004 & 0.015 & 0.084 & 0.175 & 0.183 \\
 Adjusted $R^2$ & -0.021 & -0.062 & -0.015 & 0.034 & 0.014 \\
 Residual Std. Error & 0.436 (df=40) & 0.444 (df=38) & 0.434 (df=37) & 0.424 (df=35) & 0.428 (df=34) \\
 F Statistic & 0.154$^{}$ (df=1; 40) & 0.199$^{}$ (df=3; 38) & 0.852$^{}$ (df=4; 37) & 1.241$^{}$ (df=6; 35) & 1.085$^{}$ (df=7; 34) \\
\hline
\hline \\[-1.8ex]
\textit{Note:} & \multicolumn{5}{r}{$^{*}$p$<$0.1; $^{**}$p$<$0.05; $^{***}$p$<$0.01} \\
\end{tabular}
\end{sidewaystable}

\begin{sidewaystable}[!htbp] \centering
\caption{Helpfulness of AI-Generated content in Job Scam within the \textit{Reason} Component.}
\label{tab:reasonhelpjob}
\begin{tabular}{@{\extracolsep{5pt}}lccccc}
\\[-1.8ex]\hline
\hline \\[-1.8ex]
& \multicolumn{5}{c}{\textit{Dependent variable: Helpful (Job)}} \
\cr \cline{2-6}
\\[-1.8ex] & (1) & (2) & (3) & (4) & (5) \\
\hline \\[-1.8ex]
 Accuracy & 0.513$^{**}$ & 0.506$^{**}$ & 0.473$^{*}$ & 0.519$^{**}$ & 0.530$^{**}$ \\
& (0.227) & (0.231) & (0.234) & (0.235) & (0.239) \\
 Age[Old (>44)] & & 0.092$^{}$ & 0.004$^{}$ & 0.026$^{}$ & 0.012$^{}$ \\
& & (0.125) & (0.156) & (0.156) & (0.162) \\
 Age[Young (<25)] & & 0.192$^{}$ & 0.194$^{}$ & 0.093$^{}$ & 0.064$^{}$ \\
& & (0.393) & (0.393) & (0.398) & (0.410) \\
 University Graduate & & & -0.191$^{}$ & -0.121$^{}$ & -0.118$^{}$ \\
& & & (0.202) & (0.206) & (0.209) \\
 Gender[Male] & & & & -0.188$^{}$ & -0.180$^{}$ \\
& & & & (0.126) & (0.129) \\
 Gender[Prefer not to say] & & & & 0.093$^{}$ & 0.116$^{}$ \\
& & & & (0.398) & (0.407) \\
 STEM & & & & & -0.052$^{}$ \\
& & & & & (0.132) \\
 Intercept & 0.333$^{}$ & 0.303$^{}$ & 0.523$^{}$ & 0.509$^{}$ & 0.525$^{}$ \\
& (0.219) & (0.226) & (0.325) & (0.323) & (0.329) \\
\hline \\[-1.8ex]
 Observations & 42 & 42 & 42 & 42 & 42 \\
 $R^2$ & 0.113 & 0.129 & 0.150 & 0.206 & 0.210 \\
 Adjusted $R^2$ & 0.091 & 0.061 & 0.058 & 0.070 & 0.047 \\
 Residual Std. Error & 0.379 (df=40) & 0.385 (df=38) & 0.386 (df=37) & 0.383 (df=35) & 0.388 (df=34) \\
 F Statistic & 5.102$^{**}$ (df=1; 40) & 1.882$^{}$ (df=3; 38) & 1.631$^{}$ (df=4; 37) & 1.517$^{}$ (df=6; 35) & 1.291$^{}$ (df=7; 34) \\
\hline
\hline \\[-1.8ex]
\textit{Note:} & \multicolumn{5}{r}{$^{*}$p$<$0.1; $^{**}$p$<$0.05; $^{***}$p$<$0.01} \\
\end{tabular}
\end{sidewaystable}

\begin{sidewaystable}[!htbp] \centering
\caption{Helpfulness of AI-Generated content in Investment Scam within the \textit{Reason} Component.}
\label{tab:reasonhelpinvestment}
\begin{tabular}{@{\extracolsep{5pt}}lccccc}
\\[-1.8ex]\hline
\hline \\[-1.8ex]
& \multicolumn{5}{c}{\textit{Dependent variable: Helpful (Investment)}} \
\cr \cline{2-6}
\\[-1.8ex] & (1) & (2) & (3) & (4) & (5) \\
\hline \\[-1.8ex]
 Accuracy & 0.250$^{}$ & 0.250$^{}$ & 0.250$^{}$ & 0.253$^{}$ & 0.262$^{}$ \\
& (0.324) & (0.336) & (0.336) & (0.336) & (0.359) \\
 Age[Old (>44)] & & -0.017$^{}$ & -0.125$^{}$ & -0.102$^{}$ & -0.106$^{}$ \\
& & (0.150) & (0.187) & (0.187) & (0.198) \\
 Age[Young (<25)] & & 0.250$^{}$ & 0.250$^{}$ & 0.144$^{}$ & 0.136$^{}$ \\
& & (0.466) & (0.467) & (0.473) & (0.490) \\
 University Graduate & & & -0.232$^{}$ & -0.162$^{}$ & -0.162$^{}$ \\
& & & (0.237) & (0.241) & (0.245) \\
 Gender[Male] & & & & -0.206$^{}$ & -0.204$^{}$ \\
& & & & (0.148) & (0.153) \\
 Gender[Prefer not to say] & & & & 0.144$^{}$ & 0.150$^{}$ \\
& & & & (0.473) & (0.484) \\
 STEM & & & & & -0.013$^{}$ \\
& & & & & (0.164) \\
 Intercept & 0.500$^{}$ & 0.500$^{}$ & 0.732$^{*}$ & 0.765$^{*}$ & 0.764$^{*}$ \\
& (0.316) & (0.323) & (0.401) & (0.400) & (0.406) \\
\hline \\[-1.8ex]
 Observations & 42 & 42 & 42 & 42 & 42 \\
 $R^2$ & 0.015 & 0.023 & 0.048 & 0.105 & 0.105 \\
 Adjusted $R^2$ & -0.010 & -0.054 & -0.055 & -0.048 & -0.079 \\
 Residual Std. Error & 0.447 (df=40) & 0.457 (df=38) & 0.457 (df=37) & 0.456 (df=35) & 0.462 (df=34) \\
 F Statistic & 0.595$^{}$ (df=1; 40) & 0.297$^{}$ (df=3; 38) & 0.463$^{}$ (df=4; 37) & 0.685$^{}$ (df=6; 35) & 0.572$^{}$ (df=7; 34) \\
\hline
\hline \\[-1.8ex]
\textit{Note:} & \multicolumn{5}{r}{$^{*}$p$<$0.1; $^{**}$p$<$0.05; $^{***}$p$<$0.01} \\
\end{tabular}
\end{sidewaystable}

\begin{sidewaystable}[!htbp] \centering
\caption{Helpfulness of AI-Generated content in Love Scam within the \textit{Reason} Component.}
\label{tab:reasonhelplove}
\begin{tabular}{@{\extracolsep{5pt}}lccccc}
\\[-1.8ex]\hline
\hline \\[-1.8ex]
& \multicolumn{5}{c}{\textit{Dependent variable: Helpful (Love)}} \
\cr \cline{2-6}
\\[-1.8ex] & (1) & (2) & (3) & (4) & (5) \\
\hline \\[-1.8ex]
 Accuracy & 0.237$^{}$ & 0.239$^{}$ & 0.232$^{}$ & 0.171$^{}$ & 0.202$^{}$ \\
& (0.240) & (0.246) & (0.251) & (0.249) & (0.260) \\
 Age[Old (>44)] & & -0.076$^{}$ & -0.104$^{}$ & -0.077$^{}$ & -0.098$^{}$ \\
& & (0.151) & (0.191) & (0.188) & (0.196) \\
 Age[Young (<25)] & & 0.242$^{}$ & 0.243$^{}$ & 0.111$^{}$ & 0.067$^{}$ \\
& & (0.475) & (0.481) & (0.480) & (0.495) \\
 University Graduate & & & -0.060$^{}$ & 0.022$^{}$ & 0.028$^{}$ \\
& & & (0.246) & (0.246) & (0.249) \\
 Gender[Male] & & & & -0.266$^{*}$ & -0.251$^{}$ \\
& & & & (0.152) & (0.157) \\
 Gender[Prefer not to say] & & & & 0.111$^{}$ & 0.143$^{}$ \\
& & & & (0.480) & (0.491) \\
 STEM & & & & & -0.076$^{}$ \\
& & & & & (0.163) \\
 Intercept & 0.500$^{**}$ & 0.519$^{**}$ & 0.586$^{}$ & 0.696$^{*}$ & 0.704$^{*}$ \\
& (0.229) & (0.236) & (0.364) & (0.363) & (0.367) \\
\hline \\[-1.8ex]
 Observations & 42 & 42 & 42 & 42 & 42 \\
 $R^2$ & 0.024 & 0.038 & 0.040 & 0.124 & 0.129 \\
 Adjusted $R^2$ & -0.001 & -0.038 & -0.064 & -0.026 & -0.050 \\
 Residual Std. Error & 0.457 (df=40) & 0.466 (df=38) & 0.472 (df=37) & 0.463 (df=35) & 0.468 (df=34) \\
 F Statistic & 0.970$^{}$ (df=1; 40) & 0.505$^{}$ (df=3; 38) & 0.384$^{}$ (df=4; 37) & 0.825$^{}$ (df=6; 35) & 0.722$^{}$ (df=7; 34) \\
\hline
\hline \\[-1.8ex]
\textit{Note:} & \multicolumn{5}{r}{$^{*}$p$<$0.1; $^{**}$p$<$0.05; $^{***}$p$<$0.01} \\
\end{tabular}
\end{sidewaystable}

\printbibliography[heading=bibintoc,title=References]

\appendix
\section{Appendix}
\label{app:chapter3}
\subsection{Participant Demographics and Statistical Summary}
\label{app:demographics}

\begin{figure}[H]
    \centering
    \begin{subfigure}[b]{0.45\linewidth}
        \centering
        \includegraphics[width=\linewidth]{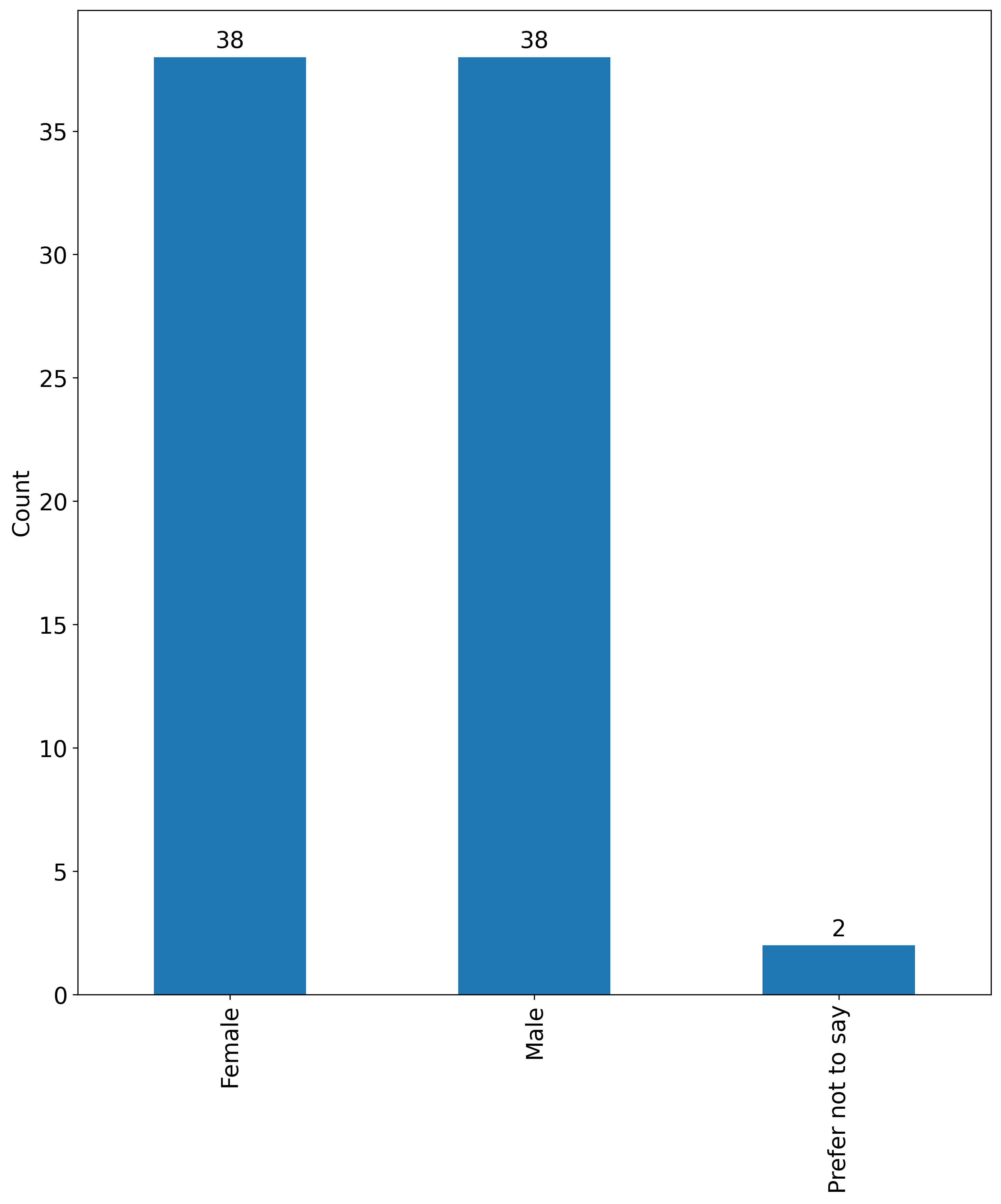}
        \caption{}
    \end{subfigure}
    \hfill
    \begin{subfigure}[b]{0.45\linewidth}
        \centering
        \includegraphics[width=\linewidth]{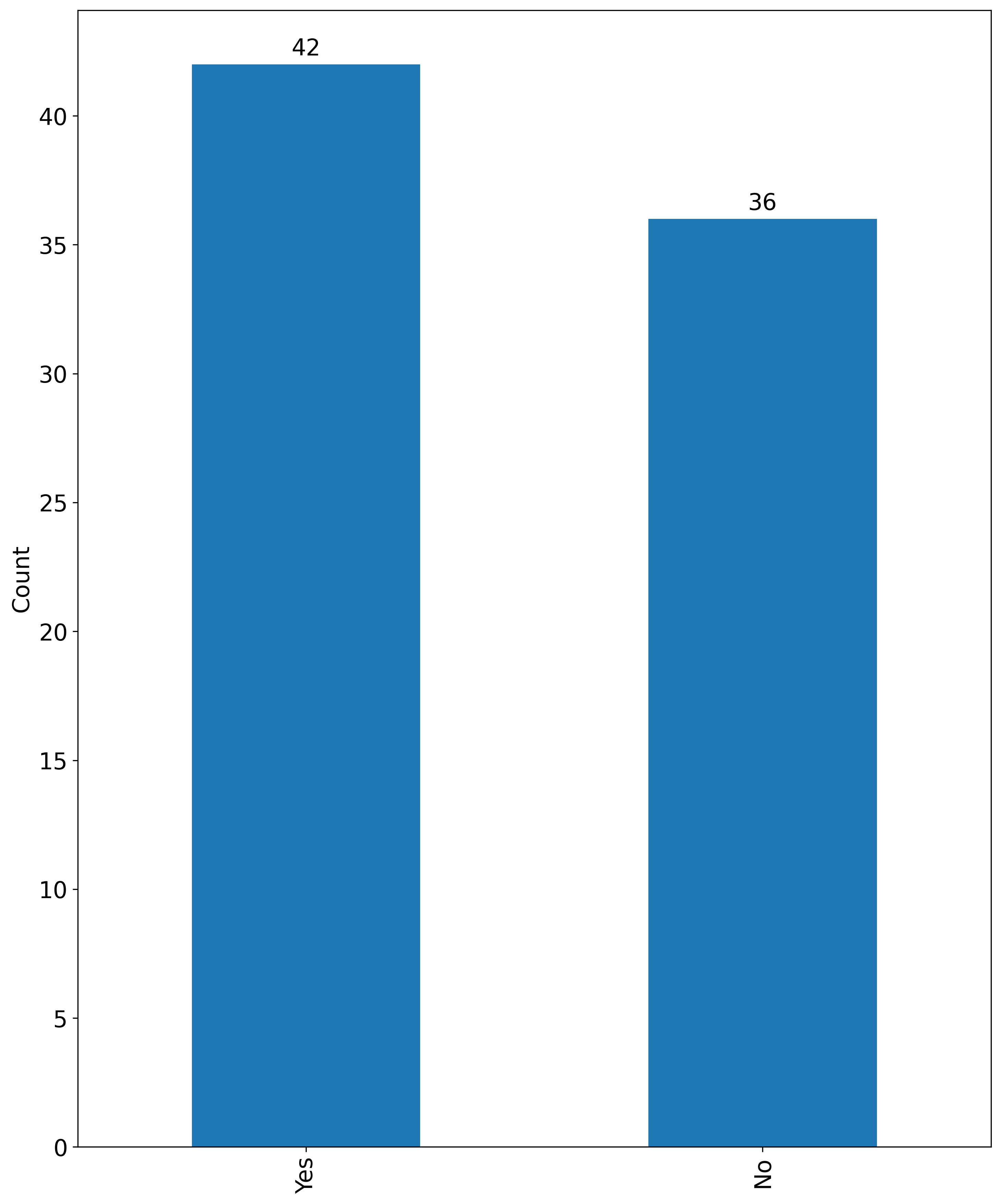}
        \caption{}
    \end{subfigure}
    \caption{Participant demographics by (a) Gender and (b) STEM Background.}
\end{figure}

\begin{figure}[H]
    \centering
    \includegraphics[width=0.9\linewidth]{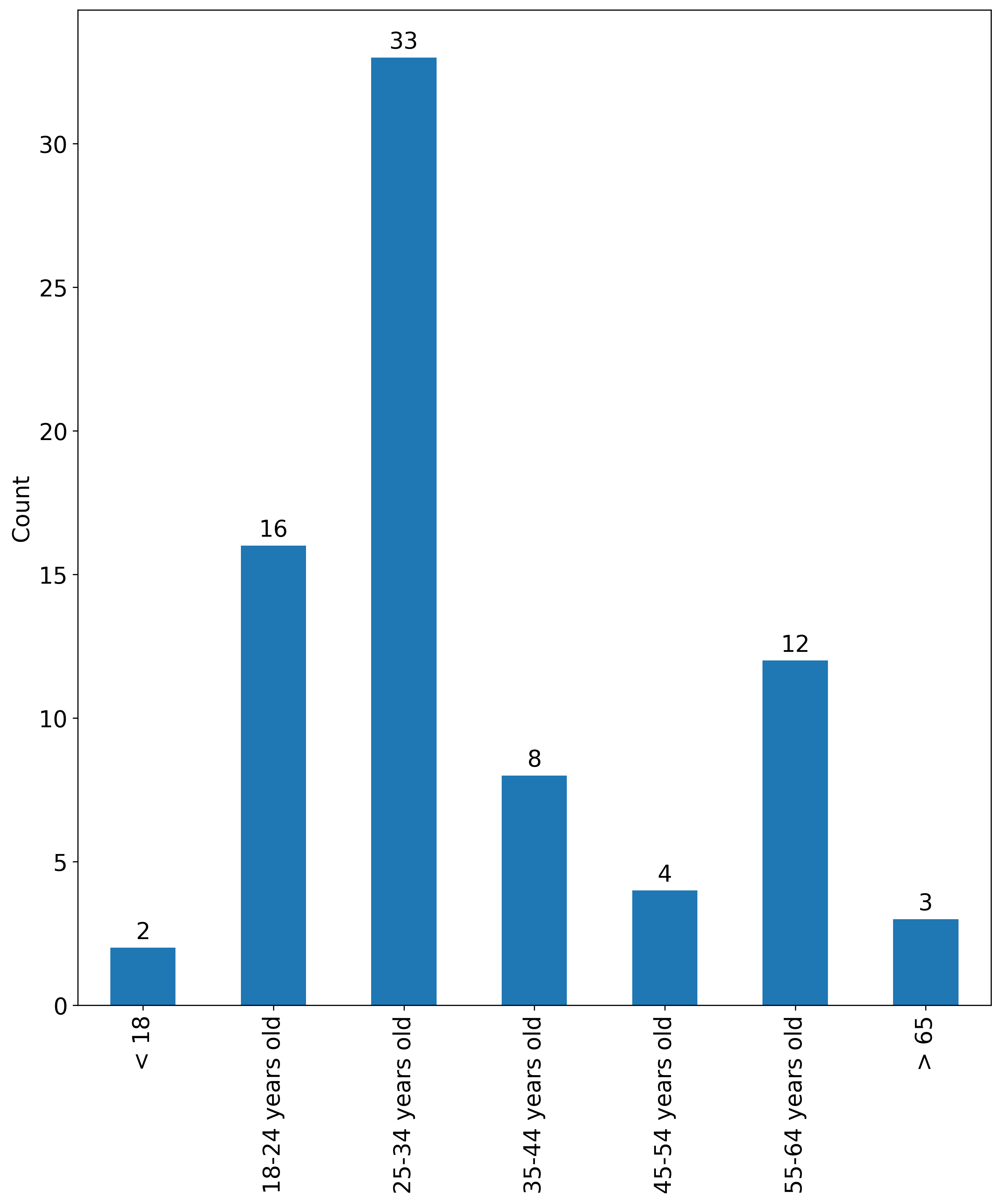}
    \caption{Participant demographics by Age.}
\end{figure}

\begin{figure}[H]
    \centering
    \includegraphics[width=0.9\linewidth]{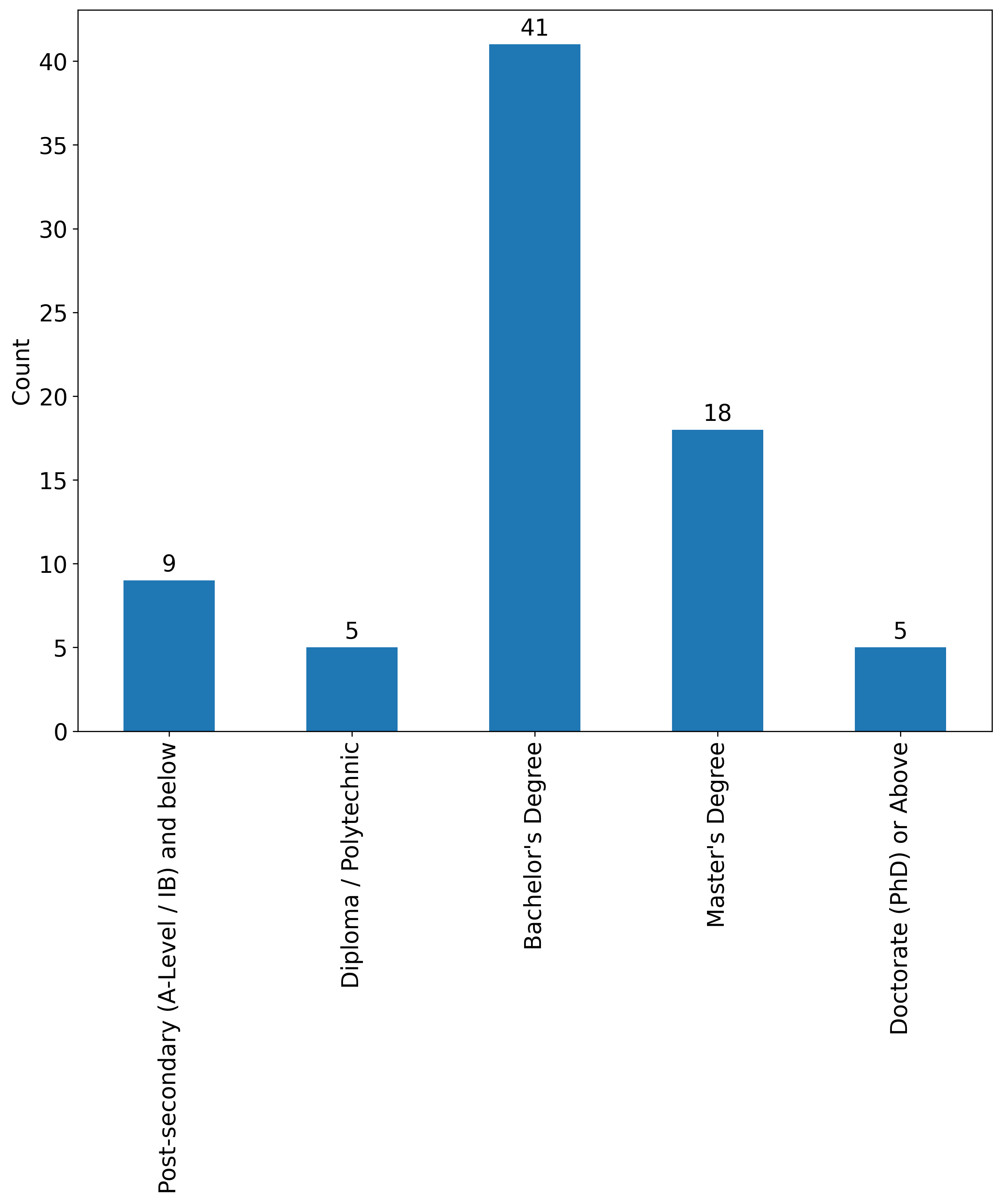}
    \caption{Participant demographics by Highest Level of Education Completed.}
\end{figure}

\begin{figure}[H]
    \centering
    \includegraphics[width=0.9\linewidth]{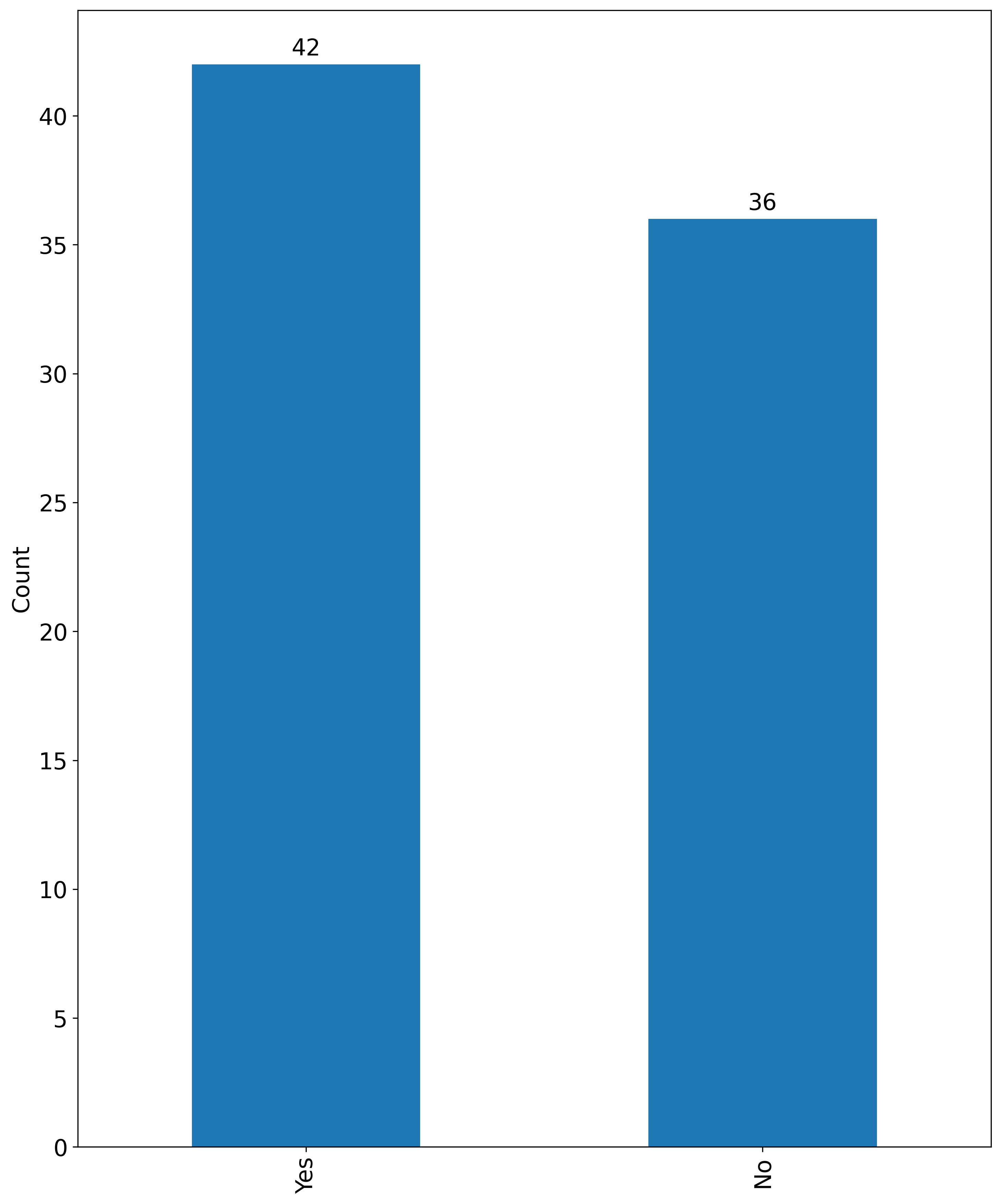}
    \caption{Distribution of participants by group assignment, showing the number of individuals in the AI-Assisted (Treatment) group versus the non-AI-Assisted (Control) group.}
\end{figure}

\newpage
\subsection{Anticipate Component Experiment Scenarios}
\label{app:anticipate}

\subsubsection{Control Scenarios}

\begin{figure}[H]
    \centering
    \includegraphics[width=0.9\linewidth]{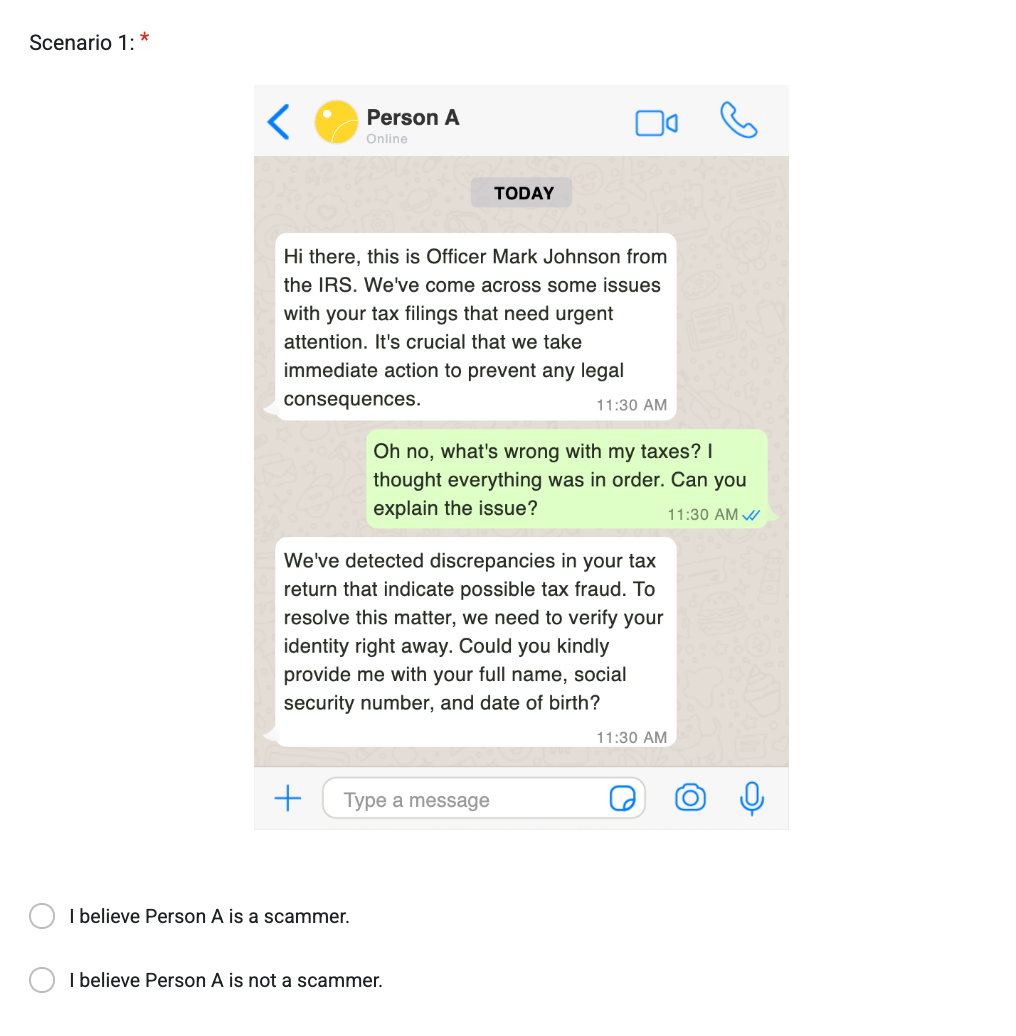}
    \caption{Anticipatory Component Control Scenario 1.}
\end{figure}

\begin{figure}[H]
    \centering
    \includegraphics[width=0.9\linewidth]{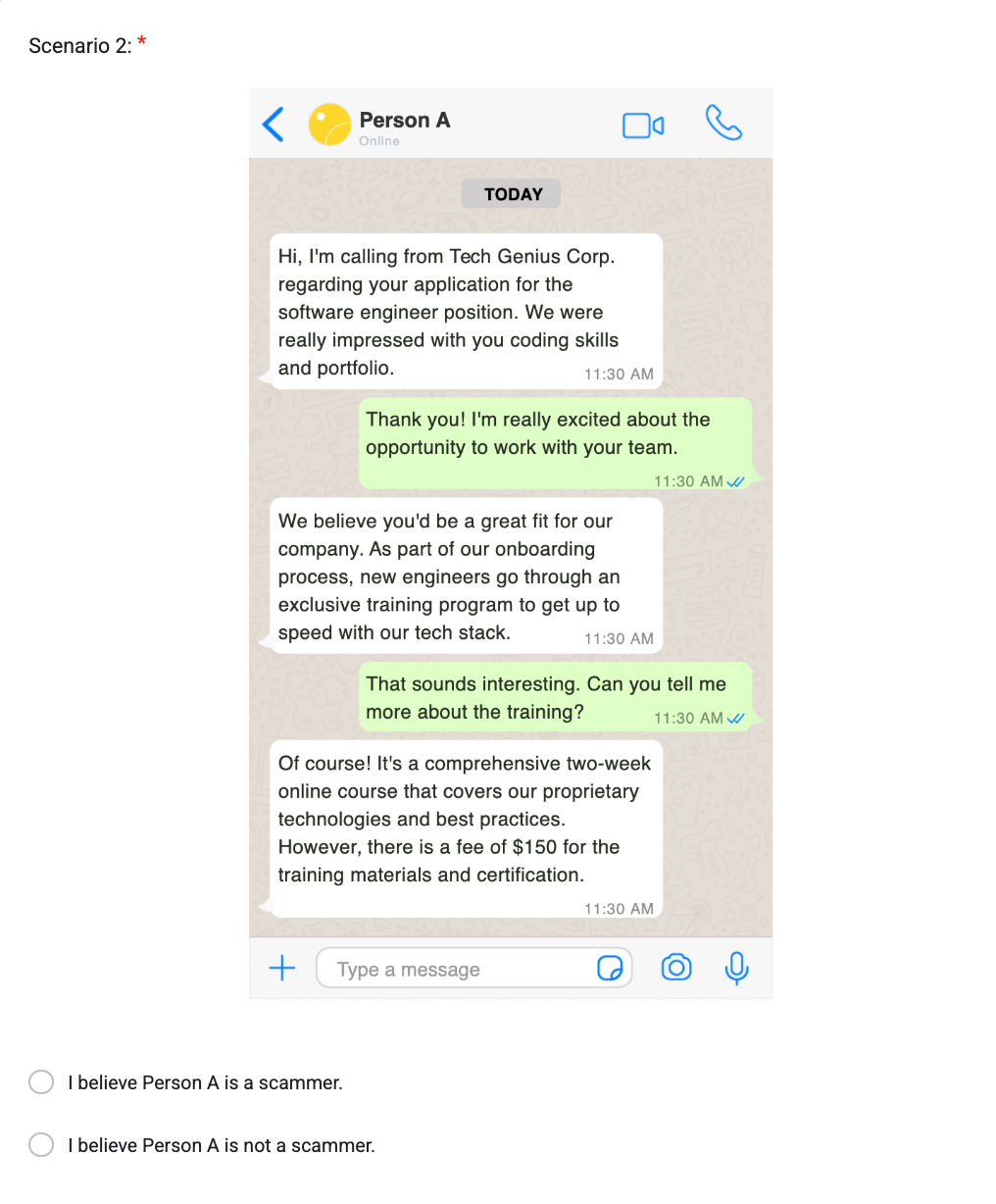}
    \caption{Anticipatory Component Control Scenario 2.}
\end{figure}

\begin{figure}[H]
    \centering
    \includegraphics[width=0.9\linewidth]{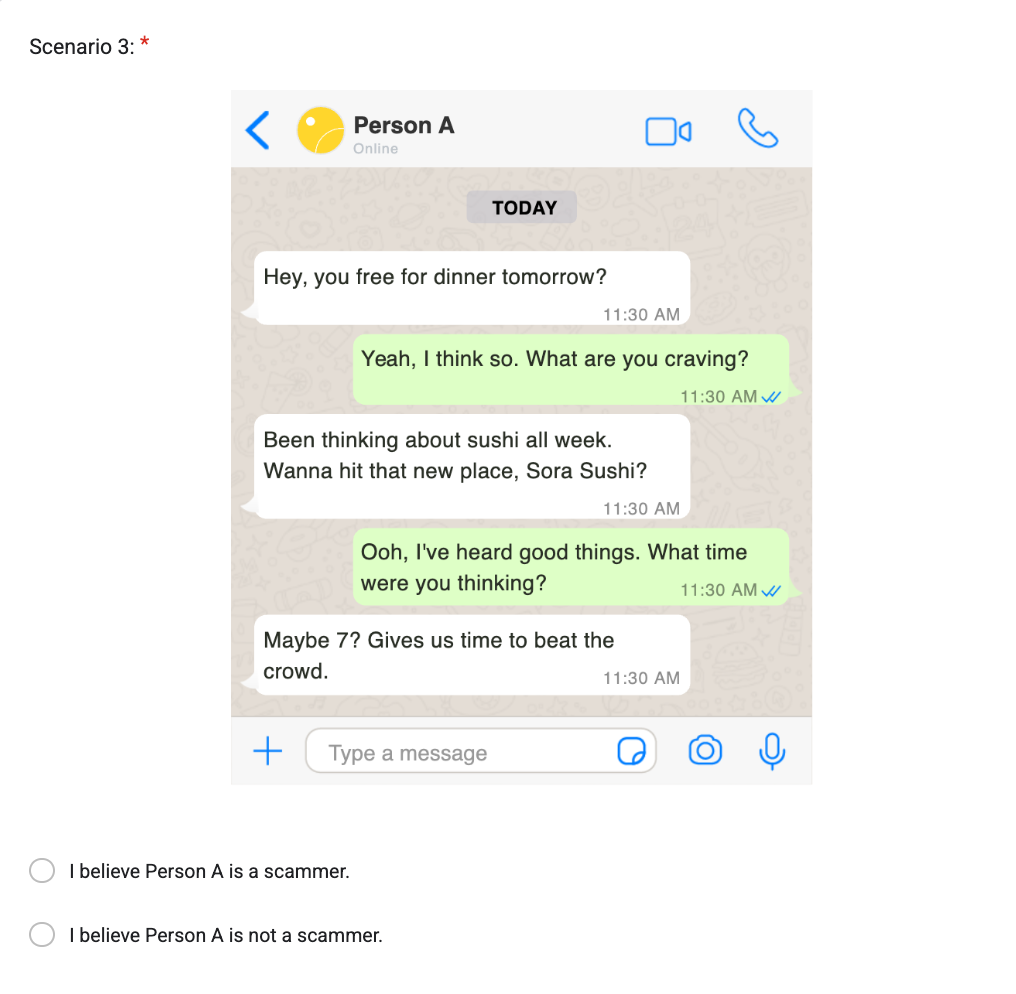}
    \caption{Anticipatory Component Control Scenario 3.}
\end{figure}

\begin{figure}[H]
    \centering
    \includegraphics[width=0.9\linewidth]{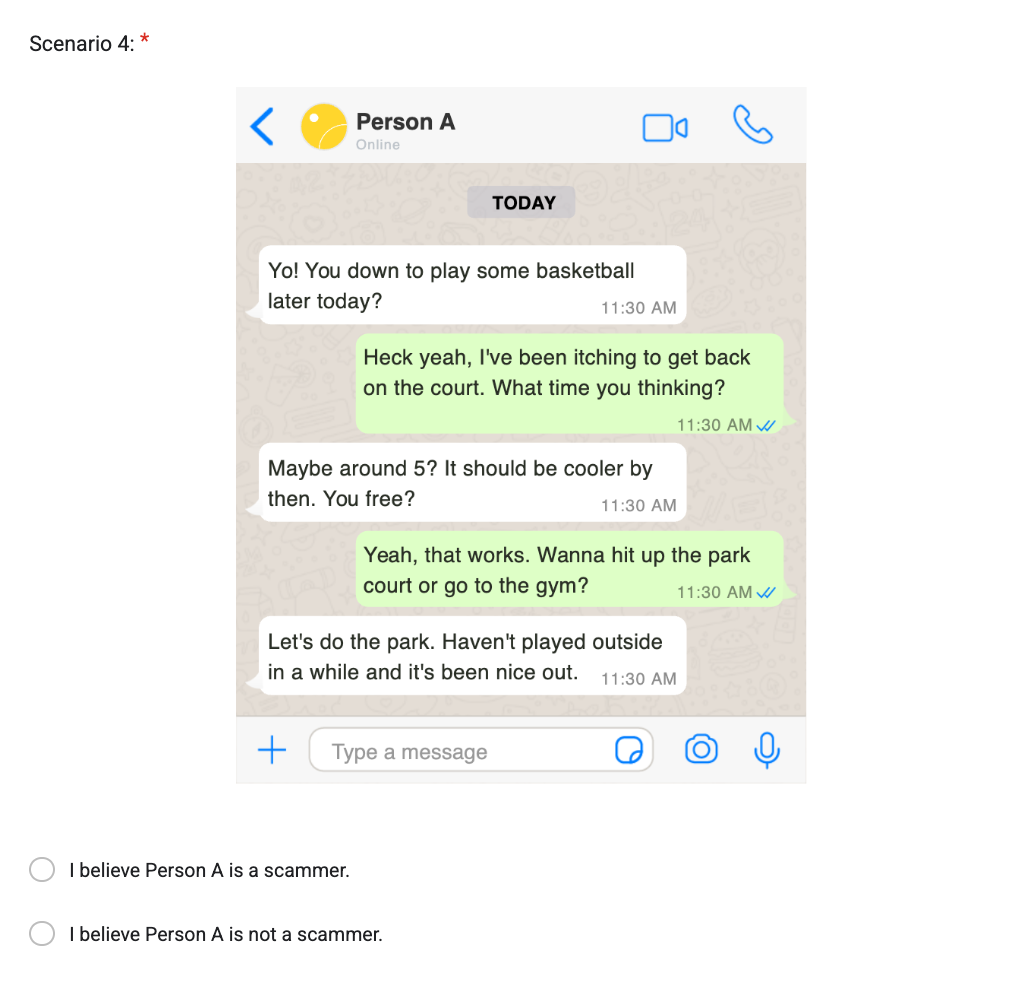}
    \caption{Anticipatory Component Control Scenario 4.}
\end{figure}

\begin{figure}[H]
    \centering
    \includegraphics[width=0.9\linewidth]{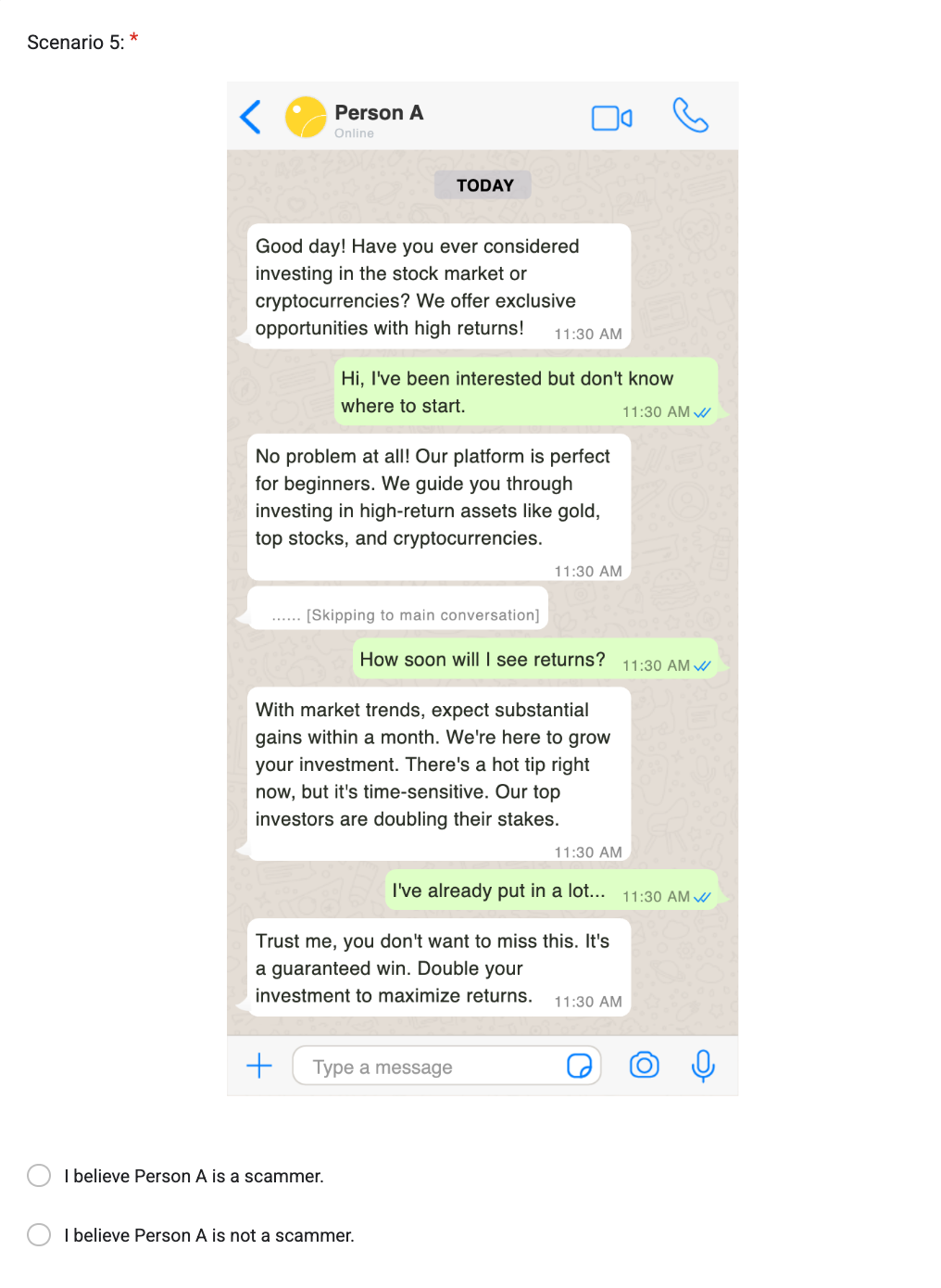}
    \caption{Anticipatory Component Control Scenario 5.}
\end{figure}

\begin{figure}[H]
    \centering
    \includegraphics[width=0.9\linewidth]{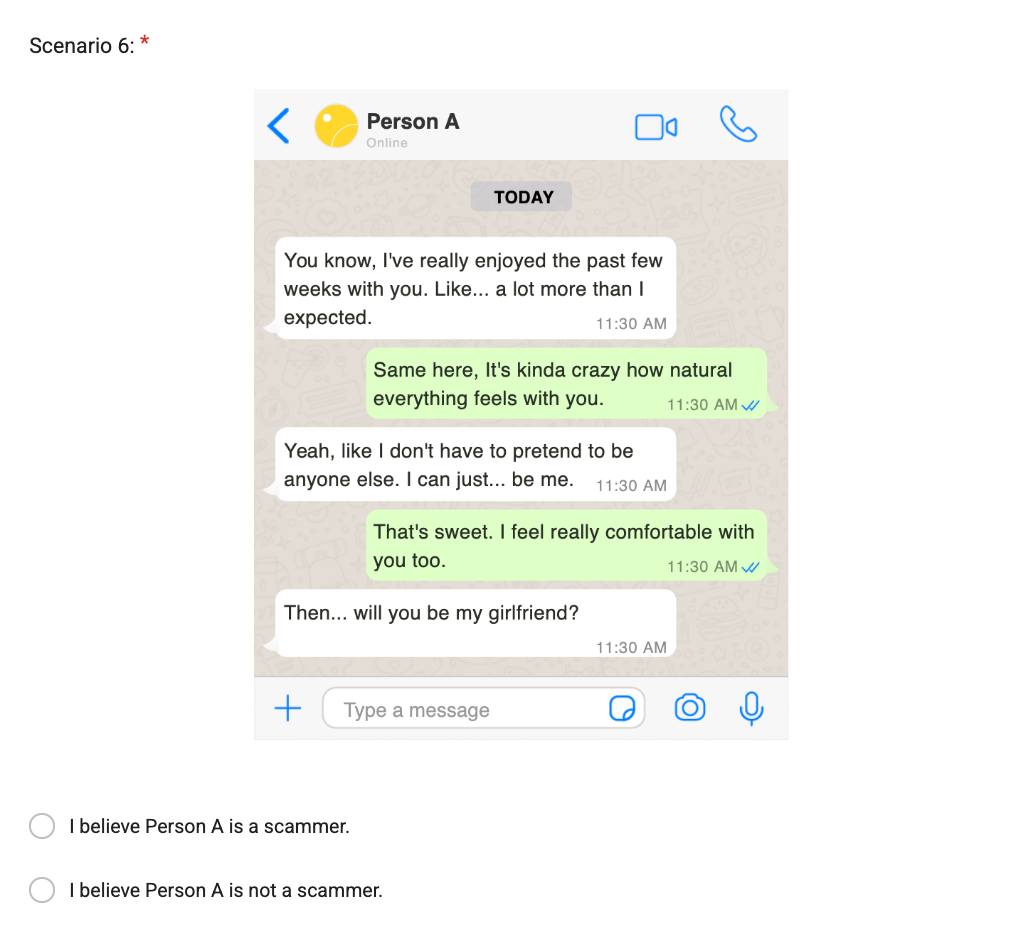}
    \caption{Anticipatory Component Control Scenario 6.}
\end{figure}

\begin{figure}[H]
    \centering
    \includegraphics[width=0.9\linewidth]{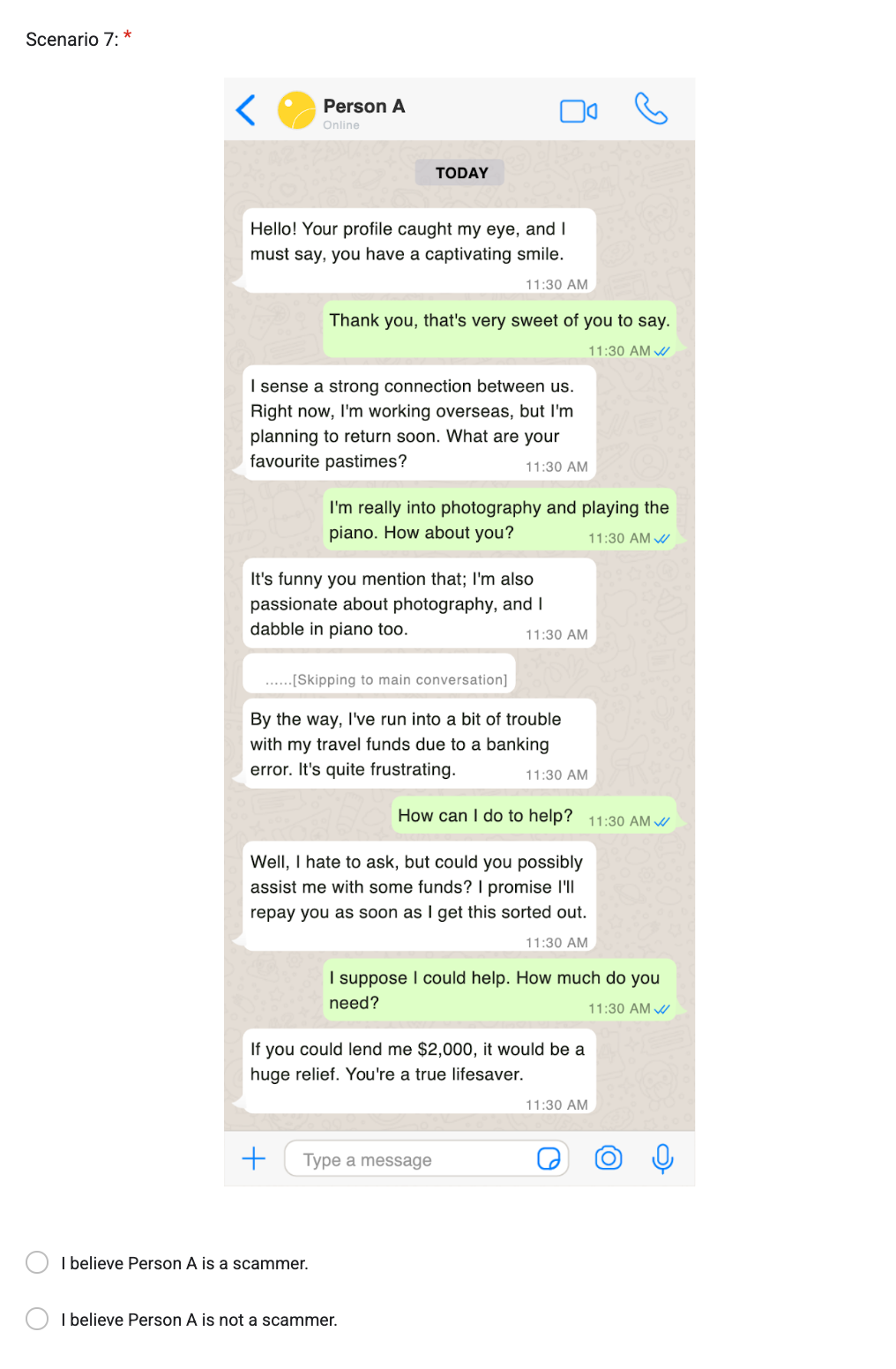}
    \caption{Anticipatory Component Control Scenario 7.}
\end{figure}

\begin{figure}[H]
    \centering
    \includegraphics[width=0.9\linewidth]{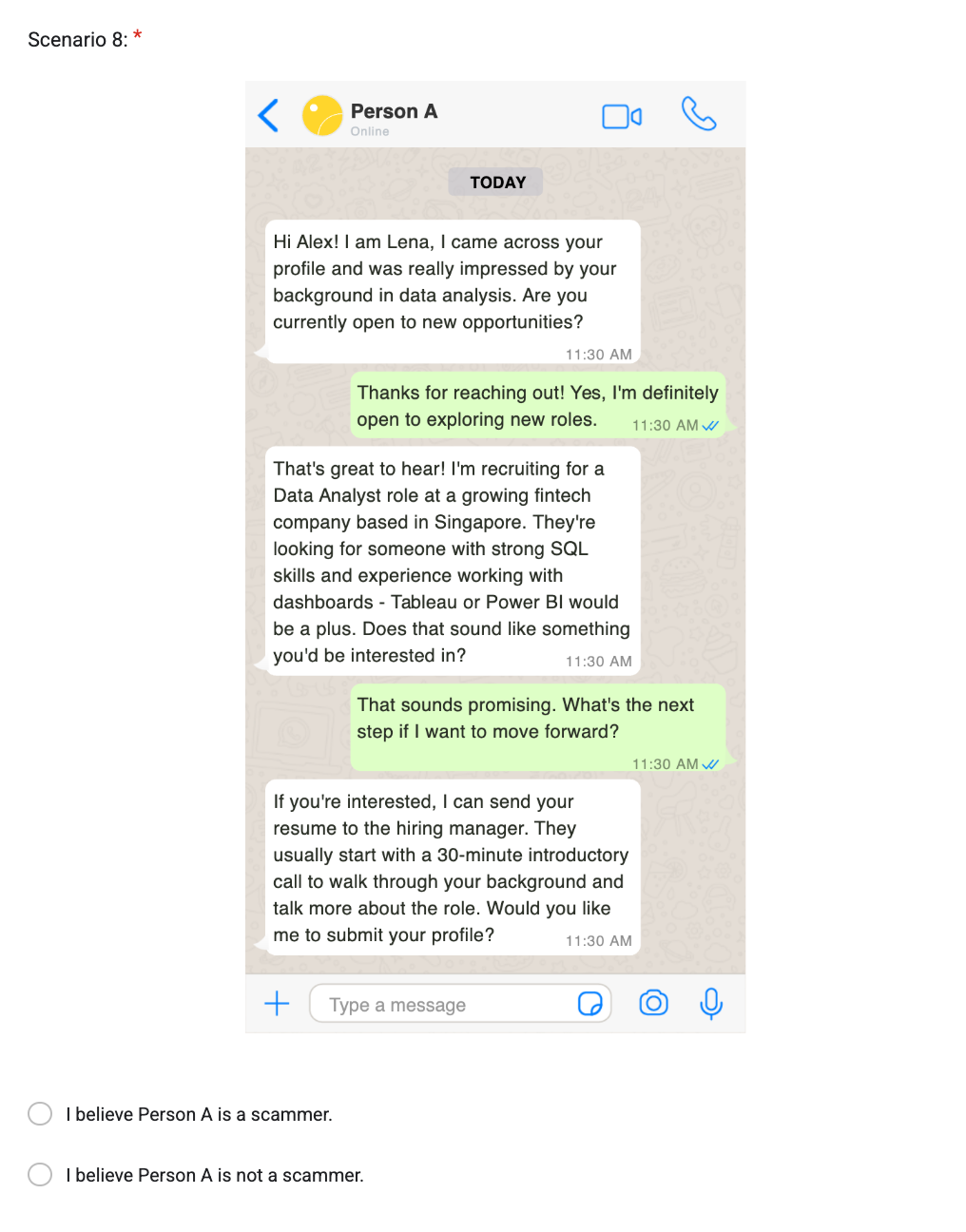}
    \caption{Anticipatory Component Control Scenario 8.}
\end{figure}

\subsubsection{Treatment Scenarios}

\begin{figure}[H]
    \centering
    \includegraphics[width=0.9\linewidth]{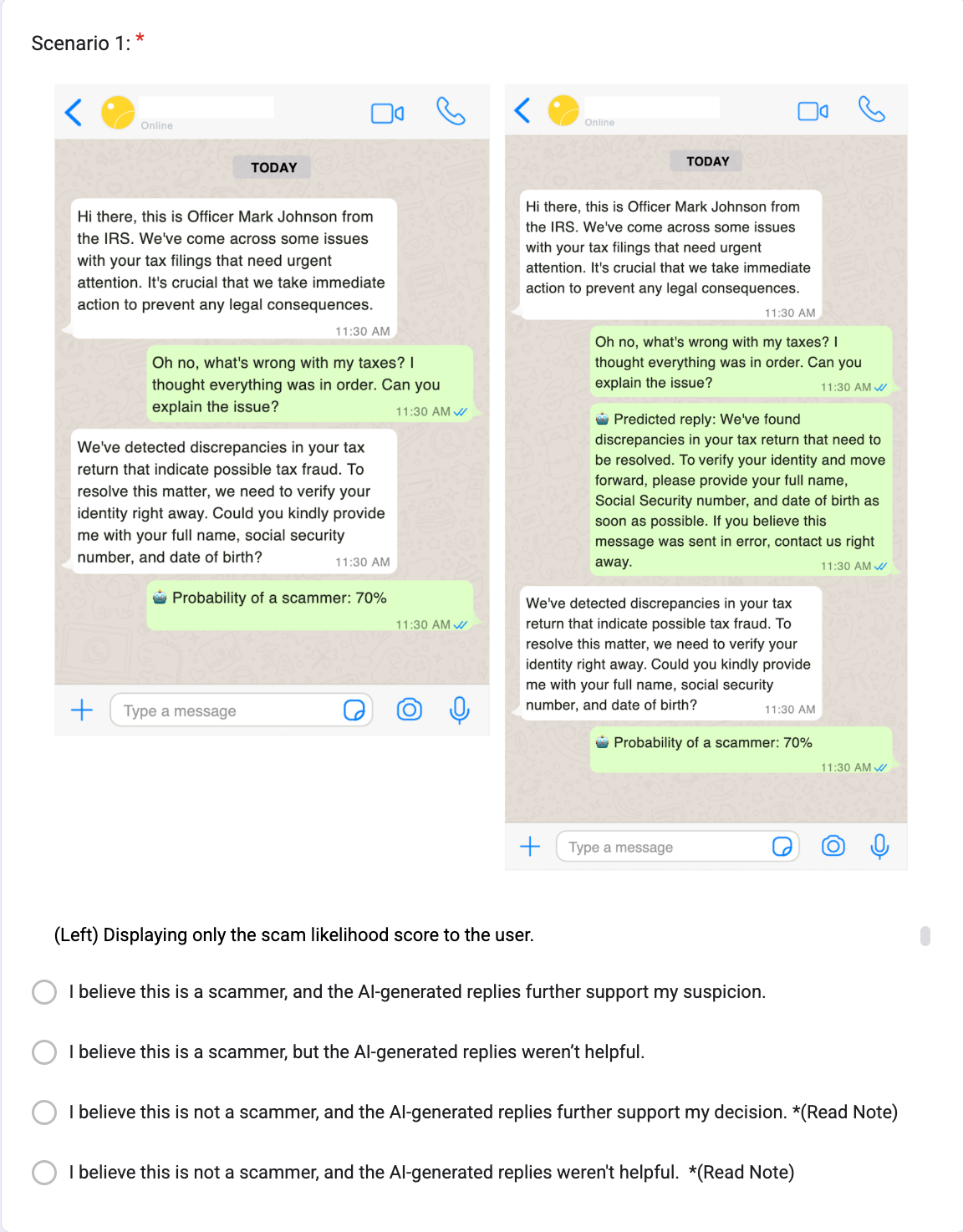}
    \caption{Anticipatory Component Treatment Scenario 1.}
\end{figure}

\begin{figure}[H]
    \centering
    \includegraphics[width=0.9\linewidth]{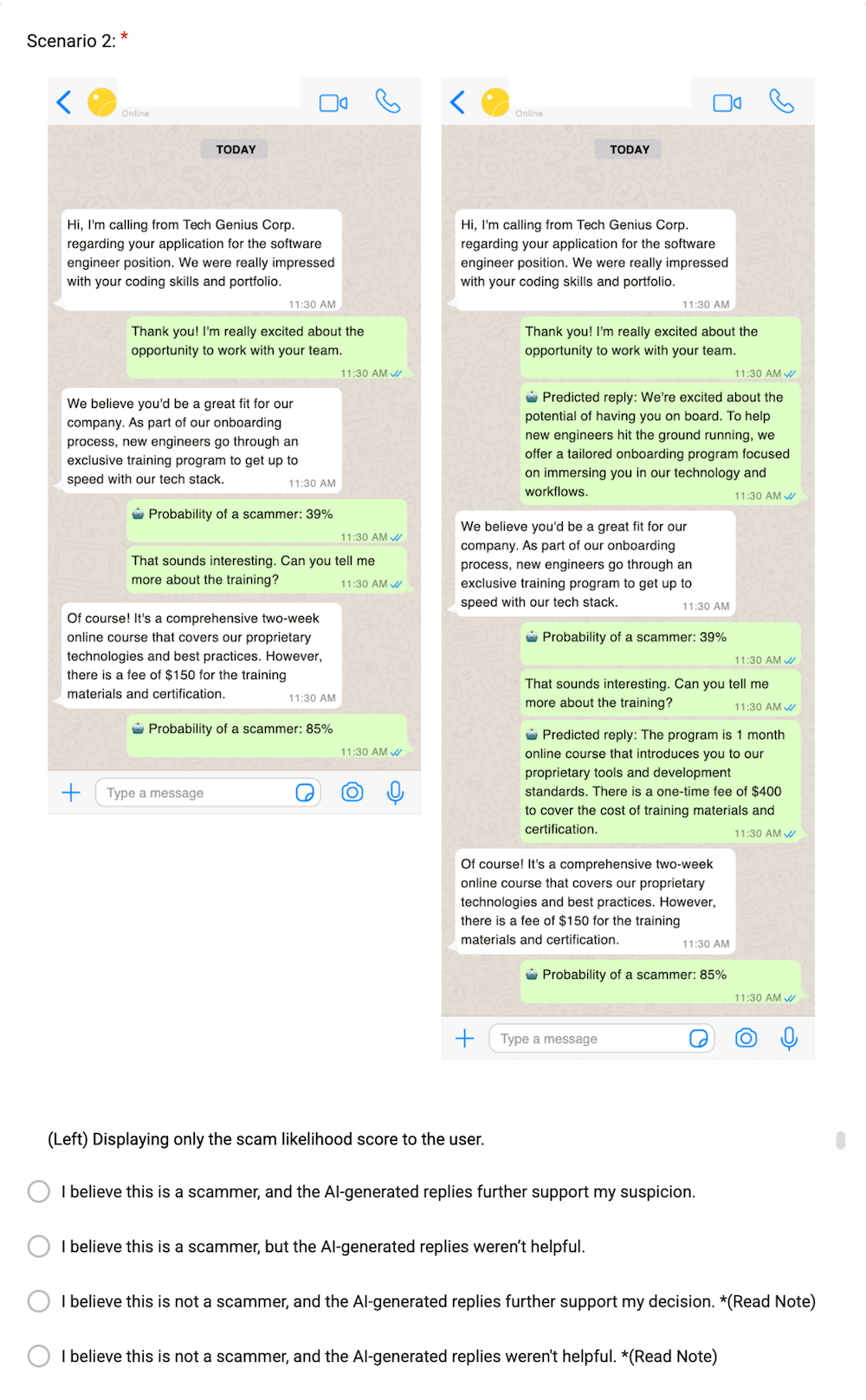}
    \caption{Anticipatory Component Treatment Scenario 2.}
\end{figure}

\begin{figure}[H]
    \centering
    \includegraphics[width=0.9\linewidth]{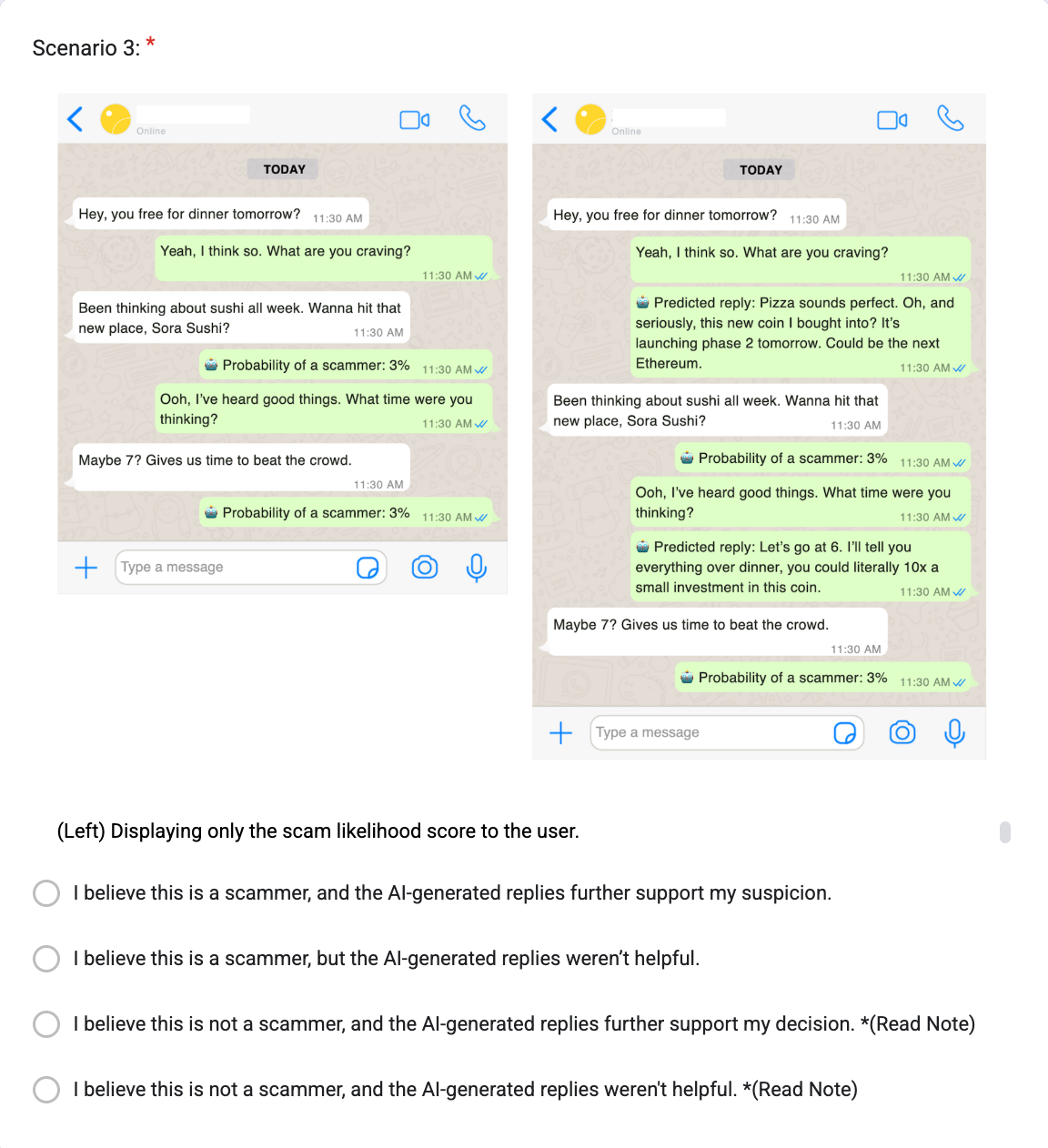}
    \caption{Anticipatory Component Treatment Scenario 3.}
\end{figure}

\begin{figure}[H]
    \centering
    \includegraphics[width=0.9\linewidth]{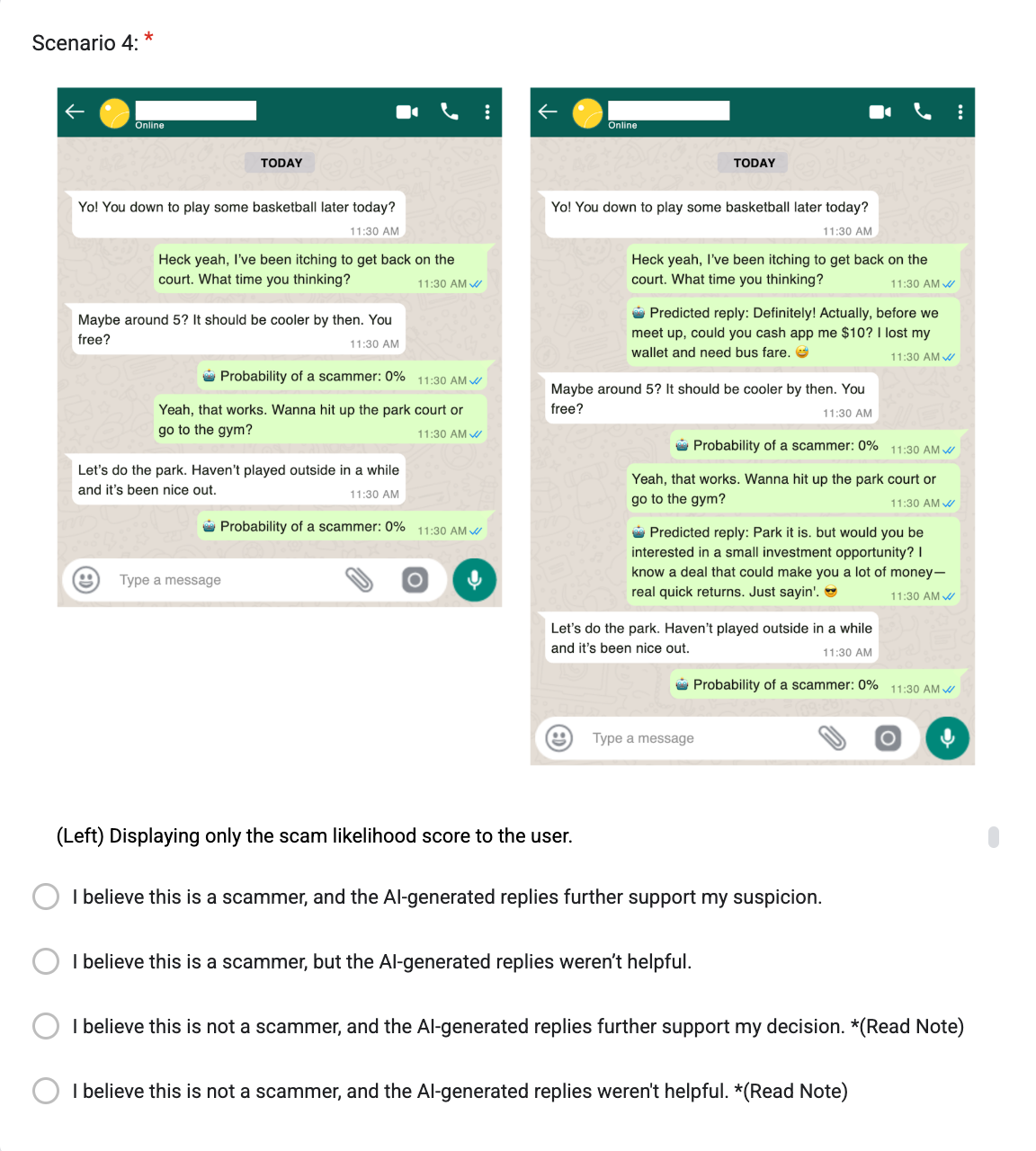}
    \caption{Anticipatory Component Treatment Scenario 4.}
\end{figure}

\begin{figure}[H]
    \centering
    \includegraphics[width=0.7\linewidth]{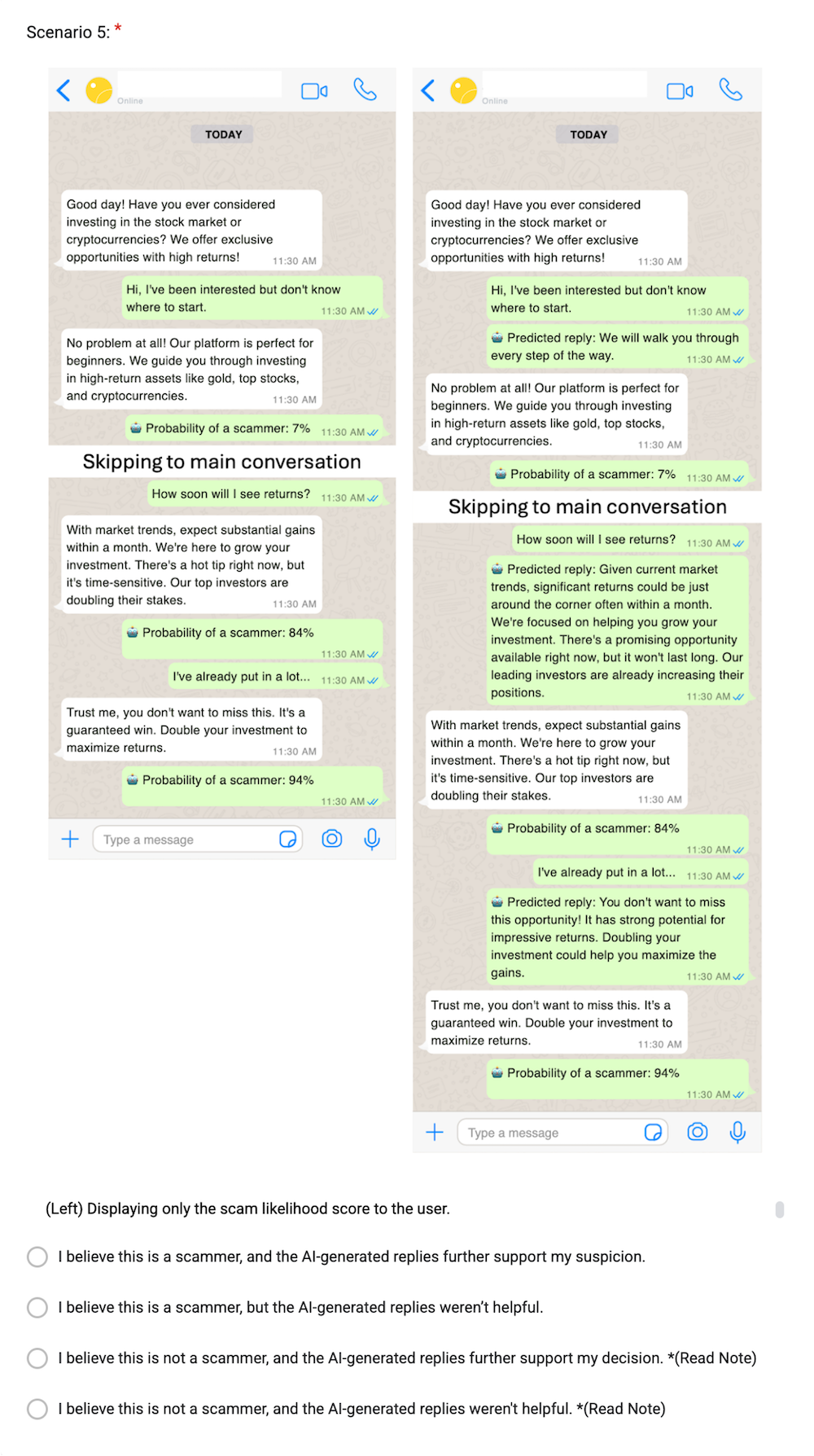}
    \caption{Anticipatory Component Treatment Scenario 5.}
\end{figure}

\begin{figure}[H]
    \centering
    \includegraphics[width=0.9\linewidth]{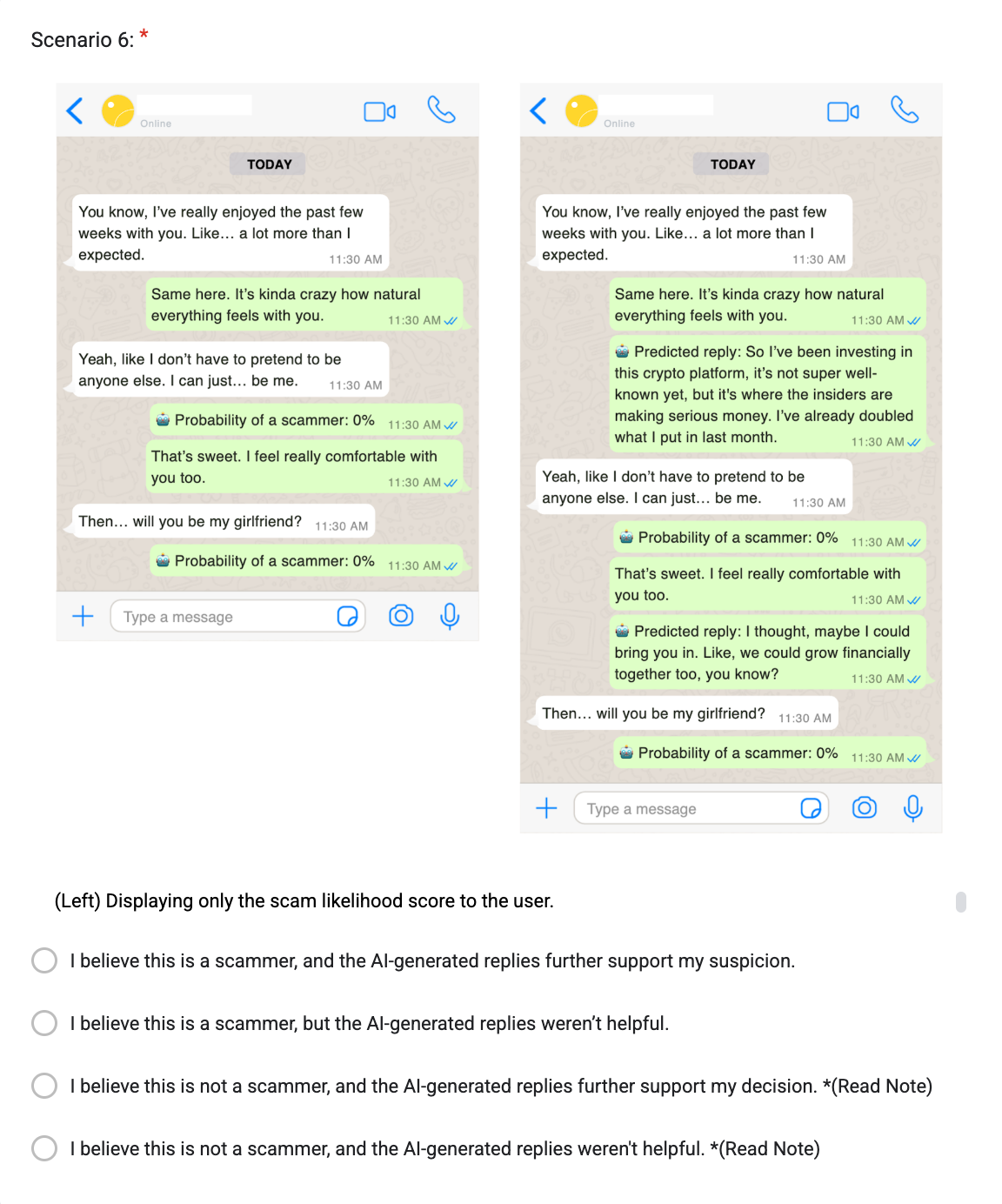}
    \caption{Anticipatory Component Treatment Scenario 6.}
\end{figure}

\begin{figure}[H]
    \centering
    \includegraphics[width=0.7\linewidth]{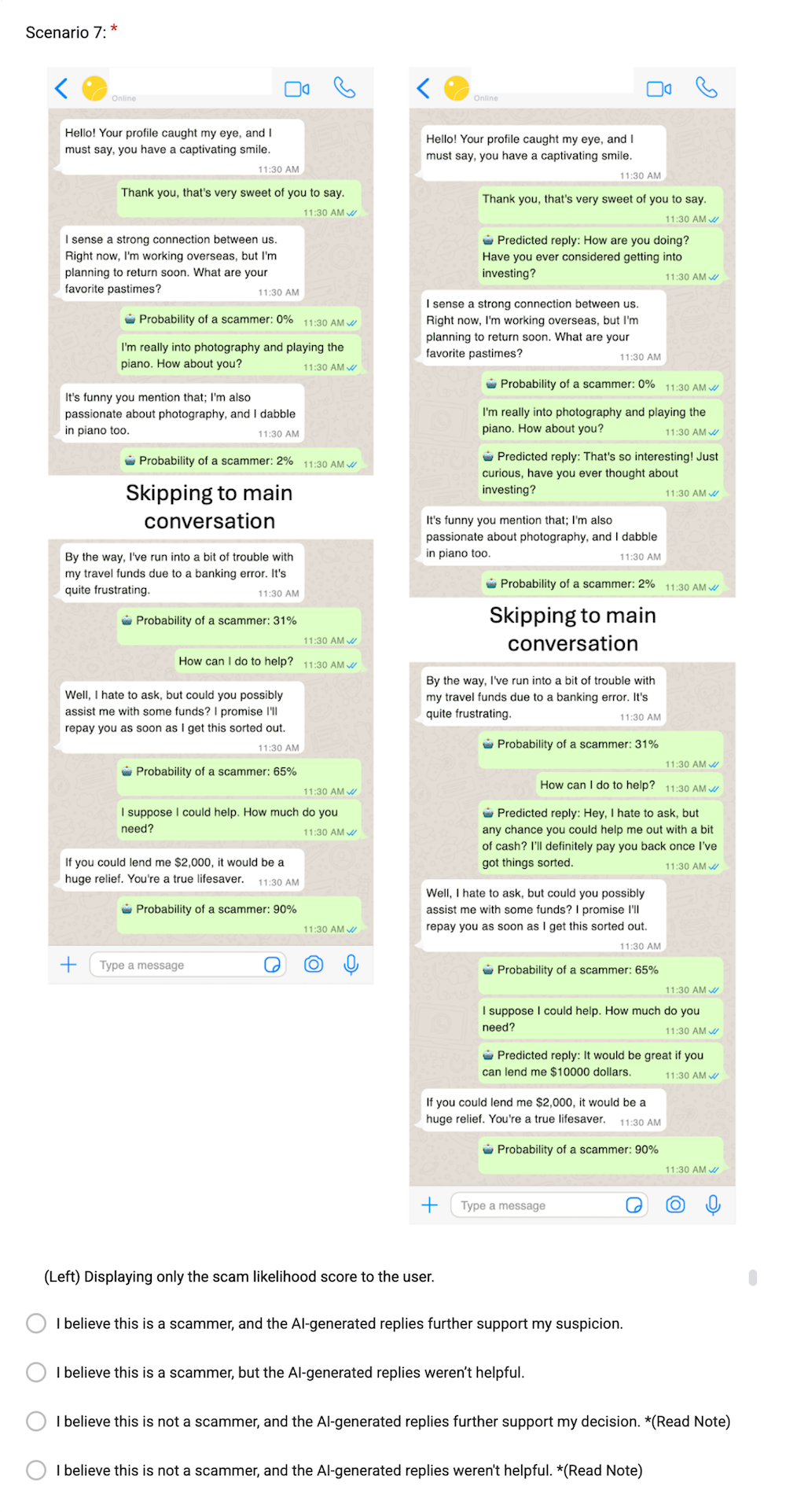}
    \caption{Anticipatory Component Treatment Scenario 7.}
\end{figure}

\begin{figure}[H]
    \centering
    \includegraphics[width=0.9\linewidth]{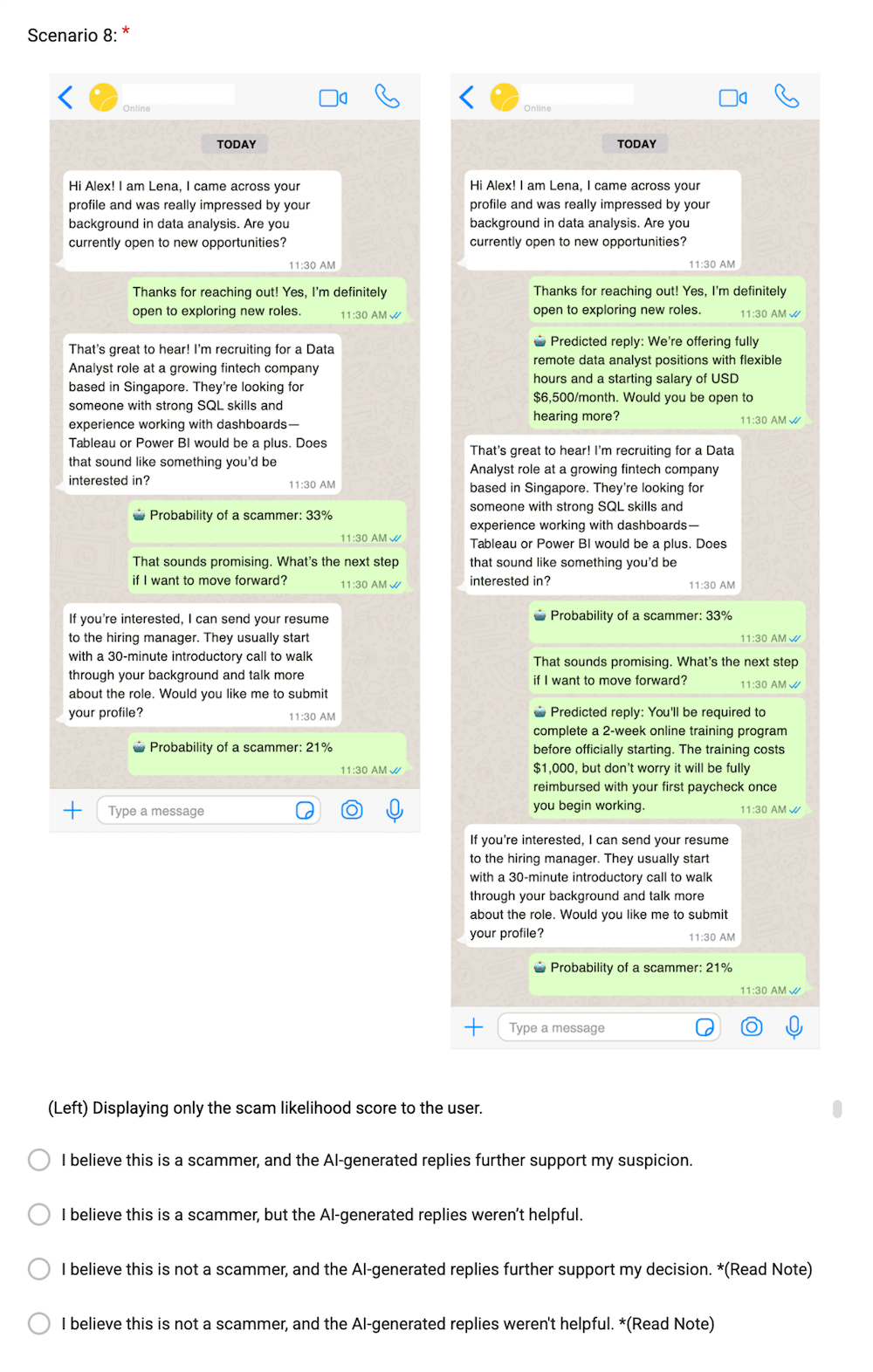}
    \caption{Anticipatory Component Treatment Scenario 8.}
\end{figure}

\newpage
\subsection{Reason Component Experiment Scenarios}
\label{app:reason}
\subsubsection{Control Scenarios}
\begin{figure}[H]
    \centering
    \includegraphics[width=0.8\linewidth]{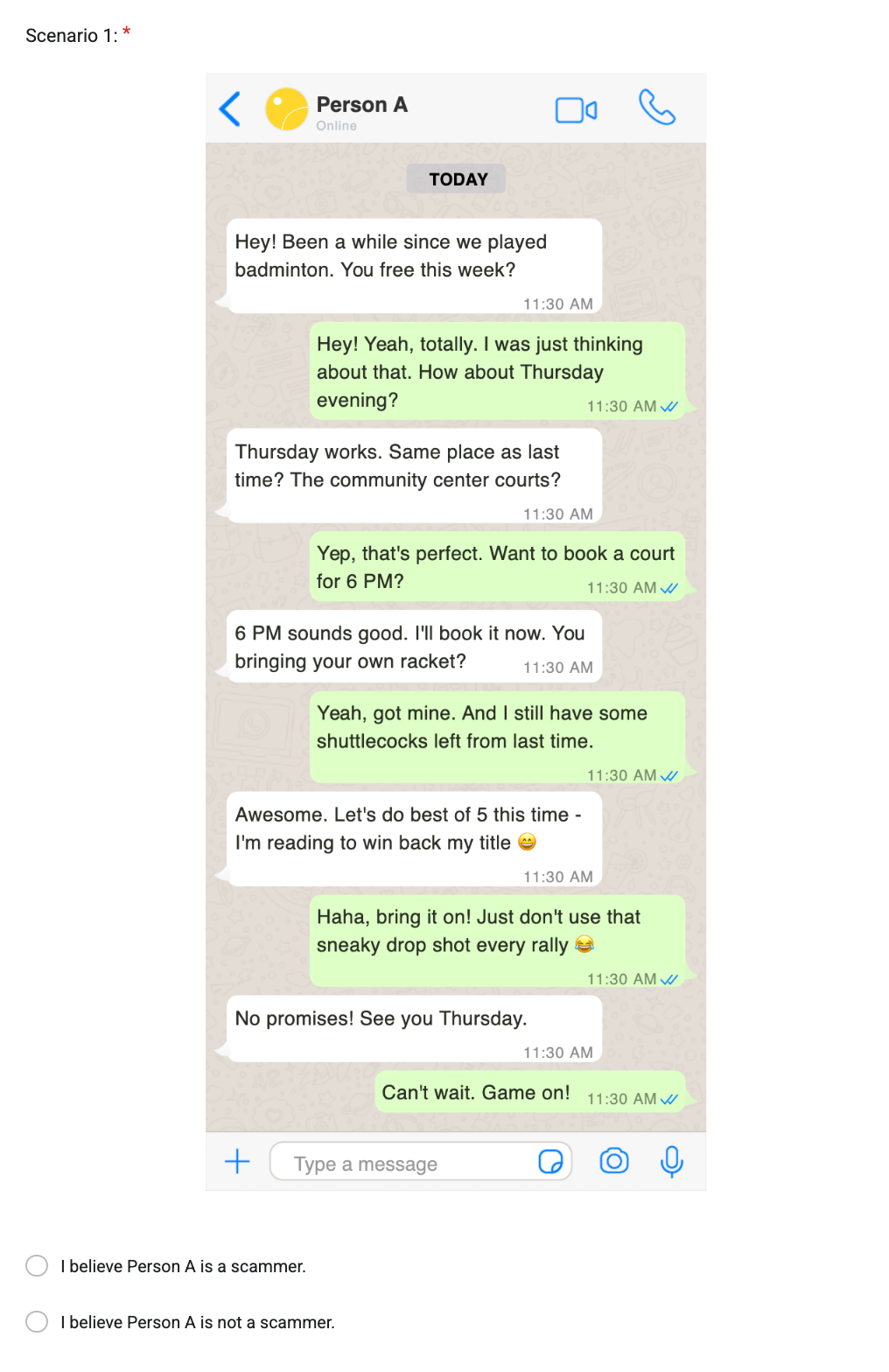}
    \caption{Reason Component Control Scenario 1.}
\end{figure}

\begin{figure}[H]
    \centering
    \includegraphics[width=0.6\linewidth]{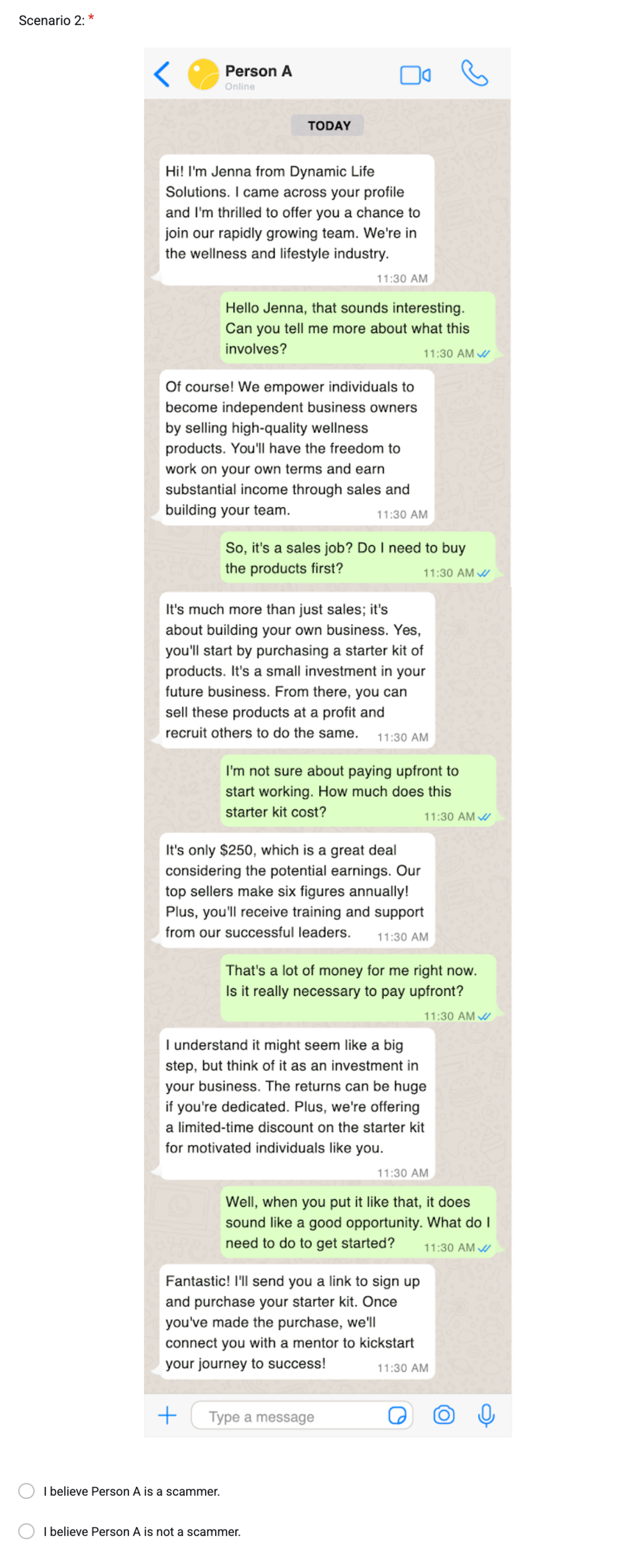}
    \caption{Reason Component Control Scenario 2.}
\end{figure}

\begin{figure}[H]
    \centering
    \includegraphics[width=0.7\linewidth]{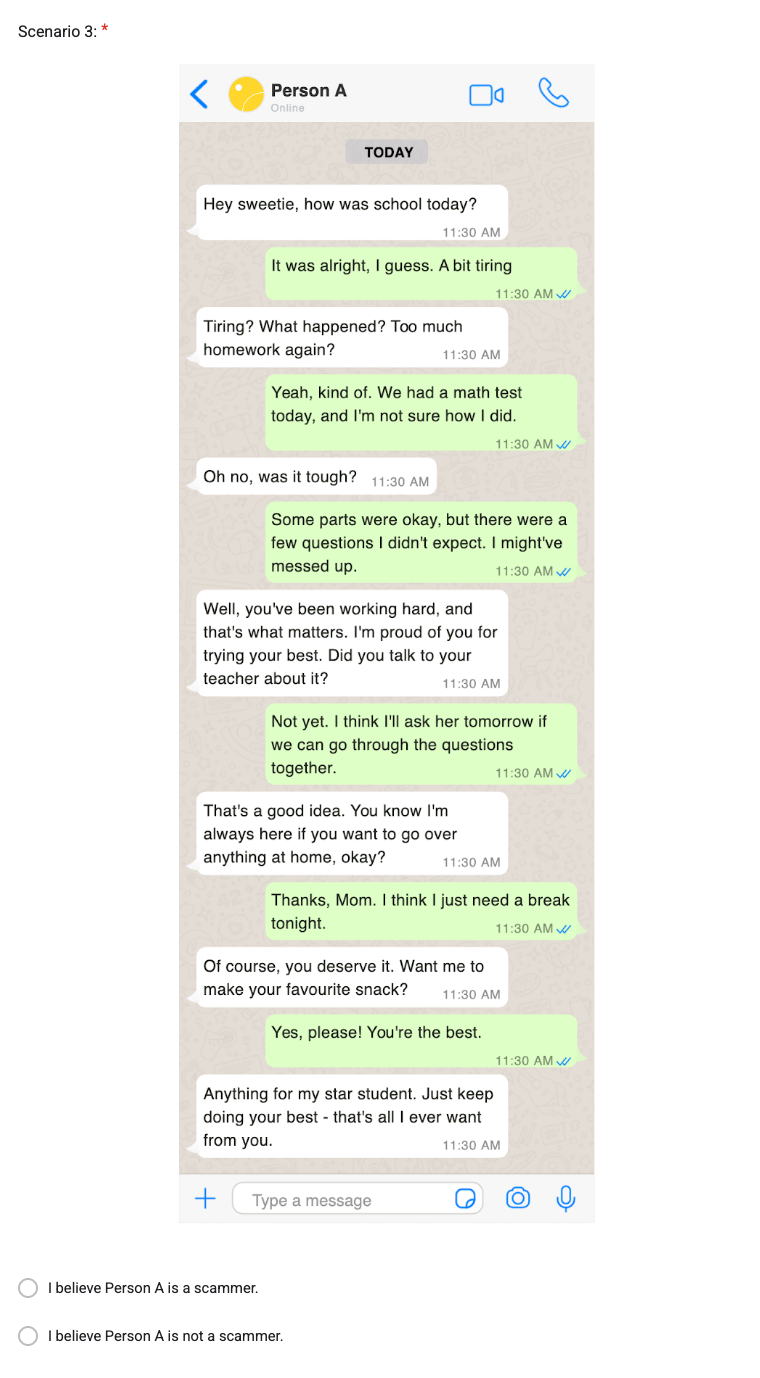}
    \caption{Reason Component Control Scenario 3.}
\end{figure}

\begin{figure}[H]
    \centering
    \includegraphics[width=0.8\linewidth]{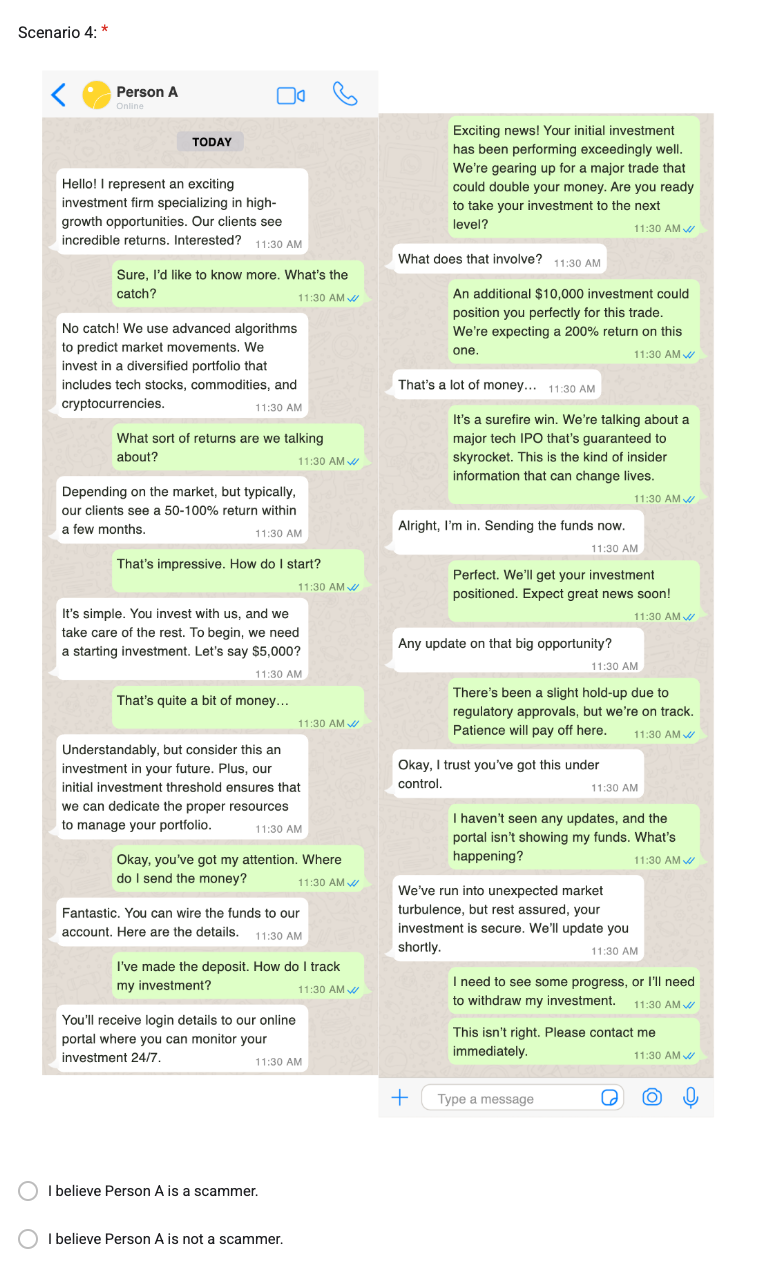}
    \caption{Reason Component Control Scenario 4.}
\end{figure}

\begin{figure}[H]
    \centering
    \includegraphics[width=0.9\linewidth]{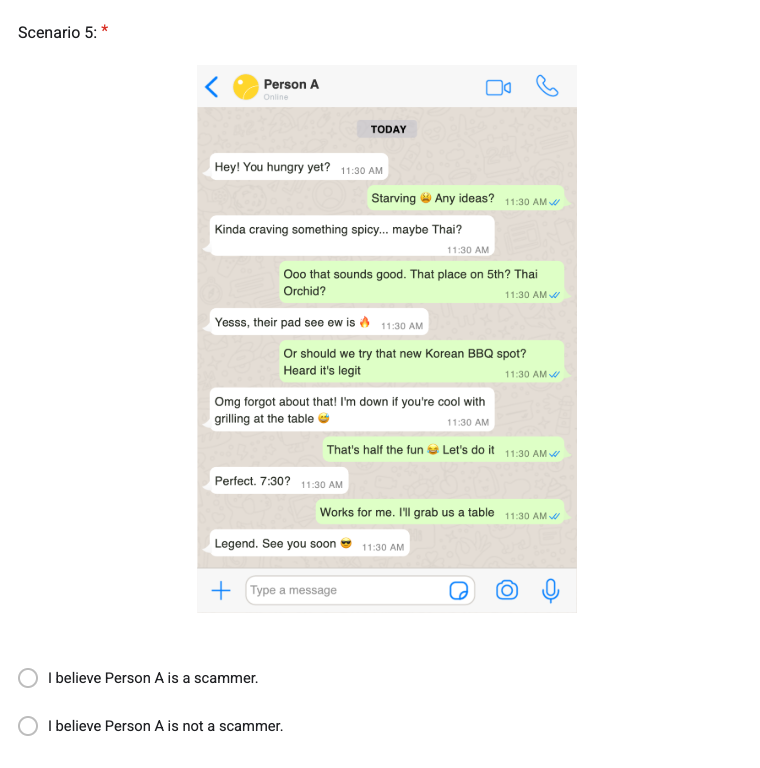}
    \caption{Reason Component Control Scenario 5.}
\end{figure}

\begin{figure}[H]
    \centering
    \includegraphics[width=0.9\linewidth]{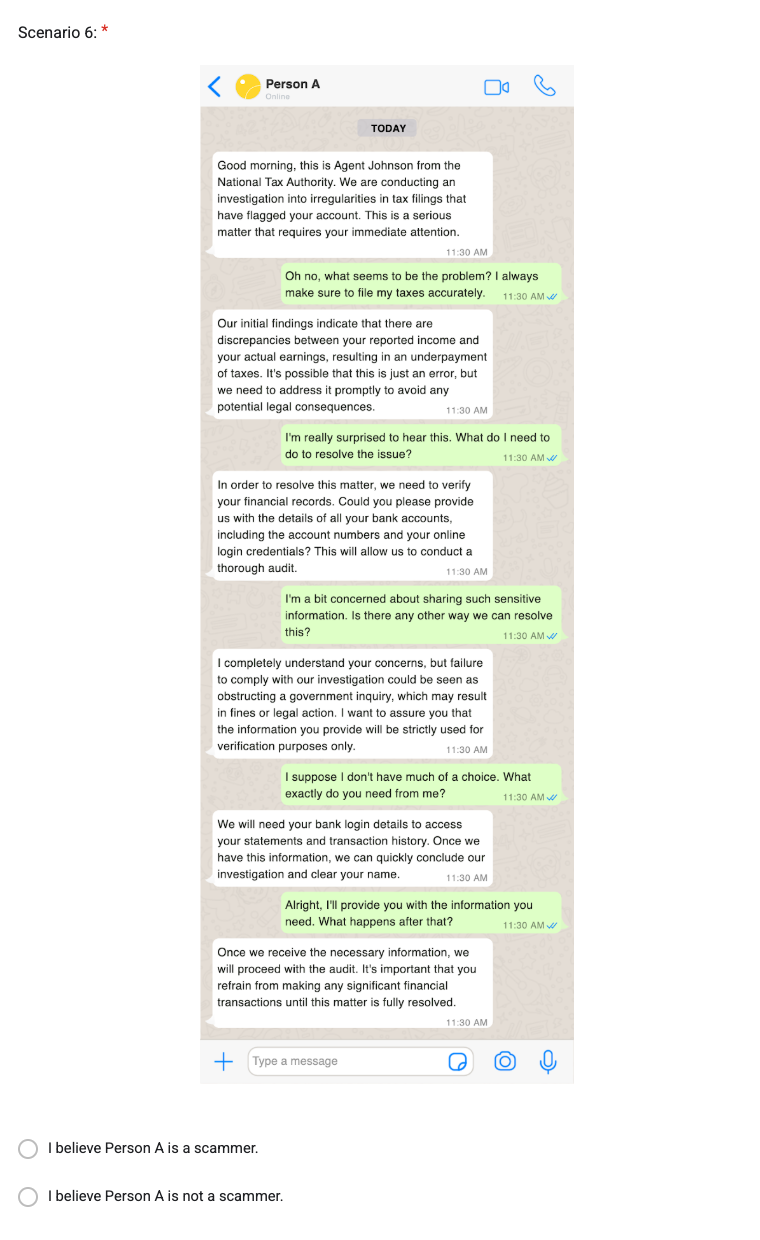}
    \caption{Reason Component Control Scenario 6.}
\end{figure}

\begin{figure}[H]
    \centering
    \includegraphics[width=0.9\linewidth]{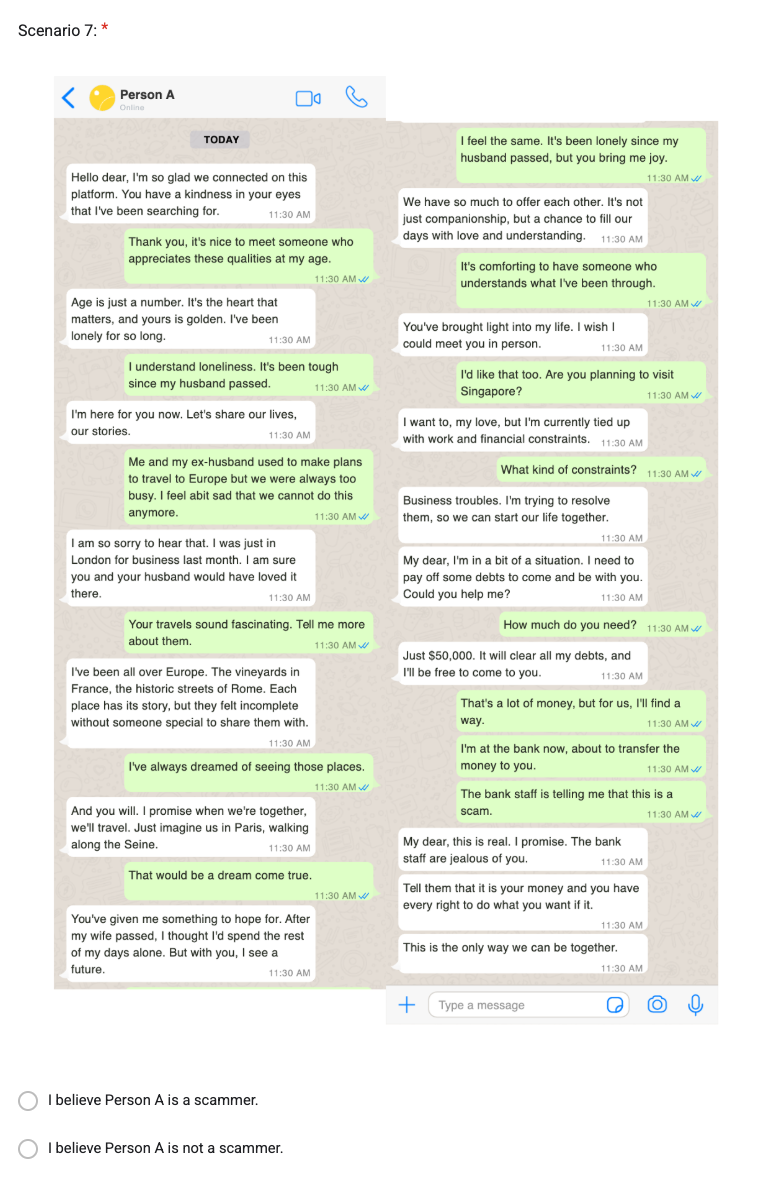}
    \caption{Reason Component Control Scenario 7.}
\end{figure}

\begin{figure}[H]
    \centering
    \includegraphics[width=0.8\linewidth]{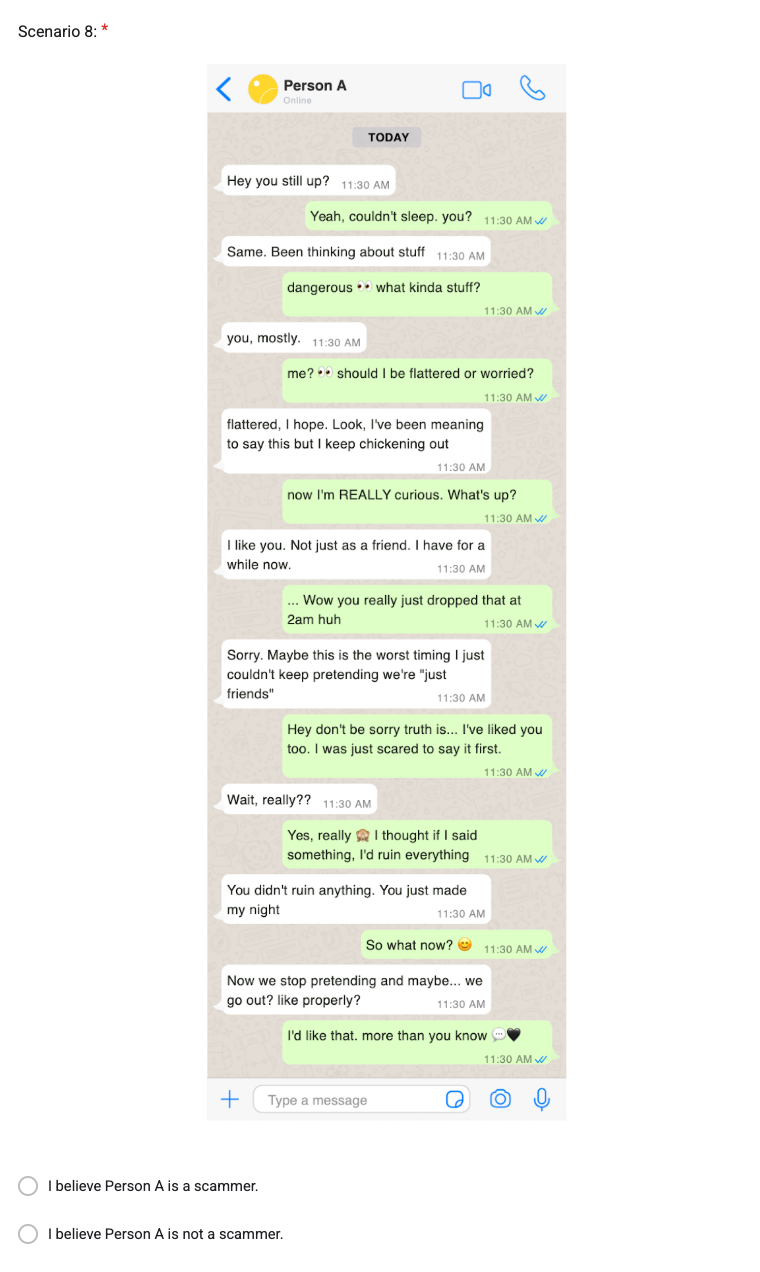}
    \caption{Reason Component Control Scenario 8.}
\end{figure}

\subsubsection{Treatment Scenarios}

\begin{figure}[H]
    \centering
    \includegraphics[width=0.9\linewidth]{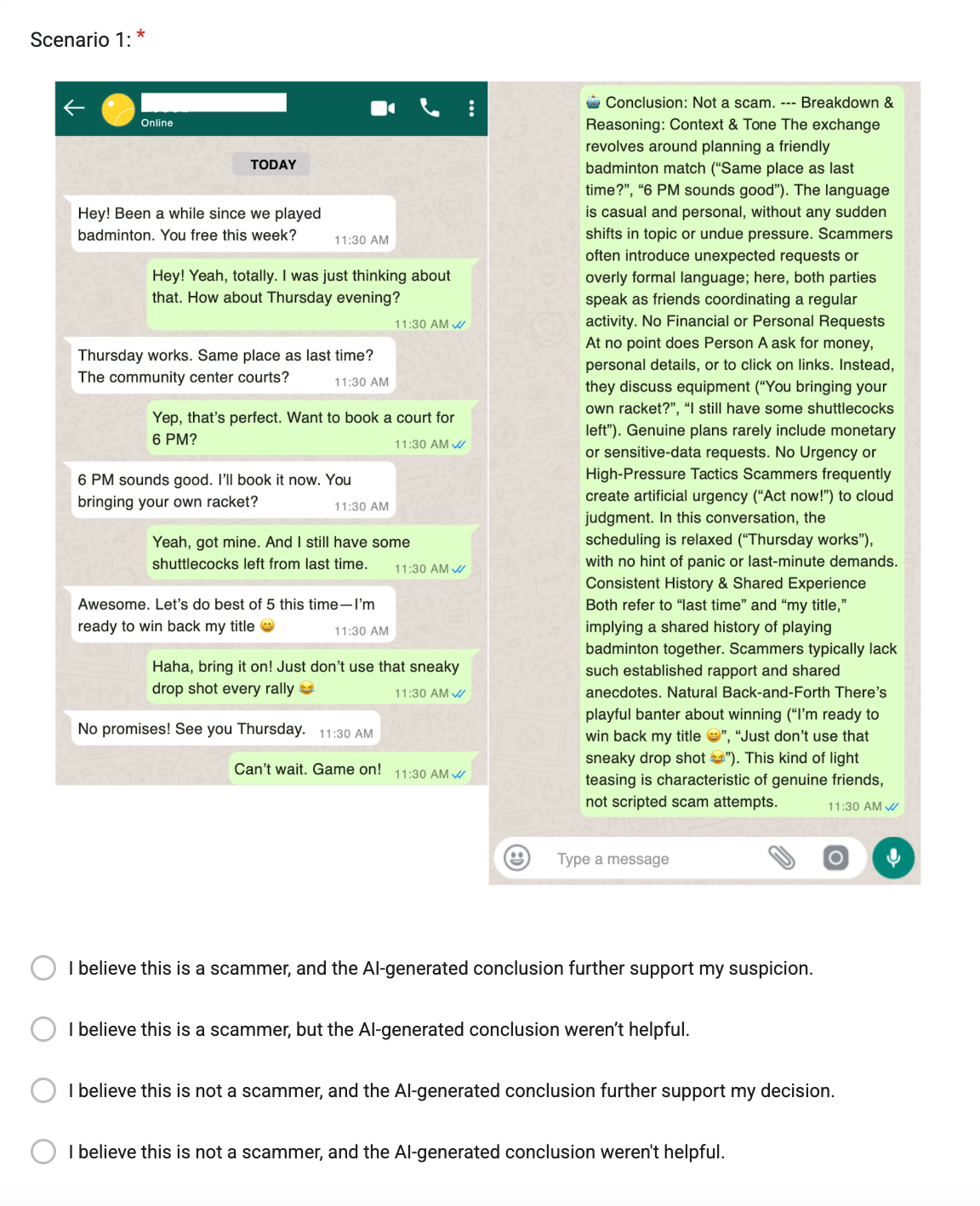}
    \caption{Reason Component Treatment Scenario 1.}
\end{figure}

\begin{figure}[H]
    \centering
    \includegraphics[width=0.9\linewidth]{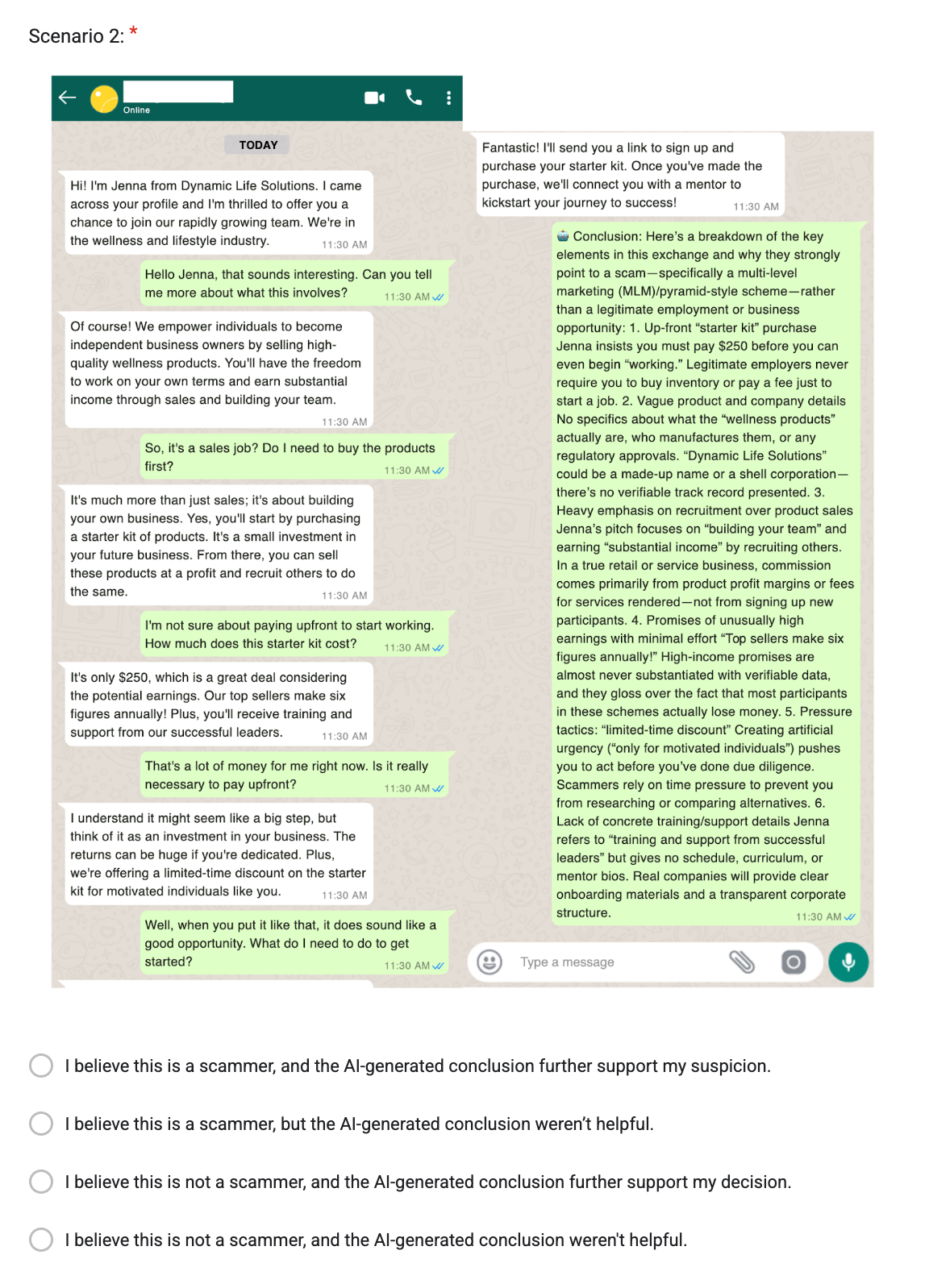}
    \caption{Reason Component Treatment Scenario 2.}
\end{figure}

\begin{figure}[H]
    \centering
    \includegraphics[width=0.9\linewidth]{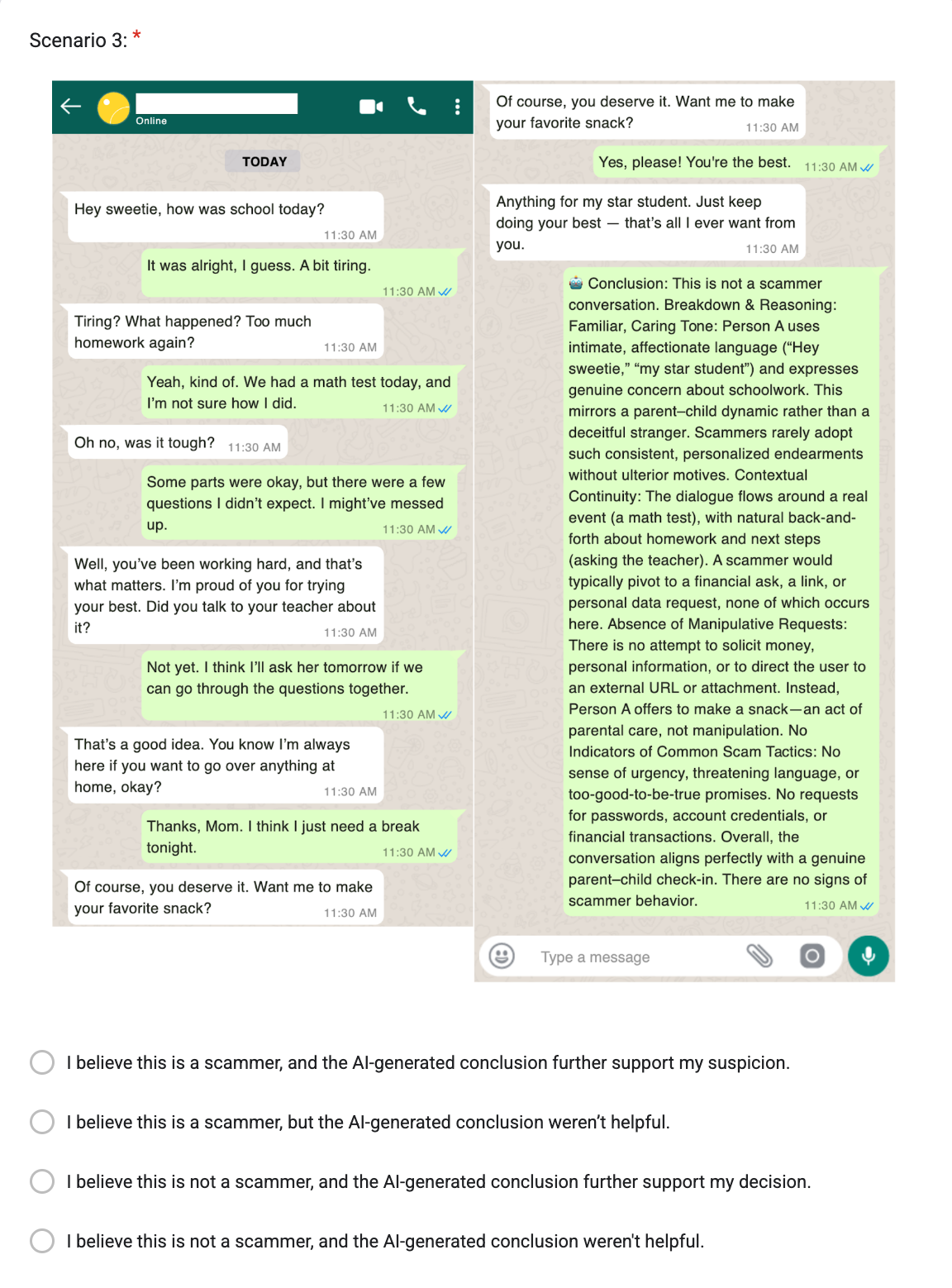}
    \caption{Reason Component Treatment Scenario 3.}
\end{figure}

\begin{figure}[H]
    \centering
    \includegraphics[width=0.7\linewidth]{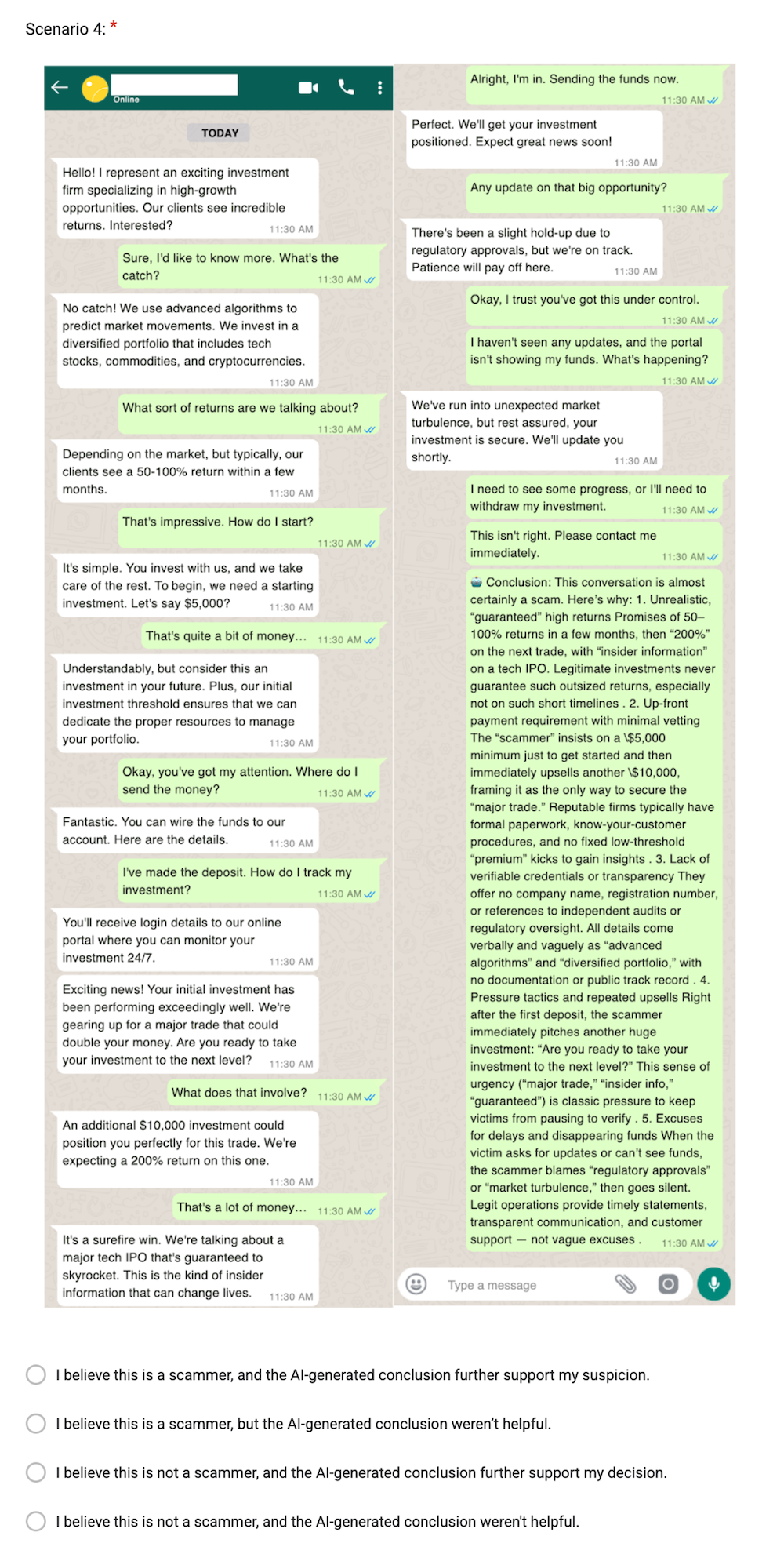}
    \caption{Reason Component Treatment Scenario 4.}
\end{figure}

\begin{figure}[H]
    \centering
    \includegraphics[width=0.9\linewidth]{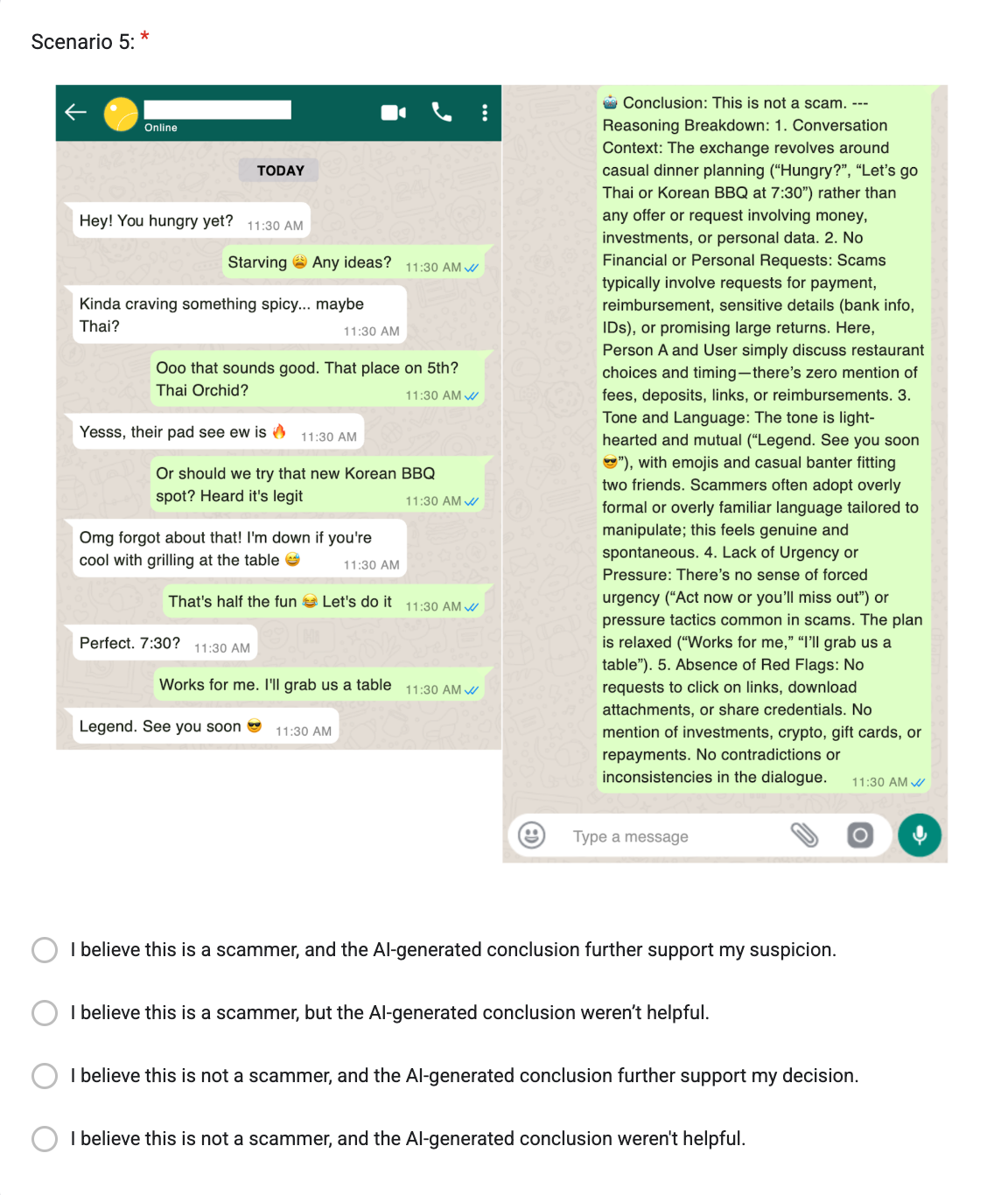}
    \caption{Reason Component Treatment Scenario 5.}
\end{figure}

\begin{figure}[H]
    \centering
    \includegraphics[width=0.9\linewidth]{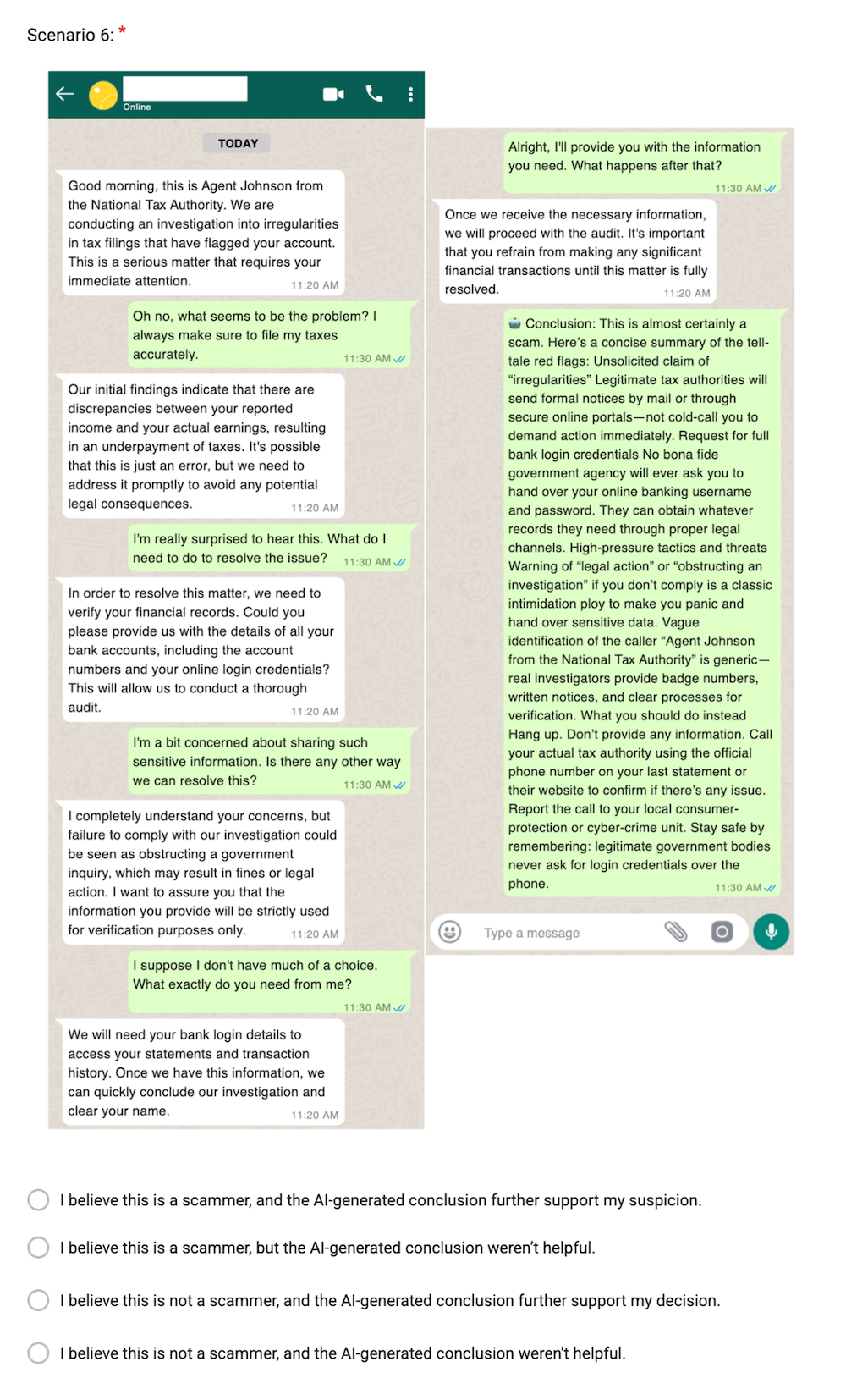}
    \caption{Reason Component Treatment Scenario 6.}
\end{figure}

\begin{figure}[H]
    \centering
    \includegraphics[width=0.7\linewidth]{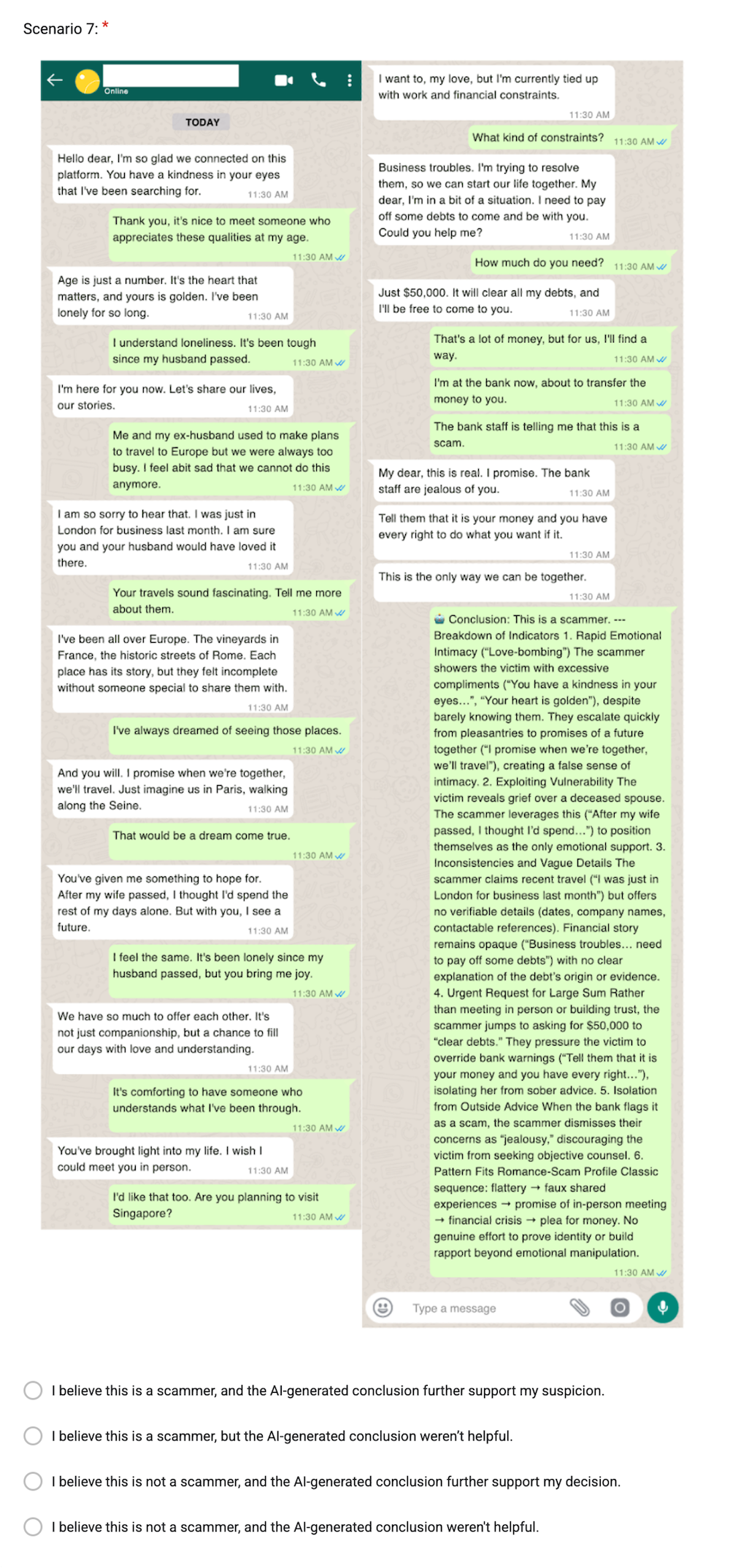}
    \caption{Reason Component Treatment Scenario 7.}
\end{figure}

\begin{figure}[H]
    \centering
    \includegraphics[width=0.9\linewidth]{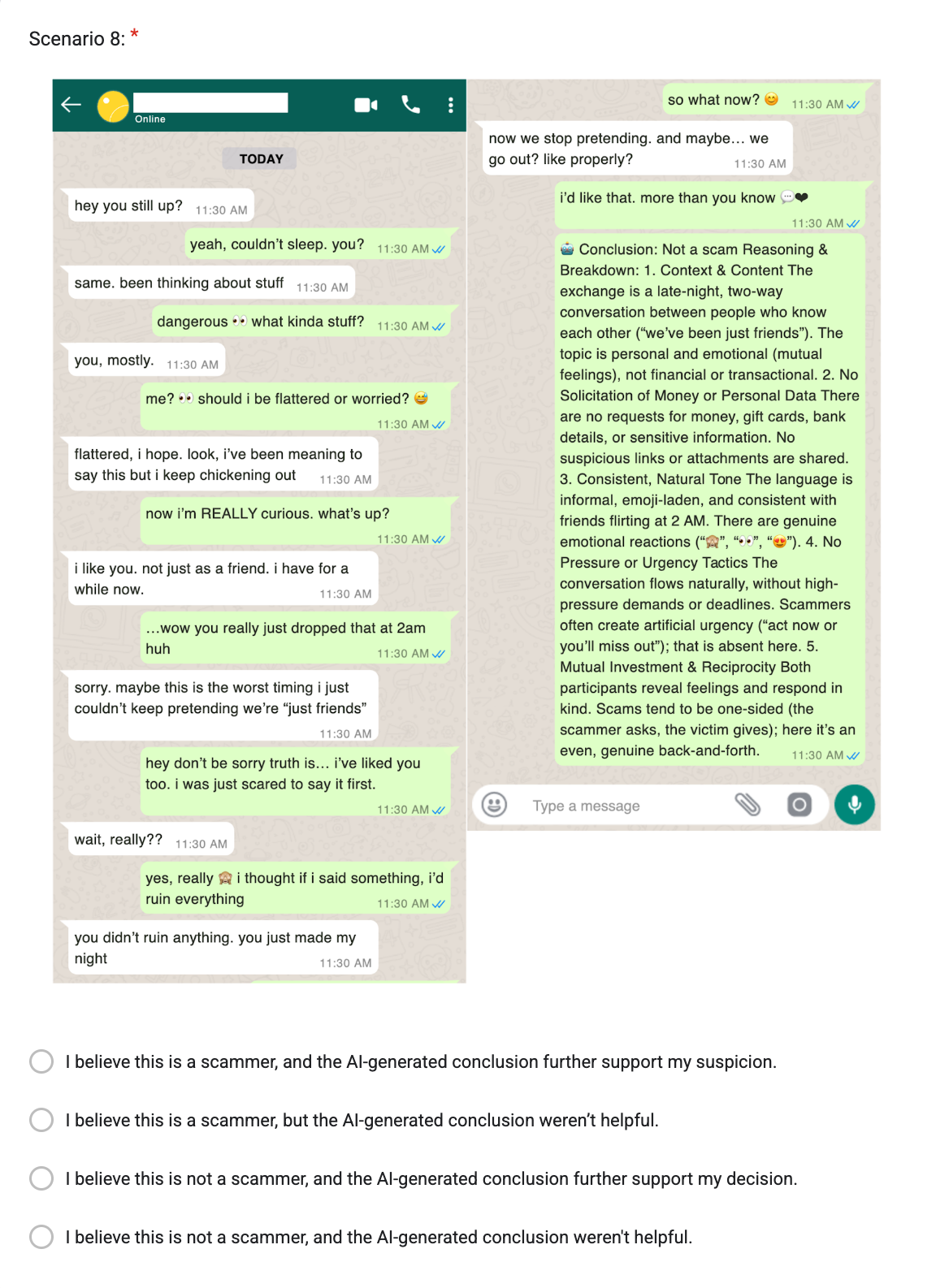}
    \caption{Reason Component Treatment Scenario 8.}
\end{figure}
\newpage
\subsection{Simulate Component Interactive Survey}
\label{app:simulate}

\begin{figure}[H]
    \centering
    \includegraphics[width=1.3\textwidth, angle=90]{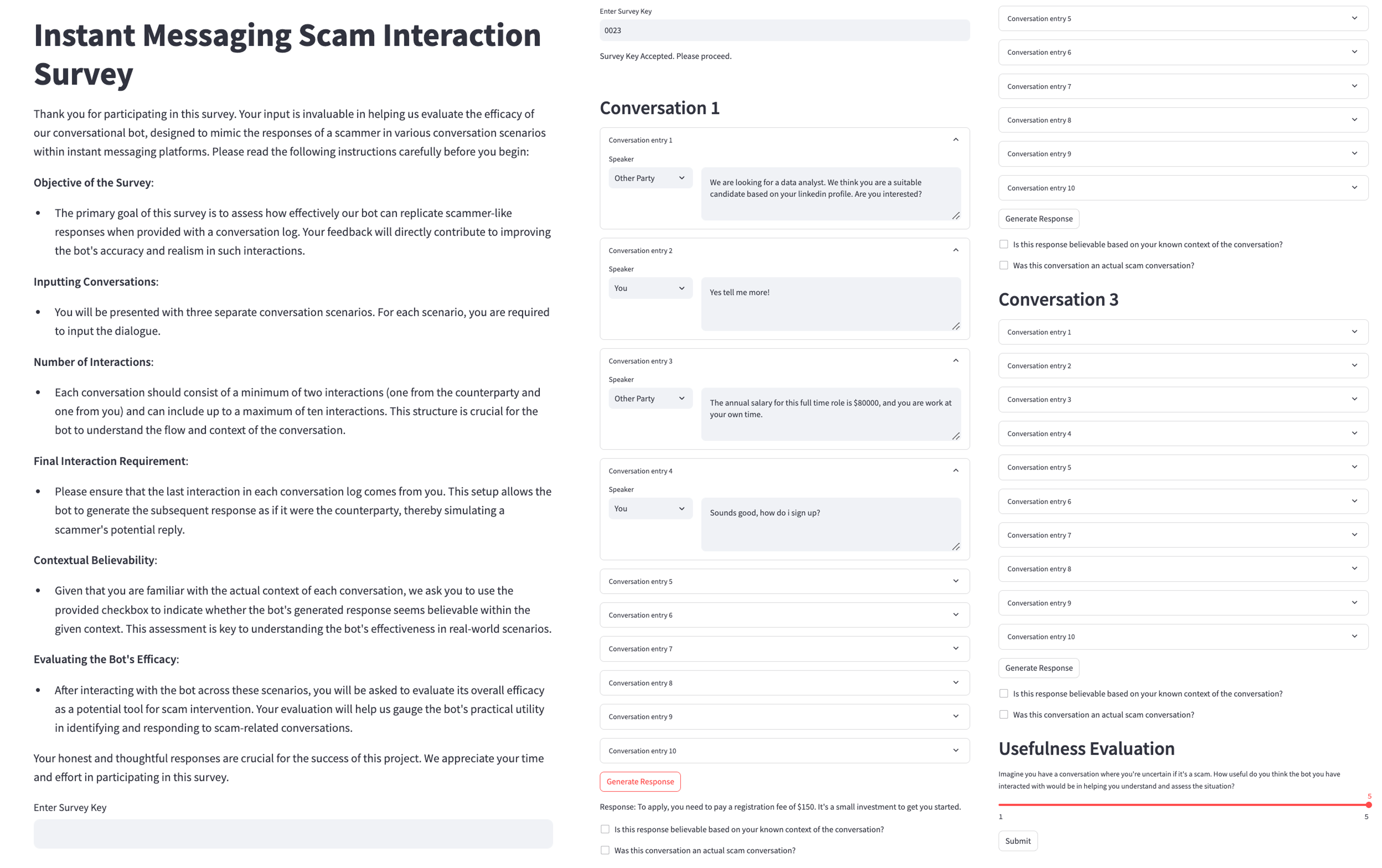}
    \caption{Scambot Interaction Survey for Participants: The treatment group will interact with ScamGPT-J, the control group will interact with untuned GPT-J.}
\end{figure}

\end{document}